\documentclass[12pt]{article}
\pdfoutput=1
\usepackage{cmap}
\usepackage{isoent,amssymb,revsymb}
\let\oldtimes\times
\let\oldomega\omega

\usepackage[utf8]{inputenc}  %  newer: inputenx
\usepackage[LAE]{fontenc}    %  persian/farsi
\usepackage[X2,T3,T2A,T1]{fontenc}
\usepackage[farsi,arabic,french,ukrainian,russian.ancient,russian,ngerman,english]{babel}
\DeclareUnicodeCharacter{0462}{\CYRYAT}
\DeclareTextSymbolDefault{\CYRYAT}{X2}
\DeclareUnicodeCharacter{0463}{\cyryat}
\DeclareTextSymbolDefault{\cyryat}{X2}

\let\times\oldtimes
\let\omega\oldomega

\usepackage[dvipsnames]{xcolor} 

\usepackage{parcolumns}
\usepackage{tikz}

\usepackage{mxedruli}

\usepackage{CJKutf8}
\newcommand{\zhs}[1]{\begin{CJK}{UTF8}{gbsn}#1\end{CJK}}
\newcommand{\zht}[1]{\begin{CJK}{UTF8}{bsmi}#1\end{CJK}}
\newcommand{\jap}[1]{\begin{CJK}{UTF8}{ipxm}#1\end{CJK}}

\newcommand*\justify{%
  \fontdimen2\font=0.4em% interword space
  \fontdimen3\font=0.2em% interword stretch
  \fontdimen4\font=0.1em% interword shrink
  \fontdimen7\font=0.1em% extra space
  \hyphenchar\font=`\-% allowing hyphenation
}

\usepackage[square,comma,numbers,compress]{natbib}

\newcommand{\origttfamily}{}
\let\origttfamily=\ttfamily
\renewcommand{\ttfamily}{\origttfamily \hyphenchar\font=`\-}

\newcounter{firstbib}

\newcounter{extrafoot}

\usepackage[pdfpagelabels=false,linktocpage=true,bookmarksnumbered=true,breaklinks=true,pagebackref=true,colorlinks=true,urlcolor=Mahogany,citecolor=Mahogany,linkcolor=Mahogany]{hyperref}

\usepackage{hypernat}

\newcounter{dummy}

\begin{document}
\bibliographystyle{plainnat}
\begin{flushright}
{\tt $\langle$\href{https://hal.archives-ouvertes.fr/hal-01638181}{hal-01638181}$\rangle$}
\end{flushright}

\vspace{3.5cm}

\renewcommand{\thefootnote}{\fnsymbol{footnote}}
\begin{center}
{\LARGE\baselineskip0.9cm
Photon-photon scattering and related phenomena. 

{\large Experimental and theoretical approaches: The early period}\\[1.5cm]}

{\large 
K. Scharnhorst\footnote[2]{\begin{minipage}[t]{12cm}
E-mail: {\tt k.scharnhorst@vu.nl},\hfill\ \linebreak
ORCID: \url{https://orcid.org/0000-0003-3355-9663}
\end{minipage}}
}\\[0.3cm]

{\small 
Vrije Universiteit Amsterdam,
Faculty of Sciences, Department of Physics and Astronomy,
De Boelelaan 1081, 1081 HV Amsterdam, The Netherlands}\\[1.5cm]

\begin{abstract}
We review the literature on possible violations of the superposition
principle for electromagnetic fields in vacuum from the earliest 
studies until the emergence of renormalized QED at the 
end of the 1940's. The exposition covers experimental work 
on photon-photon scattering and the propagation of light in 
external electromagnetic fields and relevant theoretical work
on nonlinear electrodynamic theories (Born-Infeld theory and QED)
until the year 1949. To enrich the picture, pieces of reminiscences 
from a number of (theoretical) physicists on their work in this 
field are collected and included or appended.
\end{abstract}

\end{center}

\renewcommand{\thefootnote}{[\alph{footnote}]}
\thispagestyle{empty}

\newpage
\section*{}
\addcontentsline{toc}{section}{Contents}

\tableofcontents

\newpage
\section{\label{introduction}Introduction}

\noindent
Light has catched the imagination of men since the 
earliest historic times and theories on its nature 
probably have an even long history. The physical world 
around us to a large extent is being perceived by us
by means of (visible) light and also the discourse of physical 
theory-building is strongly influenced by our anthropomorphic
capability ``to see''. A hypothetic intelligent species 
without such a visual capability probably would approach
physical theory-building in a different manner than we do
- trying to figure out the nature of that 
``dark matter'' that is visible to mankind.
To us, light and its interaction with (charged) matter is 
prototypical, and so is the most advanced to date theory describing 
it - quantum electrodynamics (QED). Our visual perception
of the world (and also many physical experiments) is resting
on two pillars: the interaction of light with charged particles whose
strength is ruled by the fine structure constant, and the (for all
practical purposes) non-interaction of light with itself 
described by the superposition principle. In the literature,
various formulations of the superposition principle can be found. 
As we are both concerned with experimental and theoretical work to us the 
superposition principle is more than a mathematical rule 
related to the linearity of the equations describing a physical 
phenomenon. Quite generally, we understand the superposition
principle as the principle that two physical influences can
be superposed without impeding each other. In this sense, in our 
context it can be understood as a specific non-interaction principle --
a non-self-interaction principle.\\

That the superposition 
principle for light is something special to be thought about has been
recognized long ago. In 1604, Johannes Kepler wrote in his book
{\it Ad Vitellionem Paralipomena, Quibus
Astronomiae Pars Optica Traditur}\footnote{Proposition 26 (\citep{1604kepler},
p.\ 23, reprint (1859) p.\ 142, reprint (1939) p.\ 32: cf.\ our 
App.\ \ref{originals}, p.\ \refstepcounter{dummy}\pageref{akepler};
English transl.\ \citep{2000kepler}, p.\ 37)}\label{kepler}:
``{\it The rays of light neither mutually color each other, 
nor mutually illuminate each other, nor mutually impede each other 
in any way.} $\ldots$ This is just like 
one physical motion's not impeding another.'' It seems that
Kepler is the first to explicitly mention this property of light:
His intellectual predecessor, Erazmus Witelo (Vitellio, Vitello;
13th century) Kepler refers to
in the title of his book does not mention this property in book II
of his {\it Perspectiva} \citep{1535witelo,1991unguru} where
one would expect such an observation to occur\footnote{It is rather 
Ibn al-Haytham (Alhazen; ``the father of optics'') who might be considered
as having observed and described earlier the superposition principle
for light in some rudimentary form. A.\ I.\ Sabra notes
(\citep{1967sabra}, p.\ 207, footnote 29): ``Ibn al-Haytham 
described an experiment (the well-known
{\it camera obscura experiment}) to show that beams of coloured light 
do not mix, and therefore do
not affect one another, when they meet in space 
(at the opening of `the dark place').'' Ibn al-Haytham described his
conclusions the following way
(For the Arabic text see \citep{1983sabra}, Latin: \citep{1572risner},
book I, chap.\ 5, item 29, p.\ 17,
cf.\ our App.\ \ref{originals}, p.\ \refstepcounter{dummy}\pageref{aalhaytham}.
For a reprint of the Latin text see \citep{2001smith1}, book I, chap.\ 7,
p.\ 57, item [6.87].
We quote here from the English translation \citep{2001smith2}, 
book I, chap.\ 7, p.\ 380, item [6.87];
for another English translation see \citep{1989sabra}, vol.\ I, 
book I, chap.\ 6, item I 116b, p.\ 91.):\label{alhaytham}
``Therefore, the lights do not mix in air; instead, each of them extends 
along straight lines; and those lines are parallel, or they intersect, 
or they have various [other] orientations. And the form of each 
light-source radiates along all the [straight] lines 
that can be extended from it through the air, and in accord with 
this [the resulting forms of light] do not mingle in the air, 
nor is the air tinted by them; rather, they merely pass through 
its transparency, and the air does not thereby become transformed.''.
Witelo covers the camera obscura experiment by al-Haytham in 
an abbreviated manner without clearly expressing 
the superposition principle aspect we are interested in 
(cf.\ \citep{1535witelo}, book II, proposition 5, p.\ 40, 
reprint \citep{1572risner}, p.\ 64, 
reprint \citep{1991unguru}, pp.\ 242/243, 
English translation: \citep{1991unguru}, p.\ 47).}. 
Somewhat later than Kepler,
in 1690, Christiaan Huygens commented in his
{\it Trait\'e de la Lumi\`ere}\footnote{\citep{1690huygens}, p.\ 20:
cf.\ our App.\ \ref{originals}, p.\ \refstepcounter{dummy}\pageref{ahuygens};
English transl. \citep{1912huygens}, pp.\ 21/22 (We owe this quote
Vavilov \citep{1928vavilov}.).}\label{huygens}:
``Another property of waves of light, and one of the most marvellous,
is that when some of them come from different or even from opposing
sides, they produce their effect across one another without any
hindrance.'' At the beginning of the 19th century the 
superposition principle led Thomas Young to the formulation of the principle 
of interference within the framework of the wave theory of
light \citep{1991kipnis}. However, while within the framework of the wave
theory of light the superposition principle must have been 
more or less natural, for 
the proponents of the rival model -- the corpuscular model of light 
developed by Descartes and Newton --
the superposition principle required special consideration. 
Serge\u\i\ I.\  Vavilov
writes on this account\footnote{\citep{1928vavilov}, p.\ 555, 
reprint \citep{1954vavilov1}, p.\ 234: 
cf.\ our App.\ \ref{originals}, p.\ \refstepcounter{dummy}\pageref{avavilov1};
English transl.\ \citep{1928vavilov}, 
arXiv:1708.06817/ $\langle$hal-01599214$\rangle$,
p.\ T-1 (= P-3).  Also see
\citep{1950vavilov}, part 2, \S\ 2.}\label{vavilov1}:
\setcounter{extrafoot}{\value{footnote}}
\renewcommand{\thefootnote}{\arabic{footnote}}
\setcounter{footnote}{2}
``The reproach often made in the XVIII.\
century to the Newtonian theory 
consisted just in saying that the collisions of the light
corpuscles, i.e., a violation of the superposition,
should be observed.
An answer to this difficulty was the admission of the extreme
smallness of the corpuscles: \guillemotleft I know
-- $\,$L$\,$o$\,$m$\,$o$\,$n$\,$o$\,$s$\,$o$\,$v\footnote{$^{\it 
[K.S.:\ orig.\ footn.]}\,$ 
\begin{otherlanguage}{russian}
М.\ В.\ Ломоносов, Слово о происхождении света. Собрание разных 
сочинений, ч.\
\end{otherlanguage}
%{\cyrrm M.\ V.\ Lomonosov, Slovo o proiskhozhdenii sveta. Sobranie raznykh 
%sochineni\u i, ch}.\ 
III [M.\ V.\ Lomonosov, Slovo o 
proiskhozhdenii sveta. Sobranie raznykh sochineni\u{\i}, ch.\ III], 
1803, p.\ 155.
(K.S.: \citep{1756lomonosov})}$^{\it [K.S.:\ orig.\ footn.]}\,$
wrote addressing the defenders of the corpuscular hypothesis
-- that you divide 
the material of light into such fine 
particles and place them in universal space with so little
density that the whole quantity can be compressed and packed
in the porous crevices of one grain of sand.\guillemotright''.\\

\renewcommand{\thefootnote}{[\alph{footnote}]}
\setcounter{footnote}{\value{extrafoot}}
In the 19th century, the wave theory of light got the upper hand 
over the corpuscular model in 
explaining optical phenomena and James Clerk Maxwell with his 
equations finally brought mathematical clarity to electromagnetic phenomena,
including the propagation of electromagnetic waves and specifically 
visible light. The phenomenon of the non-self-interaction of light
(in vacuum) described, for example, by Kepler and Huygens received 
its mathematical explanation in terms of the linearity of the Maxwell 
equations in vacuum. However, as Faraday discovered in material 
(polarizable) media electromagnetic fields may exert an influence
on each other: In a magnetic field, a rotation of the 
plane of polarization of linearly polarized light propagating along the 
direction of the magnetic field can be observed (Faraday effect). 
Consequently, the linearity of the Maxwell equations has a restricted 
range of validity. Besides the Faraday effect, the discovery
of the (electro-optic) Pockels and Kerr effects and the (magneto-optic)
Voigt, Majorana, and Cotton-Mouton effects (birefringence in a 
magnetic field for propagation of light perpendicular to it) 
showed that material media can be polarized by 
electromagnetic fields in such a manner that the superposition principle 
for electromagnetic fields does not apply for material
(polarizable) media in general (For references concerning these effects see 
\citep{1967palik,1932beams}.). However, matterless space, the vacuum,
remained immune from such effects in the view of physicists.\\

With the advent of the quantum theory at the beginning of the 
20th century the corpuscular model of light
made a surprising return with the emergence of the concept of the photon.
The particle-wave dualism characteristic for all quantum phenomena
led to a dialectical synthesis of the wave and corpuscular models
of light. However, once the particle-like aspects of light had been
recognized the century-old problem of the corpuscular model of light 
reemerged. What about the scattering of corpuscles of light -- 
photons -- among each other? Qualitative investigations of this problem
began to emerge. The first theoretical attempt to study this problem 
was published in 1925 by Konstantin N.\ Shaposhnikov \citep{1925schaposchnikow}
by writing down the quantum mechanical equations for energy-momentum
conservation for the photon-photon scattering process.
Shaposhnikov's main purpose was to emphasize that these equations
allow solutions which correspond to a non-violation of
the superposition principle. In 1926,
Louis de Broglie then pointed out (\citep{1926debroglie}, Chap.\ XI, Sec.\ 2,
pp.\ 96-98; cf.\ our Appendix \ref{appbrillouin}, 
p.\ \refstepcounter{dummy}\pageref{appbrillouin}) 
that these equations also allow
solutions which correspond to nontrivial scattering events between
two photons in deviation from the superposition principle. This
point has further been elaborated in a short contribution by
Arthur L.\ Hughes and George E.\ M.\ Jauncey \citep{1928hughes} in 1928. 
Vavilov, in 1928, discussed the status of the superposition principle for light
from an experimental point of view by reporting on laboratory experiments
of his own and invoking extraterrestrial considerations by means of
an analysis of the problem of the solar corona (\citep{1928vavilov},
cf.\ our subsecs.\ \ref{scatteringexp} and \ref{astro}). The article
by Vavilov was followed by a public exchange of comment and response
between Shaposhnikov and Vavilov \citep{1929shaposhnikov,1929vavilov}.
In his reply (\citep{1929vavilov}, p.\ 395, 
p.\ T-3 (= P-5) of the English transl.), 
giving credit for this consideration to Yakov I.\ Frenkel Vavilov pointed out 
that the superposition principle for light should be violated, in 
principle, in any case due to the (negligibly small) 
gravitational interaction of two quanta of light.
The same argument had been made in 1928 already by 
L\'eon\ Rosenfeld and Enos E.\ Witmer (\citep{1928rosenfeld}, p.\ 521)
in an article in which they considered from a qualitative point of view
the role of collisions among photons for the discussion of black 
body radiation.\\

However, the physical mechanism
for photon-photon interaction processes and their
quantitative details remained beyond consideration for a couple of further
years. In 1933, Otto Halpern \citep{1933halpern} finally proposed that
virtual electron-positron pairs could be at the origin of photon-photon
collisions. While this clarified the qualitative picture to be
applied for the description of photon-photon scattering, only the 
early development of quantum 
electrodynamics (QED) provided physicists with the theoretical tools for
a quantitative answer: It was finally given by two students 
of Werner Heisenberg,
Hans Euler and Bernhard Kockel, who calculated in 1935 the leading nonlinear 
corrections to the Maxwell equations in vacuum \citep{1935euler}. 
Within the framework of QED, it
turned out that photon-photon scattering as a characteristic feature of
a nonlinear electrodynamic theory has a very low probability for
all practical purposes. Once the theoretical picture had been
established the challenge emerged to demonstrate experimentally
consequences of the violation of the linearity of Maxwell equations
in vacuum, i.e., of the violation of the superposition principle for light
(or, speaking more generally, for electromagnetic fields).
The processes of Delbr\"uck scattering (elastic scattering of a photon
in the Coulomb field of a nucleus) and photon 
splitting (in an external field) have meanwhile experimentally 
been confirmed (cf.\ \citep{1994milstein,2003lee}). 
The phenomenon of photon-photon scattering itself, however,
has escaped experimental observation for many decades and has been
demonstrated with sufficiently high statistics by 
the ATLAS collaboration in 2019 only \citep{2019atlas}
(after evidence for it had been seen for the first time at CERN
by the ATLAS collaboration in 2017 \citep{2017atlas} and then 
the CMS collaboration in 2018 \citep{2018cms}). This remarkable success
underscores that the demonstration of such a fundamental 
physical phenomenon as photon-photon scattering is still 
a problem at the edge of current experimental and observational 
capabilities (for some other recent development see 
\citep{2017mignani}).\\

There is a continuous current, varying in intensity over time, of
experimental and theoretical research in the field of nonlinear 
electrodynamic phenomena in vacuum. The main motivation for experimental
work derives from the prominent role electromagnetic phenomena play
in the physical world we live in. Theoretical studies often have
a wider set of motivations which are often linked to the changing 
directions of theoretical thinking. New research can successfully 
only be based on the knowledge of the achievements and failures of 
past generations of scientists. While the main publications from the
past in the field of nonlinear electrodynamics in general and of QED 
in particular are well known many details of the early thinking
and experimentation concerning photon-photon scattering and 
related phenomena seem to be largely forgotten. It is the purpose of the 
present review to collect as completely as possible the early
literature on this subject as a source and inspiration for 
further research. The time span covered reaches from the earliest 
publications that can be linked to the problem of nonlinear
electrodynamics up to 1949, i.e., the end of the 1940's,  
which roughly marks the begin of a new period  
in the history of theoretical physics with the emergence of 
renormalized quantum electrodynamics. Of course, the year 1949 is somewhat
arbitrarily chosen based on an analysis of the occurrence of relevant
literature references. In a certain way, the short review article 
by Kunze in 1949 \citep{1949kunze} marks the end of the period
the present exposition is concerned with.\\

The present review has a mixed character. On one hand, 
primarily it is written as a literature review. 
On the other hand, it is also a piece 
of science history. To illuminate various social and science 
history aspects we quote extensively from a number of different
sources. This rather unconventional set-up serves a twofold purpose.
First of all, it provides the reader with a more complex and
first-hand picture of the early developments. The second purpose consists
in collecting the widely scattered pieces of text for the convenience
of the reader in one place. Some of the early pieces of text have been 
translated into English to make them more easily accessible to the current
and future generations of physicists.\\

The review is split in a section on early experimental work and
a section on theoretical studies in nonlinear electrodynamics. 
Corresponding to the historic course of research, 
we cover both Born-Infeld electrodynamics as a (mainly) classical field
theory and quantum electrodynamics which gives rise to an effective
action by means of which contact can be made with a classical 
theory as Born-Infeld electrodynamics. However, the relation between
both theories is more than a historic one. Also modern research benefits
from viewing both theories as special instances of nonlinear 
electrodynamics because this allows to elucidate more easily general
principles. The early literature on Born-Infeld electrodynamics
has been included in the list of references somewhat more broadly 
than that from QED. As there does not seem to exist any comprehensive 
review of the early developments of Born-Infeld theory this 
choice seemed to be appropriate.\\

The list of references is split into two parts. One part
arranged according to the year of publication
contains the main body of references related to photon-photon scattering
and related nonlinear electrodynamic phenomena up to 1949. Most of 
these references are cited and commented in the main body of our text, 
very few references are only listed in the bibliography 
for completeness and are not mentioned elsewhere 
\citep{1935schroedinger1,1936blokhintsev,undatedschroedinger1,undatedschroedinger2,undatedschroedinger3,1933smirnov}.
The references in the main bibliography 
represent the present literature review in the narrow sense.
All further references cited are collected in a separate list of (auxiliary) 
references. We deliberately cite literature after 1949 only once this
seems absolutely to be necessary for the presentation or discussion.
Consequently, most of the more recent relevant literature from 1950 
onwards is absent. The main list of references is supplemented 
by a list of author names where for identification purposes the first
names and, if known, the dates of birth and death are given. If 
available, we provide for each name a link to further information.
For the convenience of the reader, we rely primarily on 
Wikipedia articles. Preference has been given 
to English language articles. If an English language articles was not 
existent we have tried to make some other reasonable language choice.
If no Wikipedia article was available at all, a link to some other
information page on the Internet has been selected. 
In a number of cases, no
appropriate information on the Internet was available. We then make
reference to some printed obituary. In a small number of cases, we had to
restrict ourselves to links to VIAF (Virtual International Authority File, OCLC)
or some related source. Finally, for very few author names, we were 
unable to trace any further author information.\\

\section{Experiment}

In the period under consideration, experimental work concerning 
the study of pho\-ton-photon scattering and related phenomena has been fueled
by a number of different developments in physics. There have been
two main directions of experimental work: 

\begin{itemize}
\item[1.] 
Experimental work 
concerning the study of photon-photon scattering in the narrow sense,
i.e., direct searches for experimental signatures of the scattering
of light by light. 
\item[2.] 
The experimental study of the propagation of light (photons) in strong 
(electromagnetic) fields. This main direction of research 
can historically be divided into 3 subcategories:

\begin{itemize}
\item[A.]
The dependence of the propagation of light on the intensity of the light wave. 
\item[B.]
The propagation of light (photons) in macroscopic, constant 
magnetic and electrical fields. 
\item[C.]
The propagation/scattering of photons in the Coulomb field of nuclei. 
\end{itemize}
\end{itemize}
\noindent
Finally, it seems to be appropriate to add as a further direction of 
experimental work:
\begin{itemize}
\item[3.]
Considerations related to extraterrestrial and astronomical observations.
\end{itemize}

\noindent
In the following subsections arranged according to the above classification
we will give an overview over the experimental and observational 
work up to the end of the 1940's. Early experimental and observational work
on photon-photon scattering and related phenomena has been scarce.
We have tried to cover the relevant literature as completely as possible
and in the subsections below we will refrain from making any further
statements to the effect that no further relevant work is known to us.\\

\subsection{\label{scatteringexp}Photon-photon scattering}

The first attempt to directly observe the scattering 
of photons by photons in an experiment seems to have been undertaken
in 1928 in the Soviet Union by S.\ I.\ Vavilov (Institute of Biological
Physics, Moscow). In a contribution read to
the 6.\ Congress of Russian Physicist in Moscow on August 6, 1928
Vavilov reports\footnote{\citep{1928vavilov}, \S\ 2, pp.\ 556/557, 
reprint \citep{1954vavilov1}, p.\ 236:
cf.\ our App.\ \ref{originals}, p.\ \refstepcounter{dummy}\pageref{avavilov2};
English transl.\ pp.\ T-2/T-3 (= P-4/P-5). Also cf.\ his note 
\citep{1930vavilov}.}\label{vavilov2}:
``Under laboratory conditions, the highest\  
$\,$i$\,$n$\,$s$\,$t$\,$a$\,$n$\,$t$\,$a$\,$n$\,$e$\,$o$\,$u$\,$s$\,$\ 
radiation densities can certainly be obtained
by means of the light of a condensed spark. 
Concentrating this light with a lense,
instantaneous values of the light energy density 
that exceed the corresponding value on the surface of the sun
can be achieved easily. 
In this case, the average density is small due to the short duration and
rareness of the sparks, but the hypothetical effect of the ``collisions''
of light quanta must be proportional to the\  
$\,$s$\,$q$\,$u$\,$a$\,$r$\,$e$\,$\ of 
the instantaneous density, therefore, a spark turns out to be 
considerably more advantageous than, for example, an arc. 
Preliminary experiments with a spark the light of which met in an
evacuated container\
$\,$d$\,$i$\,$d\ $\,$n$\,$o$\,$t\
$\,$u$\,$n$\,$c$\,$o$\,$v$\,$e$\,$r\ $\,$a$\,$n$\,$y\ 
$\,$n$\,$o$\,$t$\,$i$\,$c$\,$e$\,$a$\,$b$\,$l$\,$e\
$\,$s$\,$c$\,$a$\,$t$\,$t$\,$e$\,$r$\,$i$\,$n$\,$g.
These observations have been carried out with the usual
precautions, in front of distant container walls covered with 
black velvet; for control, the experiments have been repeated
with the light of an incandescent lamp that delivered the same\ 
$\,$a$\,$v$\,$e$\,$r$\,$a$\,$g$\,$e$\,$\ radiation 
density; in both cases the result has been equally negative.''.\\

Shortly thereafter, in 1930, A.\ L.\ Hughes and 
G.\ E.\ M.\ Jauncey (Washington U., St.\ Louis, USA)
published a somewhat more elaborate description of a similar
experiment to detect collisions of photons \citep{1930jauncey,1930hughes}.
In difference to Vavilov, they used (intense) light from the 
sun to perform the experiment. Two light beams were allowed to
meet at an angle of $120^{\circ}$ and the observation for scattered
light was performed with the dark-adapted eye
in the plane of the scattering beams at the 
(forward) direction of the bisector of the angle between them. 
In their experiment, Hughes and Jauncey took into account that 
with the chosen geometric set-up light resulting from photon-photon
collisions would emerge with a higher frequency according to
the quantum-mechanical laws of energy-momentum conservation
in the collisions. In the experiment, no experimental sign 
of photon-photon collisions was found and Hughes and Jauncey
give as bound for the cross section $\sigma$ of photon-photon 
scattering $\sigma < 3\times 10^{-20}\ cm^2$.\\

Interestingly, the article by Hughes and Jauncey \citep{1930hughes}
is not only followed by a short note by Vavilov \citep{1930vavilov} pointing
out the results of his earlier experiments \citep{1928vavilov} but also
by a short note by A.\ K.\ Das (Alipore Observatory, Calcutta, India)
\citep{1931das} giving details of an earlier (unpublished) idea 
for performing an experiment to study photon-photon scattering. 
Noting that due to the (likely) miniscule size of the effect photon-photon
scattering cannot be observed by optical methods 
(cf.\ the analogous comment by Vavilov: \citep{1928vavilov}, \S\ 2, p.\ 558, 
p.\ 237 of the reprint, p.\ T-4 (= P-6) of the English translation), 
Das proposed to observe single (scattered) photons 
by means of ``Elektronenz\"ahlrohren''
(electron counter tubes) he had constructed for the investigation of 
gamma and cosmic rays. From a modern point of view, it also interesting
to note that Das envisioned the experiment to be performed with X-rays
or gamma-rays (and not with visible light for which the cross section
for photon-photon scattering is smaller).\\

The experiment in 1930 by Hughes and Jauncey was followed in 1933 
by an analogous experiment performed by F.\ L.\ Mohler (Bureau of
Standards, Washington, D.\ C., USA) \citep{1933mohler} which differed
from that of Hughes and Jauncey geometrically by the angle between the 
scattering light beams ($180^{\circ}$ in the experiment by Mohler
versus $120^{\circ}$ in the experiment by the former; also see the
corresponding comment by Hughes and Jauncey \citep{1934hughes}).
Mohler relied in his experiment on the light of projection lamps
and the observation for scattered photons was performed 
with the human eye in the 
direction perpendicular to the joint axis of the two scattering
light beams. The result of the experiment was negative and Mohler
expressed the result in his article in terms of a limit on the 
cross section $\sigma$ for photon-photon scattering 
($\sigma < 6\times 10^{-17}\ cm^2$).\\

The following logical step in the experimental search for the existence of
photon-photon collisions has been made in two Master theses: the human
eye as detector of light (radiation) is being replaced by a photographic
film. In 1937, B.\ Castaldi (Clark University, Worcester, MA, USA)
investigates the problem of photon-photon scattering for visible light
\citep{1937castaldi} (thesis advisor: P.\ M.\ Roope).
In the experiment of Castaldi, a sodium Lab-arc is used as source
of light (with a wavelength of approximately 589 nm). 
Two beams of light met at an angle of $60^{\circ}$
and the photographic film used as detector of scattered light was placed 
in the plane of the scattering beams at the (forward) direction of 
the bisector of the angle between them (as in the 
experiment by Hughes and Jauncey). In the (final) trial (after the
design phase), the photographic plate was 
exposed for 28 days (672 hours) and then developed. No trace of 
scattered light was found. Castaldi estimates that his method was
661.6 times as sensitive as that of Hughes and Jauncey relying on
the human eye as detector. In 1940, to study the problem of
photon-photon collisions R.\ V.\ Wiegand (Montana State
College, Bozeman, MT, USA) uses X-rays for which, according to
Euler and Kockel \citep{1935euler}, the cross section ($\sim\omega^6$, where
$\omega$ is the frequency of the scattering light quanta)
is larger than for visible light 
\citep{1940wiegand} (thesis advisor: A.\ J.\ M.\ Johnson). 
The experiment has been performed with X-rays 
of a wavelength of 1.473 \AA ( = 0.1473 nm)
and for an collision angle of $28^{\circ}\ 8,6^\prime$.
``X-ray film was placed on the six sides of the collision point
and long exposures taken to determine if radiation was given off 
in any direction.'' The maximal exposure
time was 25 hours but ``no ``collision'' radiation of sufficient 
amount to register on the photographic film was found''
(both quotes are from the abstract of \citep{1940wiegand}).\\ 

\subsection{Propagation of light in strong electromagnetic fields}
\subsubsection{\label{intensity}Influence of the intensity on the 
propagation of light}

In the late 19th/early 20th century the question has been studied if 
the speed of light might depend on the intensity of the propagating
light wave. Experiments by J.\ J.\ M\"uller 
(University of Leipzig, Germany) performed with visible light 
and published in 1872 \citep{1872mueller} indicated that the speed of
light (in air) increases with increasing frequency and intensity
(cf.\ the tables on pp.\ 107, 120
of \citep{1872mueller}). The intensity variations studied by M\"uller
were not larger than 1:10. M\"uller used interference experiments
to investigate the dependence of the wavelength $\lambda$
of an incident (monochromatic, with frequency $f$) 
light beam on its intensity and inferred from it a corresponding
variation in the speed of light $c = f \lambda$.
Such a phenomenon of the intensity dependence of 
the speed of light (in vacuum) would clearly 
represent a violation of the superposition principle and would
-- on the theoretical side -- correspond to a Lagrangian for the (gauge)
field describing the light wave which is not quadratic in this field.\\

Contemporary physicists apparently had been aware of the potential 
significance of the results of M\"uller.
Not long after the publication of his work \citep{1872mueller}
they have checked his assertion in other experiments with higher
precision and found no indication for any intensity dependence of the 
speed of light hereby ruling out any noticeable self-interaction 
of a light beam with itself (F.\ Lippich, University of Prague, Austria:
\citep{1875lippich}; H.\ Ebert, University of Erlangen, Germany:  
\citep{1887ebert}; Th.\ E.\ Doubt, University of Chicago, USA: 
\citep{1904doubt}). Ebert \citep{1887ebert} besides giving the results 
of his experiments also uses astronomical arguments (pp.\ 381-383) 
to infer that the speed of light cannot depend on the intensity 
of the light wave in any significant way (cf.\ our subsection \ref{astro}).
Doubt achieving in his experiments light intensity variations 
between 1:43000 and 1:290000 concluded
that any corresponding variation in the speed of light, if any at all, 
must be smaller than roughly one part in one billion.\\

\subsubsection{\label{expconst}Propagation of light in macroscopic, constant 
magnetic and electrical fields}

\vspace{-0.8cm}

\hspace{3.8cm}\footnote{For a different discussion of
the experiments by Watson \citep{1929watson} 
and Farr and Banwell \citep{1932farr,1940banwell}
dealt with in this subsection see p.\ 11 of
\citep{2013battesti}.}

\vspace{0.8cm}

The earliest experiment to be mentioned in this section belongs
to a time before the advent of the special theory of relativity.
In the year 1900, R. A. Fessenden (Western University of Pennsylvania, 
Pittsburgh, USA) relied on an aether model to reason 
(\citep{1900fessenden}, pp.\ 87/88) that an
electrostatic field should lead to an increase in the speed of light in
vacuo in the direction of the applied electric field. Two years later,
Fessenden reported a positive result obtained by him in a
preliminary experiment to check his assertion \citep{1902fessenden}. 
It seems that neither
Fessenden nor any other researcher has later returned to this 
subject. This is not very surprising in view of the fact that 
in 1905 Albert Einstein started to revolutionize thinking
on fundamental physics. While the aether as a vacuum model
had to be abolished even Albert Einstein seems to have
further thought about the possibility that the speed of light in vacuo might
depend on external electromagnetic fields.\\

Information in this respect comes from a reminiscence by P.\ L.\ Kapitsa 
of a conversation he has had with Albert Einstein
on the occasion of a visit\footnote{This meeting took place
at an unknown date, somewhen 
during the period between 1922 and 1934 when Kapitsa worked at the 
University of Cambridge - probably, after 1929 when the experiment
by W.\ H.\ Watson \citep{1929watson} had been performed 
(see further below). The question when this conversation
took place precisely must rest with historians of science.
For an account of the visits of Einstein to Oxford see \citep{2018fox}.}
of the latter to the Cavendish Laboratory of the 
University of Cambridge. His report provides us with an interesting glimpse
into the thinking of Albert Einstein. Kapitsa 
recalls\footnote{\citep{1980kapitsa1}, p.\ 31, 
\citep{1980kapitsa2}, p.\ 40, reprints: 1. \citep{1980kapitsa2}, p.\ 40.
2. \citep{1981kapitsa}, 3.\ ed.\ p.\ 375, 4.\ ed.\ p.\ 374,
English transl.\ \citep{1980kapitsa3}, p.\ 9, 
reprint \citep{1986kapitsa}, p.\ 317; we quote here the English
translation, for the original Russian text see our App.\ 
\ref{originals}, p.\ \refstepcounter{dummy}\pageref{akapitsa}. I owe this 
quote Yu.\ M.\ Poluektov \citep{2017poluektov}, p.\ 6.}\label{kapitsa}:
``In the 1930s, in Cavendish's
laboratory, I developed a method of obtaining magnetic fields
one order stronger than had previously been attained. In a conversation
Einstein tried to persuade me to study experimentally
the influence of a magnetic field upon the velocity of light.
Such experiments had been conducted, and no effect was discovered.
In my magnetic fields it was possible to raise the limit
of accuracy of measurement by two orders of magnitude, because
the effect was dependent on the square of the intensity of
the magnetic field. I protested to Einstein that according to the
existing picture of electromagnetic phenomena, I could not see
from whence such a measurable phenomenon would come. Having
found it impossible to prove the need for such experiments,
Einstein finally said, ``I think that der liebe Gott could not have
created the world in such a fashion that a magnetic field would
be unable to influence the velocity of light.'' Of course, it is
hard to counter that kind of argument.''\\

A first experiment investigating the propagation of light in a 
(constant homogeneous) transverse magnetic field had been carried 
out in 1929 by W.\ H.\ Watson \citep{1929watson} at the Cavendish
Laboratory in Cambridge, UK. This experiment predates the
development of the theory of quantum electrodynamics. The main
motivation for the experiment has been to look for the existence of 
a magnetic moment of the photon. The experiment has been carried
out with a magnetic field strength (magnetic induction) 
of 10 000 gauss (= 1 tesla = 1 T) (for reference, the QED critical 
field strength is $B_{\rm cr} = \frac{c^2 m_e^2}{e\hbar}\sim 10^9\ T$, $m_e$ 
is the electron mass).
The interaction region between the strong magnetic field and
the light beam used for observation was located in a Fabry-P\'{e}rot cavity
from which air had been pumped out (The quality of the obtained vacuum
is not described in \citep{1929watson}.). The interference pattern
whose potential change has been observed was produced by polarized,
monochromatic (visible) light of a wavelength of $585.2$ nm. 
The null result that was obtained in the experiment yields a bound on the
magnetic moment of the photon $\mu_\gamma$ of 
$\mu_\gamma < 1.4\cdot 10^{-22}$ emu (= $1.4\cdot 10^{-25}$ A${\rm m}^2$). 
Equivalently, the null result entails
that the change in the refractive index of the vacuum under the 
influence of the magnetic field is less than $4\cdot 10^{-7}/$T.\\

Three years later, in 1932, C.\ C.\ Farr and C.\ J.\ Banwell of the Canterbury 
University College, Christchurch, New Zealand, published the description
of an experiment designed to study the ``Velocity of propagation of 
light in vacuo in a transverse magnetic field'' \citep{1932farr}.
This experiment relied on a different experimental set-up than that
performed by Watson. A Jamin refractometer was used where one 
unpolarized, monochromatic light beam (of a wavelength of $546.1$ nm)
passed through the strong magnetic field while the other (parallel, 
of the same wavelength)
beam only experienced a weaker leakage field. Using this set-up 
Farr and Banwell achieved a roughly one order of magnitude
higher sensitivity than Watson. The strength
of the magnetic field amounted to 17 992 gauss (= 1.7992 tesla).
The quality of the obtained vacuum in the interaction region is given
as 0.005 mm of mercury (= 0.005 torr = 0.67 Pa) and the authors conclude from
the null result they obtained that 
the change in the refractive index of the vacuum under the 
influence of the magnetic field is less than $2.7\cdot 10^{-8}/$T.\\

Further eight years later, in 1940, Banwell and Farr published an account of 
an improved experiment of the same type performed using a yet different 
set-up \citep{1940banwell}, namely, a Michelson interferometer. Here
one of the two perpendicular optical arms was placed in the strong
magnetic field while the other did not experience the influence of 
any significant magnetic field at all. In the Michelson interferometer
unpolarized, monochromatic light of a wavelength of $\lambda = 546.1$ nm 
was propagating. The intensity of the applied
magnetic field is given as 19 917 oersted (corresponding to a magnetic
induction of 1.9917 T in vacuo). The quality of the
vacuum in the Michelson interferometer is stated
as 0.05 mm of mercury (= 0.05 torr = 6.7 Pa). With the probable error
taken into account, Banwell and Farr
obtain as result of their Michelson interferometer experiment a 
non-null result. In the applied magnetic field of roughly 2 T they
find an increase in the speed of light by (0.3431 $\pm$ 0.1856) m/s.
The estimated error corresponds to a change in the refractive index
of $3.1\cdot 10^{-10}/$T. Banwell and Farr conclude their article
\citep{1940banwell}, p.\ 25, with the following sentences: ``The
authors consider that they have taken all precautions to eliminate
causes tending to give a spurious effect greater than the probable
error of observation quoted above. They would however be very hesitant
in accepting the above final result as real. All that can be said
definitely is that in such a field (20,000 oersted) the effect is less
than 1 part in about $5\times 10^8$.'' (For a discussion of this
non-null result obtained by Banwell and Farr see our
Appendix \ref{appmirror}.).\\

The influence of strong (inhomogeneous) 
electric fields on the propagation of light
has been studied by J.\ Stark (Traunstein, later Eppenstatt, Germany) in the 
years after World War II (For the role of Stark in 
fascist Germany see, e.g., \citep{1977beyerchen}, chap.\ 6, pp.\ 103-122,
\citep{2001stoecker}.). The motivation for such a study Stark derived 
from a light vortex model for photons (which, of course, implicitly entails the 
existence of nonlinearities in the corresponding electromagnetic theory);
cf.\ \citep{1950stark}, sec.\ 13, pp.\ 40-44.  
In his private laboratory 
he has conducted -- among others -- the following experiment 
(\citep{1946stark}; also see \citep{1950stark}, 
sec.\ 14, pp.\ 44-50 (reprinted from \citep{1947stark1}), \citep{1952stark}):
A razor blade is positioned in a parallel trough cut out of a metal 
plate (perpendicular to the razor blade; cf.\ fig.\ 2 on
p.\ 507 of \citep{1952stark}) at close distance. 
A narrow beam of polarized light 
passes along the razor blade in the space between blade and trough and 
consequently falls on a screen (photographic plate). The razor blade/metal
plate system is placed in a vacuum tube. The image produced by the 
light beam (mostly yellow light) is compared for the 
situation without any electric field applied and the situation where
a strong electric field is applied between razor blade and metal plate.
Stark estimates the strength of the electric field close to the 
razor blade to be 1 to $1.5 \times 10^6\ V/cm$ at maximum. He reports
on the existence of a slight difference in the images for the cases
with and without electric field applied once the electric field of the 
polarized light is perpendicular to the razor blade (cf.\ fig.\ 1 on p.\ 47 
and fig.\ 2 on p.\ 48 of \citep{1950stark} and fig.\ 3 
on p.\ 508 of \citep{1952stark}). For the case of the 
electric field of the polarized light
parallel to the razor blade Stark reports a null result. The experiment
of Stark has been repeated independently by J.\ Sperling 
(University of Kiel, Germany) \citep{1948sperling,1949sperling}
who finds no effect whatsoever. For a discussion of other, related
experiments of Stark see \citep{1947stark2,1948hohl,1948kunze}. 
From out the modern 
point of view of quantum electrodynamics a null result does not come
unexpected. Comparing the maximum value of the electric field 
given by Stark as 1 to $1.5 \times 10^6\ V/cm$ with the QED critical field
strength $E_{\rm cr} = \frac{c^3 m_e^2}{e\hbar} \sim 10^{16}\ V/cm$ ($m_e$ is the
electron mass) 
one immediately recognizes that the strong electric field 
used by Stark is too weak to give rise to any nonlinear quantum
electrodynamic effects.\\

We conclude this subsection with the description of a somewhat
speculative experimental 
proposal by H.\ Bauer (University of Kiel, Germany) made in 1931
\citep{1931bauer}. The proposal starts with 
the theoretical observation of the existence of a source-free (Poynting)
energy current ${\bf S}\ \sim\ {\bf E}\times {\bf B}$ 
in the simultaneous presence of (say, constant) magnetic ($\bf B$) 
and electric ($\bf E$) fields. Concerning the existence of such an energy
current Bauer refers (see his footnote 3 on p.\ 38), among others, to Planck  
\citep{1922planck}, second part, chap.\ 2, \S\ 13, p.\ 26 (p.\ 34 of 
the English translation). 
It should be mentioned that the literature on this subject in later
decades is controversial, however, Feynman, for example, expresses the same
view as Planck \citep{1964feynman}, chap.\ 27, sec.\ 27-5, p.\ 27-8.
Bauer notes that the existence of such an energy current (for constant
cross-fields) had not (yet)
been demonstrated experimentally.
He then points out that a momentum
current ${\bf S}/c^2$ is associated with the energy current ${\bf S}$.
He then goes on to reason that if a non-vanishing cross section
for photon-photon scattering would exist in nature (the constant 
magnetic and electric fields correspond to a coherent state of photons)
one could light of frequency $\nu$ let propagate in (opposite) 
line with the momentum current ${\bf S}/c^2$.  The phenomenon of photon-photon
scattering would then result in a (tiny) frequency change of the 
light photons which could 
be measured. He estimates that a frequency change related to a cross 
section of $2\cdot 10^{-27}\ cm^2$ should be observable.
Bauer points out that a non-negative experimental result
would entail a violation of the superposition principle (characteristic
for the Maxwell theory), i.e., it would be related (effectively) to a
nonlinear electromagnetic theory. Bauer further points out that
a non-negative experimental result would show the physical reality
of the Poynting energy current (in the constant cross-field situation)
and at the same time prove the existence of the phenomenon of 
photon-photon scattering.\\

\subsubsection{\label{delbrueck}Propagation/scattering of photons in the 
Coulomb field of nuclei}

While the experiments discussed further above have intentionally been
devised to study the properties of light, the experiments 
referred to in this subsection have been performed historically 
with another objective, namely to advance the understanding of cosmic rays,
to gain further insight into the nuclear structure of atoms, and 
to explore experimentally the predictions of (relativistic) quantum
mechanics.
In the early decades of the 20th century numerous experiments have
been made to study the absorption and scattering of $\gamma$-rays
emitted by radioactive materials by various targets. We will not
review these experiments here, the interested reader is referred to
\citep{1937gentner,1952davisson} (For a related historic account 
see \citep{1959gentner}.). Here, we are
concerned with such $\gamma$-ray experiments performed 
by L.\ Meitner and collaborators (Hupfeld, K\"osters;
Kaiser Wilhelm Institute of Chemistry, Berlin, Germany) in the early 1930s 
\citep{1930meitner1,1930meitner2,1931meitner1,1931meitner2,1932meitner,1933meitner,1933delbrueck}
(For a historic discussion of these experiments see \citep{1984brown}.).
In these $\gamma$-ray absorption and scattering experiments performed
to explore the validity of the Klein-Nishina formula (one of the 
consequences of relativistic quantum mechanics) Meitner and collaborators
found that for targets with large nuclear number (nuclear charge) 
the Klein-Nishina 
formula is not sufficient to correctly describe the experimental results.
Apparently, besides the scattering of $\gamma$-rays
off electrons (described by the Klein-Nishina formula) another physical
process related to the atomic nuclei of the target material was also 
to be considered. Referring to the just discovered creation of 
positive electrons (positrons)
by $\gamma$-rays in various materials, M.\ Delbr\"uck 
(Kaiser Wilhelm Institute of Chemistry, Berlin, Germany)
hypothesized in 1933 
(in an addendum to an article  by Lise Meitner -- whose assistant he was by 
then -- and 
H.\ K\"osters)\footnote{\citep{1933delbrueck}: cf.\ our 
App.\ \ref{originals}, p.\ \refstepcounter{dummy}\pageref{adelbrueck1}, 
p.\ \refstepcounter{dummy}\pageref{adelbrueck2};
English transl.\ \citep{1984brown}, Appendix, p.\ 135.
The English translation given here slightly deviates from the English
translation printed there in order to
make the translation more adequate.}\label{delbrueck1},
``$\ldots$ that it involves a {\it photoeffect} on one of the infinitely
many electrons in states of negative energy, which according to  
$\,$\  D$\,$i$\,$r$\,$a$\,$c'$\,$s$\,$\ theory fill the entire space
with infinite density and would well be capable of such an absorptive process,
by virtue of their interaction with the nucleus.''
He then concludes\label{delbrueck2}:
``Such a proposal has also the 
consequence that that these electrons of negative energy are capable of
{\it scattering} $\gamma$-rays, and in fact {\it coherently}, analogous
to the phenomenon of the {\it unshifted} Compton line.''.\\

\begin{figure}[ht]
\hspace{1.5cm}
\includegraphics{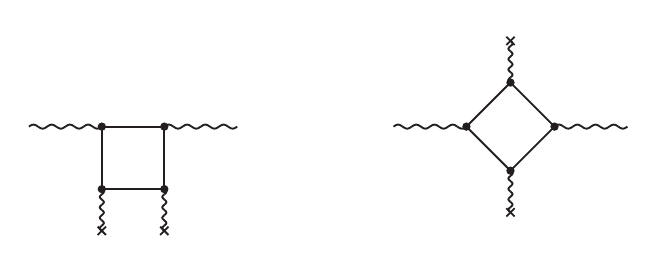}
\caption{\label{delbrueckfeynfig}Typical (lowest order) Feynman
diagrams contributing to Delbr\"uck scattering (The wavy line
ending in a cross depicts the Coulomb field of a nucleus.).}
\end{figure}

In appreciating the contribution by Delbr\"uck one has to remind 
oneself that at the time of writing of this addendum
the theory of quantum electrodynamics
was still in its birth phase but Delbr\"uck relied on the model
picture of the Dirac sea to qualitatively describe an advanced quantum field 
theoretic process whose detailed nature Delbr\"uck did not sketch.
Today, to lowest order of QED perturbation theory contributions
to Delbr\"uck scattering -- as it is now being called -- can be 
depictured by Feynman diagrams as shown in 
Fig.\ \ref{delbrueckfeynfig}.\\

\noindent
Interestingly, in 1978 Delbr\"uck has described his proposal 
and its further fate in an oral history interview of the California 
Institute of Technology Archives (\citep{1978delbrueck}, pp.\ 52-54 =
{\tt Delbr\"uck-45 - Delbr\"uck-47})
from which we quote in the following (with kind permission
by the Caltech Archives):\hfill\ \linebreak
``$\ldots$ {\tt\justify One of the
graduate students of Lise Meitner had studied the scattering by
lead of gamma rays of ThC$^{\tt 11}$; ThC$^{\tt 11}$ is a gamma ray source with
relatively hard gamma rays, as I recall, 2.6 million electron
volts. If you scatter these gamma rays on lead, then, according to
then current theory, you should find very little coherent scattering. 
Most of the scattered light should be Compton-scattered -\,- that
means scattering where the electron acts as if it were a free
electron. And after scattering you find, at right angles then, a
Compton-scattered
light quantum which is very much \underline{less} energetic
than the incoming one. This student, H.\ K\"osters had found a scattered
component which was much harder than the expected one. I put out
the conjecture that this had something to do with the new theory of
the electron that Dirac had proposed, according to which the
negative energy states of an electron (with energies below minus
mc$^{\tt 2}$) were all filled, and the electron never jumped from plus energy
to minus energy because these were filled (because of Pauli's
exclusion principle). I made the conjecture that these negative
energy electrons in the vicinity of the nucleus are not free electrons,
but that their wavefunction was distorted by the nucleus of the atom
and therefore that they could scatter. If they are free electrons
then they wouldn't scatter, but if they are disturbed by the field
of the nucleus then there could be virtual transitions from minus
to plus energy, and there would be corresponding scattering.

This problem is related to the problem of scattering of light
by light. In the classical theory, two light beams just go right
through each other and don't interact, but in quantum electrodynamics
if you take into account these negative energy electrons,
then the first light beam polarizes the vacuum, and the second
light beam then is scattered on the first one. So I made a conjecture
that these hard scattered rays should be due to this scattering of
underground electrons. The fate of this conjecture was that it
turned out that the scattered light, the scattered quanta observed by
K\"osters, were not due to that effect. Instead, they were due to the
effect that the negative energy electrons actually absorbed a
quantum, and thereby created a hole there, a positive electron. This
positive electron then could recombine with some other electron and
make annihilation radiation, and that is very much harder than the
Compton-radiation. Actually that was an obvious implication that
I had overlooked. And that came out very quickly. Nevertheless the
effect that I predicted \underline{ought} 
to be there also, and the question was
how to calculate it, and I slaved on that and it turned out to be a
nightmare to calculate that.\footnote{\label{footdelbrueck}K.S.: 
In this context, W.\ Heisenberg reported in a letter of June 16, 1935 to 
W.\ Pauli\label{heisenberg1}:
``Delbr\"uck recently was here and reported about his fruitless attempts
to calculate the coherent scattering. He certainly would be glad
if he could do this work together with somebody else, e.g., Weisskopf
(Sauter is now in G\"ottingen).'' (\citep{1985vonmeyenn}, 
letter [374] of June 16, 1934, pp.\ 331-333, specifically p.\ 332: 
cf.\ our App.\ \ref{originals}, 
p.\ \refstepcounter{dummy}\pageref{aheisenberg1}; 
English transl.: K.S.).}$^{[footn.\ K.S.]}$
\setcounter{extrafoot}{\value{footnote}}
\renewcommand{\thefootnote}{\tt\arabic{footnote}}
\setcounter{footnote}{11}
With the help of some advice by Hans Bethe
I got so far as to predict that this effect should be proportional
to the fourth power of the nuclear charge, Z$^{\tt 4}$ , and that's about all
that I predicted; it was published in a short appendix, I think,
to the paper by K\"osters.\footnote{$^{\it [K.S.:\ orig.\ footn.]}\,$
{\tt M.\ Delbr\"uck, {"}Zusatz bei der Korrektur,{"} 
in L.\ Meitner and H.\ K\"osters, 
{"}\"Uber die Streuung Kurzwelliger $\gamma$-Strahlen,{"} 
\underline{Zeitschrift f\"ur Physik}
84:137-l44 (1933), 144.} (K.S.: cf.\ \citep{1933delbrueck})}$^{\it [K.S.:\ orig.
\ footn.]}\,$ 

\hspace{0.5cm}
That's where my contribution ended to this problem, and I
never heard of it again until about 20 years later, in the fifties,
when I was long since in biology. Somebody told me that there had
been published two papers in \underline{Physical Review} 
on {"}Delbr\"uck scattering,{"}
by Bethe and some graduate students of his who had made some progress
in calculating them.\footnote{$^{\it [K.S.:\ orig.\ footn.]}\,$
{\tt F.\ Rohrlich and R.\ L.\ Gluckstern, 
{"}Forward Scattering of Light
by a Coulomb Field,{"} \underline{Physical Review} 86:1-9 (1952); H.\ A.\ Bethe
and F.\ Rohrlich, {"}Small Angle Scattering of Light by a Coulomb
Field,{"} \underline{Physical Review} 86:10-16 (1952.)} 
(K.S.: cf.\ \citep{1952rohrlich,1952bethe})}$^{\it [K.S.:\ orig. \ footn.]}\,$
So since then this name, {"}Delbr\"uck scattering{"}
exists, and if you ask theoretical physicists then I am known
scurrilously for that little incident. I understand that the actual
calculation of this effect, and experimental verification of it,
\underline{still} 
has been lingering on for the next 20 years after that, because
it turned out to be just very, very difficult to calculate; also,
in order to observe it you need to go to much higher energies -\,- I
think the optimal energy is about 10 million electron volts rather
than 2.7 -\,- and I think now it has been confirmed to exist.} 
$\ldots$''.\\[-0.4cm]

\renewcommand{\thefootnote}{[\alph{footnote}]}
\setcounter{footnote}{\value{extrafoot}}

\noindent
It remains to add that L.\ Meitner in 1934 \citep{1934meitner} 
refers to a qualitative calculation by Delbr\"uck himself of the 
particular scattering process described by him in \citep{1933delbrueck}
\footnote{The memory of Max Delbr\"uck in his above quoted interview 
apparently slightly mixes the course of events. His note 
\citep{1933delbrueck} does not contain any calculation.}.
She quotes Delbr\"uck as having calculated the angular distribution
of the Delbr\"uck scattering process and mentions that this calculation will
be published in the near future. However, apparently this calculation
never got published and also a preserved manuscript of it is not known.
Related results have been published in 1937 by N.\ Kemmer and G.\ Ludwig
(University of Zurich and ETH Zurich, Switzerland)
 \citep{1937kemmer2}. Kemmer and Ludwig (\citep{1937kemmer2}, p.\ 184)
give for the cross section $q$ for the scattering of light of (long) 
wavelength $\lambda\ \left(\gg \hbar/mc,\ m = m_e\right)$
off a nucleus with charge $Z \vert e\vert$ the formula ($c_1$, $c_2$ are
certain unknown constants and $\theta$ is the scattering angle)
\begin{eqnarray}
\label{kemmerludwigcross}
q&=&\left(\frac{e^2}{mc^2}\right)^6\cdot
\left(\frac{Z}{\lambda}\right)^4\cdot
\left[\left(c_1^2 + c_2^2\right)\left(1 + \cos^2\theta\right)
\ +\ 2 c_1 c_2 \cos\theta\right]\ \ .
\end{eqnarray}
Related results have been published in 1937 by Akhiezer and Pomeranchuk
\citep{1937akhiezer} (cf.\ subsec.\ \ref{qed}).
As M.\ Delbr\"uck in his Caltech Archives oral history 
interview recalls (see above quote) the further study of the Delbr\"uck
scattering process continued in the 1950's only.\\

\subsection{\label{astro}Considerations related to extraterrestrial and 
astronomical observations}

Since the earliest times of mankind, light has 
naturally played a prominent role
in developing an understanding of the physical world we live in. It comes
as no surprise that also the problem of photon-photon interactions
(or, nonlinear electrodynamics, or possible violations of the 
superposition principle) has been approached by means of results 
of extraterrestrial and astronomical observations. 
In the period under consideration,
three comments that belong to this line of research are known to us.
One comment has been made by H.\ Ebert 
in 1887 \citep{1887ebert} concerning the possible dependence of the speed
of light on the intensity of the propagating beam, a second consideration
on the subject of using the solar corona for obtaining limits on
the violation of the (optical) superposition principle
has been published in 1928 by S.\ I.\ Vavilov, and a third comment
concerning a possible frequency dependence of the speed of light is
due to F.\ Zwicky (California Institute of Technology, Pasadena, USA) in 1937 
\citep{1937zwicky}. In the following, we will let Ebert and Zwicky 
speak for themselves by simply quoting from their articles while the 
somewhat longer discussion by Vavilov is being summarized only.\\

Let us first quote the relevant part of the paper by
Ebert\footnote{\citep{1887ebert}, pp.\ 381-383: cf.\ our 
App.\ \ref{originals}, p.\ \refstepcounter{dummy}\pageref{aebert}; 
English transl.: K.S..
To some extent the original German text exhibits 
a historic style used at the 
end of the 19th century, the present English translation does not 
deliberately attempt to imitate this historic style.} 
in which he makes contact between his laboratory experiments 
(cf.\ subsection \ref{intensity}) and astronomical observations 
(an early example of astroparticle physics $\ldots$)\label{ebert}:

\noindent
``At the end, I am allowing myself to apply the above result to
some astrophysical questions.

\noindent
Is the velocity of the propagation of light dependent on its
intensity to some noticeable degree, beyond the circumstance that
the phenomena in the sky occurring to us at the same moment 
belong to very different times in reality,
the further complication would step in that these times even for
equal spatial distance would be very different for the different
objects. In view of the large differences in intensity that are 
being met here and in view of the large distances which the light
rays have to pass before they can reach us, even small differences
in the velocity of the propagation of light can make themselves 
noticeable to a large degree. For example, this should occur for
physical double stars; here, we have got two sources of light
that, in some rough approximation, are equally far away from us 
but whose intensity of light, in general however, considerably differs
from each other. Despite this we find that - if we succeeded in 
calculating the trajectories of double stars - the two components 
indeed show up at the same time at corresponding points of their
trajectories and together with their joint center of gravity 
always lie on a straight line. Right from this fact, we can also draw new
support for the experimental result obtained by us.\footnote{K.S.: Cf.\ 
our subsec.\ \ref{intensity}.}$^{[footn.\ K.S.]}$
One example might incidentally show yet that the bound proved by me
for the independence of the two quantities in 
question\footnote{K.S.: i.e., the velocity of light and the 
intensity of light.}$^{[footn.\ K.S.]}$  is completely
sufficient for the astronomical practice.

\setcounter{extrafoot}{\value{footnote}}
\renewcommand{\thefootnote}{\arabic{footnote}}
\setcounter{footnote}{0}
According to calculations by  
$\,$A$\,$u$\,$w$\,$e$\,$r$\,$s$\,$ and $\,$P$\,$e$\,$t$\,$e$\,$r$\,$$\,$s,
the companion of Sirius, 
calculated by $\,$B$\,$e$\,$s$\,$s$\,$e$\,$l$\,$ and discovered by
$\,$C$\,$l$\,$a$\,$r$\,$k$\,$, has an extraordinary large mass
($\sfrac{1}{2}$ to $\sfrac{2}{3}$ of the mass of Sirius itself). 
At the same time, it occurs to us as a star of about class 9 only,
also compared with its main star, its brightness is a very small one,
according to 
$\,$S$\,$c$\,$h$\,$\"o$\,$n$\,$f$\,$e$\,$l$\,$d \footnotemark)$^{\it 
[K.S.:\ orig.\ footn.]}\,$ 
\renewcommand{\thefootnote}{}
\footnotetext{1)$^{\it [K.S.:\ orig.\ footn.]}\,$ 
$\,$S$\,$c$\,$h$\,$\"o$\,$n$\,$f$\,$e$\,$l$\,$d,
Die dunklen Fixsternbegleiter. Mannheimer Verein f\"ur Naturkunde.
30.\ Jahresbericht. (K.S.: cf.\ \citep{1868schoenfeld})}
\renewcommand{\thefootnote}{[\alph{footnote}]}
appr.\ $\sfrac{l}{1000}$ of that of Sirius only.
Further above, for rays of 
medium wavelength (for sodium light) it was found that for an attenuation 
of light from intensity 1 to intensity $\sfrac{1}{33}$ 
the velocity of the propagation
of light does not change even by $\sfrac{1}{500\ 000}$. Relying on this limit,
i.e.\ assuming that for differences in brightness of 33:1 the velocity 
of propagation could differ just yet by this amount and that the latter would
grow proportionally with the difference in intensity, in the present example,
in each second the light of the companion would stay behind the light of 
the main star by $30 \times 0,6$ or 18 km. In light-time,
the distance of the Sirius system amounts to about 30 years. Assuming
it were precisely 30 years, it would take the light of the companion:
\begin{eqnarray*}
\frac{300\ 000}{299\ 982}\ \times\  30 &=& 30,0018
\end{eqnarray*}
years to reach us, i.e., of two simultaneously emanated light rays
the one emitted by the companion would always arrive about 0.7 days later
only than the one emitted by Sirius itself. As, however, the 
period of revolution of both bodies around their joint 
center of gravity is 50 years,
this difference in time and the difference in location corresponding to it
can be neglected. The assumptions made, however, are very unfavourable ones;
for all orbit determinations, 
we can disregard any dependence, existent yet perhaps, 
of the velocity of propagation on any intensity whose value lies below
the limits established by the experiment. 

\noindent
Further, the obtained result is of relevance for applications of the
$\,$D$\,$o$\,$p$\,$p$\,$l$\,$e$\,$r$\,$ principle where the relative
velocities of the stars in the direction of the line of sight 
are determined from changes in the average refractivity
of isolated spectral lines. Here, in the different cases very large
differences in brightness exist and, consequently, for the 
applicability of this principle it is important to have shown directly 
that these differences in brightness do not also have an influence.''.\\

\setcounter{footnote}{\value{extrafoot}}
As described in subsection \ref{scatteringexp}, 
Vavilov \citep{1928vavilov,1930vavilov}
had performed in 1928 a laboratory experiment to attempt the direct
(optical) observation of photon-photon scattering in vacuum.
This experiment ended with a negative result: no photon-photon
scattering has been observed. It should be noted that in 1928 
the experiment had been based on a qualitative understanding 
of the quantum nature of light only, the later qualitative 
predictions of QED for the scattering of photons by photons
were not yet made. To Vavilov it was clear that the superposition
principle for electromagnetic radiation is fulfilled to a high
degree, consequently, the experimental search for any signatures
of photon-photon scattering is confronted with a tiny effect (if any
at all). To Vavilov, a high radiation density was therefore 
the key for a possible observation of photon-photon scattering
(Only the later QED calculations, see subsection \ref{qed}, further clarified
the situation: the size of the photon-photon scattering cross section 
is critically dependent on the photon energy.). 
In his article of 1928 Vavilov, therefore,
also turned his attention to the sun with its large radiation density.
From data concerning the solar corona he concluded that in 
any photon-photon scattering at most $1.8\cdot 10^{-17}$ of the beam
energy is being scattered (\citep{1928vavilov}, \S\ 2, p.\ 558,
p.\ 237 of the reprint, pp.\ T-4 (= P-6) of the English translation; also see
\citep{1930vavilov}). From this estimate he concludes that
even if photon-photon scattering existed it could not be observed
in any terrestrial experiment because the human eye would not 
be sensible enough to detect the low intensity of scattered photons
(At that time, the human eye was the most sensitive photon 
detector in such a type of optical experiments, 
cf.\ e.g.\ \citep{1930hughes}, p.\ 777, last sentence.).
Finally, noting that there was no satisfactory theory of the 
solar corona (in 1928) he then tried to explain (in a somewhat 
more hypothetical way) certain known features of the solar corona
by means of photon-photon collisions (For a current treatment of
the solar corona see, e.g., \citep{2004aschwanden}.).\\
 
Zwicky in 1937 \citep{1937zwicky} is concerned with a 
frequency dependence of the speed of light possibly arising
within theories of nonlinear electrodynamics. He writes: 

\setcounter{extrafoot}{\value{footnote}}
\renewcommand{\thefootnote}{\arabic{footnote}}
\setcounter{footnote}{4}
\noindent
``Certain effects have recently come to our
knowledge which suggest that the differential equations governing the
propagation of light through empty space are not strictly linear. 
{\it Nonlinearity} of these equations results in a {\it dependence} 
of the {\it velocity of light on frequency}. 
One reason for the existence of slight deviations from the
superposition principle of light lies in the potential 
possibility of the formation
of pairs of positive and negative electrons by interacting photons.
This interaction necessitates a generalization of Maxwell's field equations
through the introduction of non-linear terms in the 
field strengths.\footnote{$^{\it [K.S.:\ orig.\ footn.]}\,$
Euler and Kockel, {\it Naturwiss.}, {\bf 23}, 246 (1935).
(K.S.: cf.\ \citep{1935euler})}$^{\it [K.S.:\ orig.\ footn.]}\,$ As
a result, light traveling through space which is free of matter but filled
with radiation will have a velocity depending on its frequency. In addition
similar effects arise from the gravitational interaction of light with
light and matter. Although the effects to be expected are in all probability
small the possibility of an experimental test may be kept in mind.

\noindent
An obvious way of investigating effects of the kind mentioned lies in
the observation of light signals of different frequency which have traveled
through space for a long time. If we knew of any signals which have
started {\it simultaneously} from a very distant source the dependence of the
velocity on frequency could perhaps be demonstrated by checking up on
the times of arrival of these signals. A unique case to perform this test
is afforded by the observation of distant nova outbursts. For such an
outburst we may safely assume that photons in the various emission
lines of, say, hydrogen have started out simultaneously on their long
journey. We must therefore attempt to determine whether or not the
intervals in which the different hydrogen lines flare up in the spectrum of
a distant super-nova are zero or not. With the present telescopic equipment
it is probably possible to find super-novae to distances as great as
$10^8$ light-years. With reasonable luck it should be possible to detect
differences in the time of arrival of various emission lines amounting to as
little as a few days. Observations of this kind will therefore enable us in
the most favorable cases to determine the ratio of the velocity of violet
light to that of red light traveling through internebular space with a relative
accuracy of $10^{-10}$.''.

\renewcommand{\thefootnote}{[\alph{footnote}]}
\setcounter{footnote}{\value{extrafoot}}

\noindent
He then goes on to describe certain astronomical observations 
concerning novae in the Andromeda galaxy M31 which - at the time
of writing - might be considered being related to a possible frequency
dependence of the speed of light.\\

\section{Theory}

In classical electrodynamics the superposition principle holds true,
consequently, direct interaction processes (without mediation by
charged particles) of electromagnetic fields such as photon-photon 
scattering do not occur within this framework. Its theoretical 
description, Maxwell electrodynamics, is based on the following 
Lagrange density for the electromagnetic field
\begin{eqnarray}
\label{freelagrangianBI}
{\cal L}_0&=&-\frac{\displaystyle 1}{\displaystyle 4\mu_0}\ 
F^{\mu\nu} F_{\mu\nu}\ =\
\frac{\displaystyle 1}{\displaystyle 2}\ \epsilon_0 
\left({\bf E}^2 - c^2 {\bf B}^2\right)
\ =\ \frac{\displaystyle 1}{\displaystyle 2}\
\left(\epsilon_0 {\bf E}^2 
- \frac{\displaystyle 1}{\displaystyle\mu_0} {\bf B}^2\right)
\end{eqnarray}
where $F^{\mu\nu}$ is the electromagnetic field strength tensor and
the velocity of light (squared) is given
by $c^2 = 1/(\epsilon_0\mu_0)$ with $\epsilon_0$, $\mu_0$ being the dielectric
permeability and the magnetic permittivity of the vacuum, respectively.
A characteristic feature of the Maxwell theory is that the 
Lagrange density of the electromagnetic field is quadratic in 
the electric (${\bf E}$) and magnetic (${\bf B}$) fields. Electromagnetic
theories which are to model direct interaction processes of 
electromagnetic fields have to be described by Lagrange densities which
contain higher powers of the electromagnetic fields. The first 
theoretical study which considered Lagrange densities of the 
electromagnetic field of a more general form has been published by 
G.\ Mie (University of Greifswald, Germany) in 1912 \citep{1912mie}.
However, Mie considered not only a dependence of the Lagrange density
on the electric and magnetic fields themselves but also explicitely and 
separately on the electromagnetic potentials. In his famous review of
the theory of general relativity \citep{1920pauli} in 1920, W.\ Pauli 
(University of Munich, Germany) pointed out (\S\ 64, pp.\ 754-769, 
pp.\ 188-192 of the English translation) that
the non-gauge invariant ansatz of Mie is physically not viable.
The work of Mie does not contain any explicit reference to such
processes as photon-photon scattering we are interested in in the present
review but it has to be considered as the earliest theoretical approach
in principle entailing such processes.\\

\subsection{Born-Infeld electrodynamics}

The following step in the further development of nonlinear 
Lagrangians for electromagnetism is closely related to 
a political-historic situation -- the rise to power of the Nazi
party in Germany in 1933. M.\ Born (University of G\"ottingen, Germany) 
describes in his autobiography \citep{1978born} the situation
he found himself in. After he had been dismissed from his chair
at the University of G\"ottingen as a result of the fascist Law
for the Restoration of the Career Civil Service [Gesetz \"uber 
die Wiederherstellung des Berufsbeamtentums] of April 7, 1933
he immediately left Germany for Italy in May 1933. He writes
about this time in Italy
(\citep{1978born}, part 2, chapter IV, pp.\ 254-255):
``I soon began to miss my accustomed work. But I had neither books
nor periodicals. So I looked for a subject for which no literature
was needed, something in quite a new line. I started from my old
favourite problem, the electromagnetic mass of the electron, which
I had treated in my 
thesis\footnote{K.S.: i.e., habilitation thesis 
\citep{1909born}.}$^{[footn.\ K.S.]}$ 
for admission as a lecturer. According
to Coulomb's law, the energy of a charged particle becomes infinite
when the radius shrinks to nothing. One has therefore to
assume that the electron has a finite size. But if it is a rigid 
sphere one gets into trouble with the theory of relativity because
there, in general, rigidity does not exist. In my thesis
(see Part 1, Chapter XI)\footnote{K.S.: Part 1, Chapter XI 
of \citep{1909born}.}$^{[footn.\ K.S.]}$, 
I had discovered that there are special
movements for which rigidity can be defined, and I had obtained for 
these a definite expression for the self-energy (electromagnetic
mass). But for arbitrary movements this was not possible.
The only way out of this difficulty
seemed to be the assumption that the ordinary Maxwellian theory of the
electromagnetic field holds only approximately for dimensions large
compared with the radius of the electron, while for smaller distances
it should be replaced by another, more subtle theory. A quite general frame 
for modifying Maxwell's equations had been given, many years before, by
Gustav Mie\footnote{K.S.: \citep{1912mie}.}$^{[footn.\ K.S.]}$. 
I remembered these and tried to fill out this frame by special,
plausible assumptions. But I was not successful though I spent many hours
on my verandah pondering about it. At last I succeeded by abandoning 
Mie's formalism and inventing another one. The usual linear theory
of the electromagnetic field was replaced by a non-linear one, in 
which point charges existed which carried a finite electrostatic
energy.

I had no means (tables, etc.) to work it out numerically,
and I remember the tension with which I waited for the letter
from a colleague whom I had asked to evaluate the decisive elliptic integral.
When it came I was very happy and proud\footnote{K.S.: The first Letter
of Max Born on this subject appeared in the journal {\it Nature} in 
August 1933 \citep{1933born1}.}$^{[footn.\ K.S.]}$; 
for I was filled with the intense
desire to discover something fundamental
to regain my self-confidence after the loss of my job in G\"ottingen, 
and to `show the Germans what they had lost'.
Alas, these ambitions were not fully satisfied.
My non-linear field theory made some stir, but did not lead to the 
solution of the problem of the structure of elementary particles, 
although I spent a great deal of time and effort on it during the 
following years, together with a number of collaborators,
in the first place the Pole Infeld. I shall return to this later.
The so-called Born-Infeld theory was one of the first attempts at 
a non-linear field theory and worked perfectly in the `classical' 
(non-quantum) domain, but failed in the quantum domain.
Today one has a wide empirical knowledge about elementary particles,
and it is clear that the electron cannot be treated separately.
The most promising theory, that of Heisenberg, also uses non-linear field
equations and follows thus the direction indicated by my work, but rests on
quantum theory right from the beginning. In any case, this work
helped me to overcome the bitter feelings produced by the loss of 
my position and my expulsion.''\\

In his Letter to {\it Nature} \citep{1933born1} (followed by a 
full length paper \citep{1934born1} submitted for publication 
at about the same time, August 1933) M.\ Born proposed as
a gauge-invariant nonlinear Lagrangian of electromagnetism the 
expression ($b$ is some dimensionful constant; 
\citep{1934born1}, p.\ 426, eq.\ (6.1))
\begin{eqnarray}
\label{bornaction}
{\cal L}&=&- \frac{b^2}{\mu_0}\ \sqrt{1 + \frac{\cal F}{b^2}},\\[0.3cm]
&&\hspace{2cm}{\cal F}\ =\ \frac{1}{2}\ F^{\mu\nu} F_{\mu\nu},\nonumber
\end{eqnarray}
which leads to a finite value of the self-energy of a point charge
(in contrast to Maxwell electrodynamics) as shown by M.\ Born 
and L.\ Infeld (both at the University of Cambridge, UK)
shortly thereafter \citep{1933born2}. Ya.\ I.\ Frenkel' 
(Physico-Technical Institute, Leningrad, USSR) immediately 
devoted two sections in volume 1 of the Russian edition (1934)
of his teaching book on classical
electrodynamics to this nonlinear electrodynamic theory by Born
(\citep{1934frenkel2}, chap.\ X, \S\ 11-12, pp.\ 407-422).\\

Once (arbitrary) nonlinear 
Lagrangians of the electromagnetic field are allowed the natural 
question arises by which principles the functional shape of 
the Lagrangian should be chosen. Born and Infeld formulated two
general principles which should rule the choice of the Lagrangian:
the {\it principle of the finite field}
\citep{1933born3} (the corresponding full length 
paper is \citep{1934born3})/{\it principle 
of finiteness} (``$\ldots$ a satisfactory
theory should avoid letting physical quantities become infinite.''
\citep{1934born3}, \S\ p.\ 427) and the postulate (principle) of the 
{\it invariance of action} (under space-time transformations,
\citep{1934born3}, \S\ 2, p.\ 429)). Following earlier considerations
by Eddington (\citep{1923eddington}, \S\ 88, p. 206, eq.\ (88$\cdot$4), \S\ 97,
p.\ 223, \S\ 101, p.\ 232),  
Born and Infeld write down the expression (an additive constant is being 
omitted) 
\begin{eqnarray}
\label{bornactiondet}
{\cal L}&=&- \frac{b^2}{\mu_0}\ 
\sqrt{1 + \frac{\cal F}{b^2} - \frac{{\cal G}^2}{b^4}},\\[0.3cm]
&&\hspace{2cm}{\cal F}\ =\ \frac{1}{2}\ F^{\mu\nu} F_{\mu\nu},\ \ 
{\cal G} = \frac{1}{4}\ F^{\mu\nu}F^\ast_{\ \ \mu\nu}\ 
=\ \frac{1}{c}\ {\bf E B}\ ,\nonumber
\end{eqnarray}
as ``$\ldots$ simplest Lagrangian 
satisfying the principle of general invariance'' (\citep{1934born3}, p.\ 432,
below of eq.\ (2.26))).
Contrasting eq.\ (\ref{bornactiondet}) with eq.\ (\ref{bornaction})
Born and Infeld state: ``Which of these action principles is the right
one can only be decided by their consequences.'' (\citep{1934born3}, 
p.\ 432, below of eq.\ (2.28)). A discussion of a large class of 
invariant Lagrangians in generalization of the eqs. (\ref{bornaction}),
(\ref{bornactiondet}) has been given in 1935 by W.\ H.\ Erskine
(Johns Hopkins University, Baltimore, USA) in his Ph.D.\ thesis
\citep{1935erskine1,1935erskine2}. In the following years, the choice of 
the functional form of the Lagrangian has further been discussed in 
\citep{1936baudot2,1936infeld1,1936infeld2,1936infeld3,1936infeld4,1936madhavarao2,1937darrieus,1937hoffmann,1937infeld1,1937madhavarao3}. 
Infeld in his autobiography reports on his own view of the problem in 
late 1935 by saying:
 ``$\ldots$ I recently discovered that Maxwell's theory can be 
generalized in many different ways. This is not astonishing. But I no 
longer believe that the generalization presented by Born is the 
simplest. I don't like the arbitrariness of the whole problem.''
(\citep{1941infeld}, book two, sec.\ 13, p.\ 223 (p.\ 177 of the 
reprint \# 1)). For the reader interested in science history we would
like to mention that both the autobiographies of Born \citep{1978born} and 
Infeld \citep{1941infeld} contain
much interesting material which provides the reader with a unique 
insight into the development of the Born-Infeld theory.\\

The introduction of nonlinear Lagrangians of the electromagnetic
field has resulted in a number of directions of thought whose 
significance is not limited to the particular Lagrangians considered
by Born and Infeld. Some of these are:

\begin{itemize}
\item
The nonlinear electromagnetic Lagrangians (\ref{bornaction}),
(\ref{bornactiondet}) contain a dimensionful constant $b$ which
sets the scale for the nonlinearity and, consequently, for the 
deviation of phenomena from the predictions of (linear) Maxwell
electrodynamics. Considerations concerning the value of $b$
have been given by Born and Infeld \citep{1934born3}, \S\ 8,
p.\ 446, eq.\ (8.8), and Born and E.\ Schr\"odinger 
(University of Oxford, UK)\citep{1935born2}.

\item
One of the physical consequences 
of choosing a nonlinear Lagrangian of the electromagnetic field consists
in modifications of the Coulomb law (potential) in the vicinity of
a charged particle. Resulting effects in spectroscopic problems 
have been discussed for the first time by G.\ Heller and L.\ Motz 
(Columbia University, New York, USA) \citep{1934heller}  right in 1934 
concluding, however, that these are ``barely observable''
(\citep{1934heller}, p. 505) in the problem under consideration
(fine structure of the Balmer series for hydrogen). Later, J.\ Meixner 
(University of Gie\ss en, Germany) \citep{1935meixner}, Z.\ Chrap\l ywyj 
(Jan Kazimierz University, Lvov, Poland) 
\citep{1936chraplywy3,1937chraplywyj2}, 
and B.\ S.\ R.\ Madhava Rao (Indian Institute
of Science, Bangalore, India) \citep{1937madhavarao1}
have further extended these investigations drawing
the same final conclusion, however.
A.\ Bramley (Bartol Research Foundation, Swarthmore, USA) 
\citep{1936bramley}
explored possible consequences of Born-Infeld electrodynamics
on the scattering of charged particles off each other.

\item
Nonlinear electromagnetic theories exhibit a rich field-theoretic structure
differing from that of standard Maxwell electrodynamics. Therefore,
formal field-theoretic studies to explore the mathematics related to
nonlinear electromagnetic Lagrangians are necessary. Besides Born
and Infeld themselves \citep{1934born1,1934born3} a group of 
other authors has studied formal aspects of nonlinear electrodynamics.
A number of investigations are dealing with the description of the dynamics
and the status of charged particles in nonlinear electrodynamics
\citep{1934frenkel1,1934born5,1935chraplywyj,1935feenberg,1936chraplywy1,1936chraplywy2,1936pryce1,1936pryce2,1936tonnelatbaudot,1937chraplywyj1,1937chraplywyj2,1937chraplywyj4,1938mirtskhulava,1938thomas,1939mirtskhulava,1942schroedinger2}.
Others studied the introduction of complex field variables 
into the formalism 
\citep{1935schroedinger2,1936baudot1,1936kreisler,1936madhavarao4,1937tonnelatbaudot,1937weiss}, 
investigated solutions of the field equations
\citep{1935pryce1,1935pryce2,1936mircxulava,1937madhavarao2,1938mirtskhulava,1939mircxulava,1943schroedinger}, 
and explored further questions
\citep{1935watson,1936born,1936pauli2,1936geheniau,1936madhavarao1,1936madhavarao3,1936watson,1937madhavarao2,1938iwatsuki,1938madhavarao,1938sibata}.

\item
The quantization of the nonlinear electromagnetic theory has received
some attention (in the time period under consideration quantum 
electrodynamics was still in its infancy). Here, besides the studies 
by Born and Infeld \citep{1934born2,1934born4,1934born6,1935born1,1937infeld2}
few further investigations exist
\citep{1935kwal,1935pryce3,1936pauli3,1936pryce3,1937pryce,1943mcconnell}.

\item
Born and Infeld have considered the electromagnetic theory
in Minkowski space. However, in the time period under consideration
in the present review the problem of the unification of gravitation 
and electromagnetism has been explored widely. In the context of the 
Born-Infeld theory, this problem has been considered by B.\ Hoffmann 
(University of Rochester, USA)
\citep{1934hoffmann}, \citep{1935hoffmann3} (this is the full length
paper related to the Letter \citep{1934hoffmann}), 
\citep{1935hoffmann1,1935hoffmann2}
and other authors 
\citep{1935awano,1938mandel,1938sibata,1947shirokov,1948fisher}.

\end{itemize}

In the following we would like to discuss in somewhat greater length
three such further subjects that are more closely related to the 
phenomenon of photon-photon scattering.\\

\noindent
1.\ The scattering of light by light\\

\noindent
The first comment on the scattering of light by light within 
the framework of Born-Infeld theory can be found in the classical article
by Euler and Kockel \citep{1935euler} on photon-photon scattering in QED
where they point out that the lowest order terms in the 
QED (effective) Lagrangian of the 
electromagnetic field responsible for this process are fairly 
similar to those in the Born-Infeld Lagrangian
(The most obvious difference lies in the numerical 
value of some coefficients involved. For the view of W.\ Pauli
in this respect cf.\ our App.\ \ref{apppauli}.).\\

In 1936, a first thorough analysis within (classical) 
nonlinear electrodynamics
of the interaction of two light waves is given in Chap.\ II, \S\ 5
(pp.\ 59-66, pp.\ T-59 - T-66 (= P-63 - P-70) of the English translation) 
of the candidates (Ph.D.) thesis of A.\ A.\ Smirnov (Ural 
Physico-Technical Institute, Sverdlovsk, USSR) \citep{1936smirnov}. 
Smirnov relies in
his analysis on the action (\ref{bornaction}) initially chosen
by Born \citep{1934born1} which only contains the invariant ${\cal F}$
of the electromagnetic field. The analysis is performed to leading 
nontrivial order in the constant $b$ (the critical field strength).
In principle, this kind of analysis is also applicable to the 
Euler-Kockel-Heisenberg Lagrangian (which however depends on both of
the field invariants ${\cal F}$ and ${\cal G}$). For counterpropagating
electromagnetic waves (for the choice of polarization see the 
original thesis \citep{1936smirnov}) he finds as main results
a change in the phase velocities and the emergence of small 
wave components with significantly different frequencies 
compared with the initial waves (e.g., for initial waves of the same 
frequency $\omega$ wave components with triple frequency  $3\omega$
emerge, i.e., higher-harmonics generation is found). The above 
analysis of Smirnov is published in 1940/41 \citep{1940smirnov1,1941smirnov} 
only in a slightly generalized form (cf.\ the footnotes on p.\ 449 of 
\citep{1940smirnov1} and on p.\ 48 of \citep{1941smirnov}).\\

In 1942, as part of a larger study, Schr\"odinger (Dublin Institute for
Advanced Studies, Ireland) also published 
an approximate analysis of the interaction of two 
light waves \citep{1942schroedinger1},
Part I, Sec.\ 4, pp.\ 85-88 (pp.\ 317-320 in \citep{1984schroedinger})
within the framework of the Born-Infeld theory (Schr\"odinger relies
on the form of the action (\ref{bornactiondet}).). He also finds as main
result, as Smirnov did, a change in the phase velocities of the two waves. 
Schr\"odinger also studied the interaction of three light waves
\citep{1942schroedinger1},
Part I, Sec.\ 5, pp.\ 88-92 (reprint \citep{1984schroedinger}, pp.\ 320-324).
Somewhat later, in 1943, Schr\"odinger also found an exact 
two-wave solution of the equations of motion of 
Born-Infeld theory \citep{1943schroedinger}.\\

It should also be mentioned that in late 1936 an erroneous analysis
of the interaction of two light waves in Born-Infeld theory has been
published by C.\ D.\ Thomas (University of Chicago, USA) 
\citep{1936thomas}. He assumed unchanged phase velocities in his Ansatz
for solving the wave equation
(e.g., cf.\ \citep{1936thomas}, p.\ 1048, eq.\ (16)). This incorrect
assumption, however, leads to (physically not acceptable) wave solutions
whose amplitude is linearly growing in time 
(cf.\ \citep{1936thomas}, p.\ 1048, eqs.\ (14), (14$^\prime$))
invalidating hereby the further analysis.
The papers of Smirnov and Schr\"odinger both contain comments on 
the danger of such unphysical solutions (Smirnov: \citep{1936smirnov}
Chap.\ II, \S\ 2, pp.\ 35-37, pp.\ T-35 - T-37 (= P-39 - P-41) 
of the English translation, 
\citep{1940smirnov1}, p.\ 449, \citep{1941smirnov}, pp.\ 47-48;
Schr\"odinger: \citep{1942schroedinger1}, Part I, Sec.\ 4, p.\ 87 
(reprint \citep{1984schroedinger}, p.\ 319), around eq.\ (4.10)).\\

Finally, in 1943 J.\ McConnell (Dublin Institute for
Advanced Studies, Ireland) published an analysis of photon-photon
scattering in the framework of the Born-Infeld theory \citep{1943mcconnell}.
Again, the analysis is performed to leading 
nontrivial order in the constant $b$ (the critical field strength)
and can, therefore, be compared to the description of photon-photon
scattering in the framework of the Euler-Kockel-Heisenberg Lagrangian
\citep{1935euler,1936heisenberg,1936euler} of QED. The quantization is 
performed on the level of an effective field theory technique (phrasing
it in modern field theoretic terms) and the resulting cross section for 
photon-photon scattering is compared with the analogous QED result
(\citep{1943mcconnell}, Sec.\ 4, pp.\ 170-171). McConnell finds
that the angular dependence of the differential cross section for 
photon-photon scattering and the dependence on the photon frequency are
the same to lowest nontrivial order in (quantized) Born-Infeld theory 
and QED.\\

\noindent
2.\ The scattering of light by electromagnetic background fields\\

\noindent
In the time period under review, the consideration of the scattering 
of light by electromagnetic background fields has been performed for
two different situations: for constant homogeneous electric and magnetic fields
and for the presence of a Coulomb field.\\

A first short note studying the case of a constant homogeneous electric
background field is published by S.\ P.\ Shubin (Ural 
Physico-Technical Institute, Sverd\-lovsk, USSR) and Smirnov in 1936 
\citep{1936shubin}. Details of the calculation are contained in the 
candidates (Ph.D.) thesis of Smirnov (\citep{1936smirnov}, 
Chap.\ II, \S\ 2, pp.\ 33-48, pp.\ T-33 - T-48 (= P-37 - P-52)
of the English translation)
which also comprises an analogous investigation of the case of 
a magnetic background field
(Chap.\ II, \S\ 4, pp.\ 55-58, pp.\ T-55 - T-58 (= P-59 - P-62)
of the English translation).
Shubin and Smirnov primarily find for a weak test wave that 
to leading order (in the critical field strength $b$) the electric
background field causes a change in the phase velocity of the test
wave, in dependence on the direction of its propagation and its polarization. 
They compare the situation found for the Lagrangian (\ref{bornaction})
to the birefringence found in an uniaxial crystal (also see 
\citep{1937shubin}, footnote
$\ast $ on p.\ 132 of the English version; this footnote does not
exist in the Russian version).
In his thesis (\citep{1936smirnov}, 
Chap.\ II, \S\ 3, pp.\ 48-55, pp.\ T-48 -T-55 (= P-52 - P-59)
of the English translation.), 
Smirnov extends the analysis to the next to leading order in $b$.
As additional effects of the electric background field he obtains
higher harmonics generation in the forward and backward directions
of the test wave and also a small component scattered backwards with
the frequency of the test wave itself. A couple of years later, in 1942
Schr\"odinger \citep{1942schroedinger1}, as one part of a broad analysis 
of the consequences of the Born-Infeld Lagrangian (\ref{bornactiondet}), 
also observes that an electric background field exerts a refractive effect 
on a weak test wave (p.\ 102; p.\ 334 of the reprint). 
In accordance with general insight obtained 
decades later \citep{1970boillat,1970plebanski}, 
in difference to Shubin and Smirnov who rely on the Lagrangian
(\ref{bornaction}) he finds no birefringence.\\

Within nonlinear electrodynamics, the first study of the situation that 
corresponds to Delbr\"uck scattering in QED, i.e., the scattering
of light by a Coulomb field, has been published in 1937 by Shubin
and Smirnov \citep{1937shubin}. For a fairly general class of 
nonlinear electrodynamic Lagrangians, they analyze the scattering of light
from a point charge and find the polarization averaged differential
cross section (in a certain approximation). Then, they specify their
result for the choice of the Born-Infeld Lagrangian (\ref{bornactiondet}).
Furthermore, they also include a discussion of their results on 
a level where comparison with the corresponding results for the 
Euler-Kockel-Heisenberg Lagrangian is possible. In 1940, Smirnov 
publishes a follow-up paper \citep{1940smirnov2} where instead of
the Born-Infeld Lagrangian the Lagrangian proposed by Hoffmann and 
Infeld \citep{1937hoffmann} is being used. The work by Shubin and
Smirnov was followed in 1938 by a paper by S.\-I.\ Tomonaga 
and M.\ Kobayasi (both at RIKEN -- Institute of Physical
and Chemical Research, Tokyo, Japan) who along analogous lines 
analyzed the low-frequency scattering of light by a point charge
in Born-Infeld theory. A couple of years later, in a comprehensive
study Schr\"odinger \citep{1942schroedinger1} takes up the subject
(Part II, pp.\ 100-116; pp.\ 332-348 of the reprint).
In part he agrees with the results of Shubin/Smirnov and 
Tomonaga/Kobayasi for the cross section of the scattering of light
from a charged point charge, in part he finds further terms (cf.\ eq.\
(10,10) on p.\ 109; p.\ 341 of the reprint).\\

\noindent
3. The splitting of photons in electromagnetic background fields\\

\noindent
Tomonaga and Kobayasi in their paper \citep{1938tomonaga} 
also considered for the first time within Born-Infeld theory the 
process of photon splitting, i.e., the splitting of one incoming
photon in the Coulomb field of a point charge into two outgoing
photons. In the case of QED, the same process had been mentioned
before by E.\ J.\ Williams (University of 
Copenhagen, Denmark) in 1935 \citep{1935williams}, \S\ 10,
pp.\ 47-49 (cf.\ the subsec.\ \ref{qed}). Tomonaga and Kobayasi
calculate the differential cross section for the photon splitting
process in some low-frequency approximation.\\

\noindent
4.\ The Stefan-Boltzmann law\\

\noindent
L.\ Rosenfeld and E.\ E.\ Witmer (University of G\"ottingen, Germany)
\citep{1928rosenfeld} have pointed out
in 1928 that interactions between photons will lead to changes
in the characteristics of black body radiation. Their discussion
amounts to a qualitative consideration without specifying 
details of the interactions among the photons being part of the
cavity radiation. They only mention (p.\ 521) 
that possibly correction terms to the Maxwell equations
could describe such interactions. Born-Infeld theory just specifies
such correction terms to the Maxwell equations. Consequently, 
in 1936 B.\ Kwal and J.\ Solomon (France) \citep{1936kwal} for 
the first time quantify
the leading correction to the Stefan-Boltzmann law for the 
energy density $u$ of black body radiation. They find
($\sigma$ is some constant):
\begin{eqnarray}
\label{stefan}
u&=&\sigma\ T^4\ \left(1 + \frac{14\pi\sigma T^4}{b^2}\right)\ .
\end{eqnarray}
Some time later, both authors present a derivation of the
Stefan-Boltzmann law within a fairly large class of nonlinear
electrodynamic theories \citep{1938kwal}. Equations analogous to
eq.\ (\ref{stefan}) have later also been obtained by E.\ 
Milkutat (Berlin, Germany) \citep{1938milkutat} and Schr\"odinger
\citep{1942schroedinger1}, p.\ 100, eq.\ (7, 18) 
(p.\ 332 of the reprint \citep{1984schroedinger}).\\

The early development of Born-Infeld electrodynamics in general has been
described by few reviews. The most comprehensive one is the 
review written by Max Born himself in 1937 \citep{1937born}. 
Less comprehensive are conference expositions of 1935 by Born
\citep{1935born3} and of 1937 by Chrap\l ywyj \citep{1937chraplywyj3}.
Finally, it also seems worth mentioning sections on Born-Infeld theory in
the monographs by Frenkel' \citep{1934frenkel2} 
\begin{otherlanguage}{russian}
(гл.\
\end{otherlanguage}
%{\cyrrm gl.}\ 
[gl.]/[chap.] X, \S 11, \S 12), Sommerfeld \citep{1948sommerfeld} 
(Teil [part] IV, \S 37), and Ivanenko/Sokolov \citep{1949ivanenko} (\S
32).\\

We conclude this section with two statements (dating from different times)
by Albert Einstein and Max Born himself concerning Born-Infeld 
electrodynamics. In a private letter of March 22, 1934, Albert Einstein
writes to Max Born (\citep{1969einstein}, letter 69, pp.\ 169/170, 
for the original German text see our App.\ \ref{originals}, 
p.\ \refstepcounter{dummy}\pageref{aeinstein}; here we quote the English 
translation \citep{1971einstein}, p.\ 122.)\label{einstein}:
``I am greatly interested in your attempt to attack the quantum
problem of the field from a new angle, but I am not exactly
convinced. I still believe that the probability interpretation
does not represent a practicable possibility for the relativistic
generalisation, in spite of its great success. Nor has the reasoning
for the choice of a Hamiltonian function for the electromagnetic
field, by analogy with the special theory of relativity, convinced
me. I am afraid that none of us will live to see the solution of
these difficult problems.''. Max Born comments this letter many years
later (in 1969) by the following words (\citep{1969einstein}, pp.\ 170/171, 
for the original German text see our App.\ \ref{originals}, 
p.\ \refstepcounter{dummy}\pageref{aborn}; here we quote 
the English translation \citep{1971einstein}, pp.\ 122/123.)\label{born}:
``Einstein's objections to my ideas were twofold. The first was based on his
rejection of the probability of quantum mechanics. This concerns a matter
of principle. It did not really apply to the theory devised by Infeld and
myself, because we ourselves did not in fact manage to make it fit in with
quantum mechanics; he judged our efforts in this direction to be wrong
in principle. Einstein's second objection concerned our original classical
field theory, which was complete in itself and free from inconsistencies. It
was based on the following analogy: in the special theory of relativity the
kinetic energy of a particle, which in classical mechanics is proportional
to the square of its velocity, 
is represented by a rather complicated expression;
for velocities which are small compared with that of light it tends to
the classical expression, but deviates from it when the velocity approaches
that of light. In Maxwell's electrodynamics the energy density is a
quadratic expression containing the field intensity. I replaced this with a
general expression which approximates to the classical expression whenever
the strength of the field is small compared with a certain field intensity,
but diverges from it when this is not the case. 
From this it followed automatically
that the total energy of the field of a point charge is finite, while
it becomes infinite in the Maxwellian field. The absolute field has to be
regarded as a new natural constant. Einstein did not find this analogical
construction convincing. Infeld and I found it attractive for a long time.
We abandoned the theory for completely different reasons, namely,
because we did not succeed in reconciling it with the principles of the
quantum field theory. In any case this constituted the first attempt to
overcome the difficulties of microphysics by means of a non-linear theory.
Heisenberg's theory of elementary particles, which is much talked about
today, is also non-linear. But I am guessing.''.\\

\subsection{\label{qed}Quantum electrodynamics}

As we have discussed in the Introduction (section \ref{introduction}), 
in the late 1920's among physicists the idea
developed that the emerging quantum theory should also allow and 
describe the scattering of photons by photons. However, the physical
mechanism for such scattering events initially remained unclear.
With the proposal by P.\ A.\ M.\ Dirac
of the equation named after him (1928), 
the discovery of the positron (C.\ D.\ Anderson, 1932), 
and the first steps towards the
theory of quantum electrodynamics elements of a possible
explanation became available. O.\ Halpern (New York University, USA) 
\citep{1933halpern} proposed in late 1933 in a short, qualitative note
that the occurrence of virtual electron-positron pairs
could be identified as the physical mechanism by means of which
photon-photon scattering occurs (a more elaborate quantitative discussion
announced by the author at the end of the note seems not to have
emerged). The note of Halpern had been preceeded by the discussion of
two-photon production in ``electron-proton'' annihilation\footnote{This 
process of ``electron-proton'' annihilation found its proper 
re-interpretation as electron-positron pair annihilation after 
the discovery of the positron in 1932 only.} by Dirac
\citep{1930dirac}, Oppenheimer \citep{1930oppenheimer}, Tamm \citep{1930tamm},
the process of electron-positron pair creation 
by a photon in the field of an atomic nucleus by Oppenheimer and
Plesset \citep{1933oppenheimer} (also see, published somewhat later,
Heitler and Sauter \citep{1933heitler}, Bethe and Heitler \citep{1934bethe},
Nishina, Tomonaga, and Sakata \citep{1933nishina,1934nishina}, 
Racah \citep{1934racah}),
and the Breit-Wheeler process \citep{1934breit1,1934breit2} of 
electron-positron creation in two-photon scattering.
Furthermore, also P.\ Debye is credited by W.\ Heisenberg
to have mentioned similar ideas as Halpern in a private discussion with 
him\footnote{\citep{1934heisenberg}, p.\ 228, footnote 2 
[also cf.\ \citep{1935euler}, p.\ 246, footnote 5, \citep{1936euler},
p.\ 398, footnote 3 (footnote 2 in the separate Thesis print)]}.
In early 1935, D.\ D.\ Ivanenko 
(Physico-Technical Institute, Leningrad, USSR) 
\citep{1935ivanenko}, with reference to the article \citep{1934heisenberg} 
(containing references to Halpern \citep{1933halpern} and Debye),
also discussed in a short qualitative note the possible role of 
the occurrence of virtual electron-photon pairs for photon-photon
scattering. He points out that the existence of photon-photon scattering
entails the violation of the superposition principle [characteristic
for the (linear) Maxwell theory of electromagnetism], i.e., the phenomenon
of photon-photon scattering is characteristic for a nonlinear
electromagnetic theory. Consequently, as Ivanenko notes, the 
Dirac theory leads to some nonlinear electromagnetic theory which
needs to identified. Somewhat formal attempts to link the Dirac equation with 
nonlinear electrodynamics, specifically with the theory proposed 
by Born, have been published in 1935 by K.\ V.\ Nikolski\u\i\ 
(Steklov Mathematical Institute, Moscow, USSR)
\citep{1935nikolskii1,1935nikolskii2}.\\

A first answer to the task described by Ivanenko 
is given almost simultaneously in time:
A completely new level in the discussion of photon-photon scattering
is reached with the appearance in print in spring 1935 of a short article 
by H.\ Euler and B.\ Kockel (both doctoral students of 
Heisenberg at the University of Leipzig, Germany; for some 
reminiscences of W.\ Heisenberg concerning this research
see our Appendix \ref{appheisenberg}) \citep{1935euler}.
The article marks the transition from qualitative considerations
concerning the phenomenon of photon-photon scattering to detailed 
quantitative calculations. The short article is followed in 1936
by the doctoral thesis of H.\ Euler \citep{1936euler} \footnote{In 
his thesis review (University of Leipzig Archive, Philosophical Faculty,
Prom.\ 769, reprinted in \citep{2001rechenberg}, pp.\ 125/126), Heisenberg
credits Debye with having initiated this thesis research by posing him
a question (quoted here after \citep{2001rechenberg}, p.\ 126:
cf.\ our App.\ \ref{originals}, 
p.\ \refstepcounter{dummy}\pageref{aheisenberg2};
English transl.: K.S.)\label{heisenberg2}: ``The subject of 
the present work originates from a question posed to me by colleague Debye.''} 
which contains a comprehensive exposition of the calculation user by Euler and 
Kockel to arrive at the results presented in \citep{1935euler}
(For a detailed discussion of certain aspects of this Thesis see sec. 2.1,
pp.\ 23-28 of \citep{2010wuethrich}.).
Euler and Kockel determined the leading
perturbative correction (for small field strengths and photon
frequencies well below the electron-positron pair production 
threshold) to the Lagrange density ${\cal L}_0$ of the Maxwell field. 
They find (cf.\ eq.\ (10,6) on p.\ 444 of \citep{1936euler},
written here in SI units):
\begin{eqnarray}
\label{eulerkockela}
{\cal L}&=&{\cal L}_0
\ +\ \frac{\displaystyle\alpha^2}{\displaystyle 360}\ 
\frac{\displaystyle\epsilon_0^2 c^3 \lambda_e^4}{\displaystyle\hbar}\
\left[ 4 \left(F^{\mu\nu} F_{\mu\nu}\right)^2\ +\ 
7 \left(F^{\mu\nu} F^\star_{\ \mu\nu}\right)^2\right]\\[0.3cm]
\label{eulerkockelb}
&=&{\cal L}_0\ + \
\frac{\displaystyle 2 \alpha^2}{\displaystyle 45}\ 
\frac{\displaystyle\epsilon_0^2 \lambda_e^4}{\displaystyle\hbar c}\
\left[\left({\bf E}^2 - c^2 {\bf B}^2\right)^2\ +\ 
7 c^2 \left({\bf E}{\bf B}\right)^2\right]\\[0.3cm]
\label{eulerkockelc}
&=&{\cal L}_0\ + \
\frac{\displaystyle 2 \alpha^2}{\displaystyle 45}\ 
\frac{\displaystyle\lambda_e^4}{\displaystyle\hbar c}\
\left[4\ {\cal L}_0^2\ +\ 
\frac{\displaystyle 7}{\displaystyle\mu_0^2} 
\left({\bf E}{\bf B}\right)^2\right]
\end{eqnarray}
with the free field Lagrange density
\begin{eqnarray}
\label{freelagrangian}
{\cal L}_0&=&-\frac{\displaystyle 1}{\displaystyle 4\mu_0}\ 
F^{\mu\nu} F_{\mu\nu}\ =\
\frac{\displaystyle 1}{\displaystyle 2}\ \epsilon_0 
\left({\bf E}^2 - c^2 {\bf B}^2\right)
\ =\ \frac{\displaystyle 1}{\displaystyle 2}\
\left(\epsilon_0 {\bf E}^2 
- \frac{\displaystyle 1}{\displaystyle\mu_0} {\bf B}^2\right)\ .
\end{eqnarray}
The used symbols are:
$\alpha$ is the fine structure
constant, $\lambda_e =\hbar/(m_e c)$ is the (reduced) Compton wavelength
of the electron with mass $m_e$, the velocity of light (squared) is given
by $c^2 = 1/(\epsilon_0\mu_0)$ with $\epsilon_0$, $\mu_0$ being the dielectric
permeability and the magnetic permittivity of the vacuum, respectively, 
$F^\star_{\ \mu\nu} = \epsilon_{\mu\nu\alpha\beta} F^{\alpha\beta}/2$ is the dual
of the electromagnetic field strength tensor $F^{\mu\nu}$.
Using the relations given in the eqs.\ (\ref{stressenergytensorsquaredb}),
(\ref{stressenergytensorsquaredc}) of Appendix \ref{appstress}, one can
write eq.\ (\ref{eulerkockelc}) the following way
\begin{eqnarray}
\label{eulerkockeld}
{\cal L}&=&{\cal L}_0\ + \
\frac{\displaystyle 2 \alpha^2}{\displaystyle 45}\ 
\frac{\displaystyle\lambda_e^4}{\displaystyle\hbar c}\
\left[-10\ {\cal L}_0^2\ +\ 
\frac{\displaystyle 7}{\displaystyle 4}\ 
T^{\mu\nu} T_{\mu\nu}\right]\\[0.3cm]
\label{eulerkockele}
&=&{\cal L}_0\ + \
\frac{\displaystyle 2 \alpha^2}{\displaystyle 45}\ 
\frac{\displaystyle\lambda_e^4}{\displaystyle\hbar c}\
\left[-10\ {\cal L}_0^2\ +\ 
7\ \left(\left(T_{00}\right)^2\ -\ 
\frac{\displaystyle 1}{\displaystyle c^2}\ {\bf S}^2 
\right)\right]
\end{eqnarray}
where $T^{\mu\nu}$ (eq.\ (\ref{stressenergytensor}))  
is the stress-energy-momentum tensor
of the free electromagnetic field, and the Poynting vector is given by
\begin{eqnarray}
\label{poynting}
{\bf S}&=&\frac{\displaystyle 1}{\displaystyle \mu_0}\ 
{\bf E}\times {\bf B}\ .
\end{eqnarray}
The found (Euler-Kockel-Heisenberg) Lagrangian (EKH Lagrangian) is a nonlinear
Lagrangian for the electromagnetic field which leads to nonlinear
corrections to the classical Maxwell equations. The resulting
(effective) Maxwell equations do not respect the superposition
principle and describe nonlinear electromagnetic phenomena such as
photon-photon scattering. One can recognize from eq.\ 
(\ref{eulerkockele}) that the term quadratic in the 
Poynting vector provides us with an interaction term between 
electromagnetic (photon) energy currents which is relevant for
the experimental proposal by Bauer \citep{1931bauer} discussed at the 
end of subsec.\  \ref{expconst}. Euler and Kockel also give
for the first time a quantitative estimate for the cross section $\sigma$
of photon-photon scattering. They find (for photons of frequency 
$\hbar\omega\ll m_e c^2$):
\begin{eqnarray}
\label{photoncrosslow}
\sigma&\sim &\alpha^4\ \lambda_e^8\ \left(\frac{\omega}{c}\right)^6
\end{eqnarray}
entailing a size of the cross section of $10^{-30}\ cm^2$ for $\gamma$-rays
and of $10^{-70}\ cm^2$ for visible light (\citep{1935euler}, 
bottom of p.\ 247, \citep{1936euler}, p.\ 446, 
eq.\ (10,10))\footnote{In a letter of June 17, 1934 quoted here
from \citep{1985vonmeyenn}, p.\ 331 (cf.\ our App.\ \ref{originals}, 
p.\ \refstepcounter{dummy}\pageref{aheisenberg3}; English transl.: K.S.), 
W.\ Heisenberg mentions to N.\ Bohr that\label{heisenberg3}:
``Debye came up with the idea that the 
solar corona originates from this scattering of light;
the above value [$(e^2/\hbar c)^4 (\hbar/m c)^2$ for the cross 
section] would excellently fit this thesis. However, as already said,
one still has to see whether all calculations are correct.'' 
In a letter one day earlier (\citep{1985vonmeyenn}, letter [374]
of June 16, 1934, pp.\ 331-333, specifically p.\ 332: 
cf.\ our App.\ \ref{originals}, 
p.\ \refstepcounter{dummy}\pageref{aheisenberg4}; English transl.: K.S.), 
W. Heisenberg also wrote to W.\ Pauli\label{heisenberg4}:
``This value seems to fit well to the idea of
Debye that the solar corona emerges through this scattering of 
light by light.'' S.\ I.\ Vavilov had discussed the possible role of 
the scattering of light by light for the emergence of the solar corona
in his article \citep{1928vavilov} in 1928 (cf.\ subsec.\ \ref{astro})
which is the written version of a talk delivered by him at the
VI Congress of Russian Physicists in the same year (cf.\ \citep{1928sezd}, 
p.\ 47, item 80). P.\ Debye has been a participant to this congress
(cf., e.g., \citep{1966frenkel}, chap.\ 6, p.\ 226, English translation:
chap.\ 4, p.\ 140).
It is well possible that he had listened to the talk of S.\ I.\ Vavilov.}.
These cross sections are fairly small and Euler immediately points
out (\citep{1936euler}, p.\ 446, below of eq.\ (10,10)) that 
this will make it difficult to experimentally prove the 
phenomenon of photon-photon scattering. This is a comment proving
its significance even today after 80 years. Furthermore, the
cross section estimate of Euler and Kockel explains the 
negative results of all experiments performed earlier (discussed in subsec.\ 
\ref{scatteringexp}) to directly detect photon-photon scattering.\\

Roughly one year after the article by Euler and Kockel \citep{1935euler}
an article by Heisenberg and Euler \citep{1936heisenberg} appeared 
in print which considerably extended the result 
obtained by the former. While Euler and Kockel limited the 
perturbative calculation of the effective Maxwell Lagrangian to
sufficiently small electromagnetic fields, Heisenberg and Euler 
solved the Dirac equation for constant, arbitrarily strong, parallel 
magnetic and electric fields and calculated on the basis of this 
information the effective Maxwell Lagrangian. This (in modern terms:
1-loop) effective Maxwell (Euler-Heisenberg) 
Lagrangian for constant electromagnetic
fields together with its weak-field limit, the EKH Lagrangian,
represents a milestone in the history of quantum field theory.\\

An independent confirmation of the results obtained by Euler and
Kockel came in 1936 with a short note by N.\ Kemmer and V.\ F.\
Weisskopf (University of Zurich and ETH Zurich, Switzerland) 
\citep{1936kemmer}. Some aspects of this work, which cannot easily be 
read from this 
short note, are pointed out by V.\ F.\ Weisskopf in an oral 
history interview in 1965\footnote{Cf.\ our Appendix 
\ref{appweisskopf} (quotation with kind permission of the 
American Institute of Physics), for some
related historic recollections by Kemmer see our Appendix \ref{appkemmer}.}:
``We connected the scattering of light
by light with the Delbr\"uck scattering. Today it's a triviality; 
one light quantum is replaced by the Coulomb field\footnote{K.\ S.:
More precisely, they consider for the scattering electric field ${\bf E}$ the
condition 
$\vert {\rm grad}\;{\bf E}\vert\ \ll\ \vert {\bf E}\vert\ m c/h$ (weakly
varying fields at a length scale of the electron Compton wavelength)
which in the strict sense is not fulfilled for the Coulomb field: 
in a region close to the center of the nucleus.}$^{[footn.\ K.S.]}$, 
but at that time it was not so trivial. $\dots$ Euler and Kockel at that 
time, under Heisenberg, calculated the scattering of light by
light, but had to do a lot of subtracting because there were a 
great many terms that were infinite. They did this in the usual 
clever way and got the result. And Kemmer and I showed that you 
can do the calculation without raking any subtractions, because 
you can show that it is equivalent to the Delbr\"uck scattering, 
replacing one light quantum by the Coulomb field, and the
Delbr\"uck scattering doesn't diverge.''
The same year, a long paper by Weisskopf \citep{1936weisskopf} presented 
a more efficient and improved re-derivation of the Euler-Heisenberg Lagrangian
(For some historic recollections
by V.\ F.\ Weisskopf also concerning this article see 
our Appendix \ref{appweisskopf}.). 
In \citep{1936weisskopf}, Weisskopf
in principle follows the path taken by Heisenberg and Euler 
\citep{1936heisenberg} but solves the Dirac equation for a constant
magnetic field accompanied by a parallel, spatially periodic
electric field. While Heisenberg and Euler, considering a constant
electric field, had to deal with the Klein paradox 
(any spatially constant electric field creates electron-positron
pairs) Weisskopf was able to sail around this difficulty by
considering a sufficiently weak, spatially periodic electric 
field for which no electron-positron pair creation occurs.
The result of his calculation confirmed the expression 
obtained by Heisenberg  and Euler for the effective Maxwell 
(Euler-Heisenberg) Lagrangian.\\

The theoretical study of photon-photon scattering was also in the
focus of research interests elsewhere: L.\ D.\ Landau (Ukrainian
Physico-Technical Institute, Kharkov, USSR) 
handed the task of studying certain aspects of photon-photon
scattering to his doctoral student A.\ I.\ Akhiezer.
As a first result a short note by Akhiezer, Landau, and I.\ Ya.\
Pomeranchuk (Ukrainian Physico-Technical Institute, Kharkov, USSR)
appeared \citep{1936akhiezer} treating for the photon-photon
scattering cross section the high-frequency case $\hbar\omega\gg m_e c^2$. 
They found:
\begin{eqnarray}
\label{photoncrosshigh}
\sigma&\sim &\alpha^4\ \left(\frac{c}{\omega}\right)^2
\end{eqnarray}
noting that the proportionality factor is difficult to compute
(In the low frequency case $\hbar\omega\ll m_e c^2$ studied by 
Euler and Kockel \citep{1935euler} the proportionality factor can be found
from the EKH Lagrangian.). The result by Akhiezer, Landau, and Pomeranchuk
showed that the cross section $\sigma$ for photon-photon scattering
must have ``a maximum value in a region $\hbar\omega\sim m c^2$\ ''
($m = m_e$) \citep{1936akhiezer}. The kand.\ diss.\ (Ph.D.\ Thesis) research
of A.\ I.\ Akhiezer giving the details of the calculation leading
to eq.\ (\ref{photoncrosshigh}) is finally published in 1937
\citep{1937achieser} (For historic recollections concerning
the thesis research of  A.\ I.\ Akhiezer see our Appendices 
\ref{appakhiezer}, \ref{apptisza}.). It should be mentioned that
in 1935 also W.\ Pauli (University of Zurich, Switzerland)
spent some time attacking the problem of high-frequency 
photon-photon scattering\footnote{See his letter
[421a], \citep{1993vonmeyenn}, pp.\ 769-771 (specifically p.\ 770)
of September 27, 1935 to Weisskopf, the corresponding section
is quoted in our Appendix \ref{apppauli}.}. However, the calculation
proved to be difficult and in a letter of December 5, 1935\footnote{Reprinted 
as letter [423b] in \citep{1993vonmeyenn}, pp.\ 777/778, specifically p.\ 778:
cf.\ our App.\ \ref{originals}, 
p.\ \refstepcounter{dummy}\pageref{apauli1}; English transl.: K.S..} 
Pauli informed Weisskopf\label{pauli1}: 
``Presently, the ``Euler-Kockel'' problem for short waves 
looks somewhat bleak to me.''\setcounter{footnote}{0}\renewcommand{\thefootnote}{[\alph{footnote}\alph{footnote}]} 
and in a later letter of January 20, 1936\footnote{Reprinted 
as letter [425c] in \citep{1993vonmeyenn}, pp.\ 780-782, specifically p.\ 782:
cf.\ our App.\ \ref{originals}, 
p.\ \refstepcounter{dummy}\pageref{apauli2}; English transl.: K.S..}
Pauli told Weisskopf\label{pauli2}:
``After a discussion with Oppenheimer, I have now definitely 
given up as too complicated the problem of the scattering of 
light by light for short waves.''
A short time later, the {\it Nature} article by 
Akhiezer, Landau, and Pomeranchuk \citep{1936akhiezer} appeared in 
print.\\

Besides the fundamental nonlinear quantum-electrodynamic 
process of photon-photon scattering related phenomena have found 
attention at about the same time. As discussed in subsec.\ \ref{delbrueck},
in 1933 Delbr\"uck \citep{1933delbrueck} had qualitatively 
considered the interaction of photons with the Coulomb field of
nuclei via their impact on the surrounding QED vacuum\footnote{For 
another discussion of the articles to be considered in this paragraph
on Delbr\"uck scattering see \citep{1975papatzacos}, subsec.\ 3.1,
pp.\ 84-87.}. However,
the cross section calculations related to this process Delbr\"uck
attempted in the time thereafter proved to be extremely difficult
(cf.\ footnote \ref{footdelbrueck} on 
p.\ \refstepcounter{dummy}\pageref{footdelbrueck}).
Weisskopf then independently took a look at the problem, but
as W.\ Pauli reported to W.\ Heisenberg in 
June 1935\footnote{\citep{1985vonmeyenn}, letter [412] of June 15, 1935,
pp.\ 402-405, specifically p.\ 402: cf.\ our App.\ \ref{originals}, 
p.\ \refstepcounter{dummy}\pageref{apauli3}; 
English transl.: K.S..}\label{pauli3}: ``Weisskopf is on the 
Delbr\"uck problem and the problems of subtraction physics 
occurring there are getting not very nice.'' Finally, in 1936
the joint work of Weisskopf and Kemmer paid off and as a first result
-- as already mentioned above -- they established a connection 
between the scattering of light by light in the low-frequency
domain and Delbr\"uck scattering (understood here in a somewhat
generalised sense as scattering of photons from a sufficiently 
slowly varying electric field)  
and confirmed this way the expression for the EKH Lagrangian \citep{1936kemmer}
(also cf.\ our Appendices \ref{appweisskopf}, \ref{appkemmer}).
Details of the calculation were given by Kemmer in 1937 in 
\citep{1937kemmer1}\footnote{For a rare glimpse into the communication 
between Kemmer and Weisskopf concerning this work see the letter
of June 8, 1936 from Kemmer to Weisskopf (letter [430b] reprinted in 
\citep{1993vonmeyenn}, pp.\ 792-794). For the interest of Pauli
in this problem see his letters of May 19, 1936 and of
June 3, 1936 to Weisskopf (letters [427f] and [430a] reprinted in
\citep{1993vonmeyenn}, pp.\ 788-790 (specifically p.\ 788) and 
 pp.\ 790-792 (specifically p.\ 791), respectively.)}. 
A short follow-up article by Kemmer and Ludwig
\citep{1937kemmer2} explored the calculational 
aspect that occur if the assumption
of a sufficiently slowly varying electric field is given up and 
the Coulomb field is considered instead. For this case, they 
derived qualitatively an expression for the low-frequency (differential)
cross section 
[cf.\ eq.\ (\ref{kemmerludwigcross}) at the end of subsec.\ \ref{delbrueck}] 
which, however, contained two unknown constants.
Simultaneously with Kemmer and Ludwig, Akhiezer and Pomeranchuk
had considered Delbr\"uck scattering \citep{1937akhiezer}.
They obtained in the low- and high-frequency limits ($\hbar\omega\ll m_e c^2$
and $\hbar\omega\gg m_e c^2$, respectively; $\omega$ is the photon
frequency) expressions for the
total cross section whose size however remained undetermined up to 
a constant. Specifically, they found for $\hbar\omega\ll m_e c^2$
[in accordance with the above eq.\ (\ref{kemmerludwigcross}) on page
\refstepcounter{dummy}\pageref{kemmerludwigcross}]:
\begin{eqnarray}
\label{delbrueckcrosslow}
\sigma&=&
b\ Z^4\ \left(\frac{e^2}{m_e c^2}\right)^6\ \left(\frac{\omega}{c}\right)^4
\end{eqnarray}
and for $\hbar\omega\gg m_e c^2$ (the absorptive contribution of 
real electron-positron pair creation is not included):
\begin{eqnarray}
\label{delbrueckcrosshigh}
\sigma&=&a\ \alpha^2\ Z^4\ \frac{c^2}{\omega^2}\ \ln\frac{\hbar\omega}{m_e c^2}
\end{eqnarray}
where $a$ and $b$ are unknown constants.  Akhiezer and Pomeranchuk
conclude again\footnote{\citep{1937akhiezer}, 
p.\ 567 in the Russian original (cf.\ our App.\ \ref{originals}, 
p.\ \refstepcounter{dummy}\pageref{aakhiezer}), p.\ 7 of the reprint,
p.\ 478 in the German transl.. English transl.: K.S..}\label{akhiezer}: 
``The effective cross section has a maximum at $\hbar\omega\sim m c^2$.''
($m = m_e$).\\

In the period under consideration, besides photon-photon scattering 
and Delbr\"uck scattering the possible phenomenon of splitting of photons 
has also been discussed. Halpern \citep{1933halpern} had mentioned 
in 1933 photon splitting as a process conceivable in principle
within quantum electrodynamics. Following Halpern, 
W.\ Heitler (University of Bristol, UK)
in 1936 in his book also mentioned this possibility 
(\citep{1936heitler}, 1.\ and 2.\ eds., Chap.\ IV, \S\ 19, pp.\ 193/194). 
Furthermore, he correctly notes that the probability for the
vacuum decay of a photon into two photons vanishes -- a modern consideration 
would refer to Furry's theorem \citep{1937furry} to infer this --
leaving as minimal possible case a decay into three photons.
Relying on the formalism used by Heisenberg in \citep{1934heisenberg},
M.\ P.\ Bronshte\u\i n (Leningrad Physico-Technical Institute, USSR) 
in 1937 concludes \citep{1937bronshtein} (preceeded by a corresponding 
short note in 1936 \citep{1936bronstein}) that in free Minkowski
space (spontaneous) photon splitting is not possible at all within
the framework of quantum electrodynamics\footnote{For a modern argument see
\citep{1970bialynickabirula}, sec.\ III, p.\ 2343. Note that in eq.\
(31) the first sum sign should be read as a product sign.}. 
On the other hand, photon splitting in the presence of 
an external field is possible and Williams discussed in 1935 
\citep{1935williams}, \S\ 10,
pp.\ 47-49, for the first time the splitting of a photon 
in the presence of an electric particle (electric field).\\

Within the framework of quantum electrodynamics, the further
investigation of such processes as photon-photon scattering,
Delbr\"uck scattering, and photon splitting continued in the 
1950's only and falls, therefore, beyond the time frame of 
the present exposition. In the 1930's and 1940's these subjects
have been considered both within Born-Infeld electrodynamics and
in quantum electrodynamics. Interestingly, however, the theoretical study 
of the propagation of light in constant homogeneous magnetic and 
electric fields has been performed in the period under review 
within Born-Infeld electrodynamics only. Corresponding studies within
QED began in the 1950's only (J.\ S.\ Toll).\\

\section*{Acknowledgements\phantomsection}
\addcontentsline{toc}{section}{Acknowledgements}

I am grateful to I.\ A.\ Smirnova (Kiev) and S.\ V.\ Smirnov
(Helsinki) for granting me access to \citep{1936smirnov} and
to A.\ Kleinert [Halle(Saale)] and A.\ P.\ Nosov (Ekaterinburg) 
for providing me with a copy of \citep{1947stark1} and
\citep{1991shubin}, respectively. Thanks also go to V.\ V.\ Nesterenko (Dubna)
for communication concerning \citep{1936blokhintsev}.
Kind hospitality at the Department of Physics and Astronomy 
of the Vrije Universiteit Amsterdam is gratefully acknowledged.\\

\pagebreak
\section*{Appendices}
\addcontentsline{toc}{section}{Appendices}
\newcounter{asubsection}
\setcounter{asubsection}{1}
\renewcommand{\thesubsection}{\Alph{asubsection}}
\refstepcounter{subsection}
\label{originals}
\section*{Appendix \Alph{asubsection}}
\addcontentsline{toc}{subsection}{Appendix \Alph{asubsection}}

Original language versions of the quotations in the main text:\\

\begin{itemize}
\item[P.\ \refstepcounter{dummy}\pageref{kepler}:]\label{akepler} Kepler
(\citep{1604kepler}, p.\ 23, reprint (1859) p.\ 142, 
reprint (1939) p.\ 32, proposition 26):
``{\it Lucis radii se mutu\`{o} neque colorant, neque illustrant, 
neque impediunt vllo modo.} $\ldots$
San\`{e} vt nec physicus motus alter alterum mouet.''

\item[P.\ \refstepcounter{dummy}\pageref{alhaytham}:]\label{aalhaytham}
Ibn al-Haytham (\citep{1572risner}, book I, chap.\ 5, item 29, p.\ 17,
reprint \citep{2001smith1}, book I, chap.\ 7, p.\ 57, item [6.87]):
``Luces ergo non admisc\'{e}tur in aere, 
sed qu\ae libet illar\'{u} extenditur super uerticationes rectas; 
\& ill\c{e} uerticati\'{o}es sunt \ae quidist\'{a}tes, 
\& sec\'{a}tes se, \& diuersi situs. 
Et forma cuiuslibet lucis ext\'{e}ditur super o\'{e}s uerticati\'{o}es, 
qu\c{e} possunt extendi in illo aere ab illa hora: 
neq; tam\'{e} admiscentur in aere, nec aer tingitur per eas, 
sed pertranseunt per ipsius diaphanitat\'{e} tantum, 
\& aer non amittit su\'{a} form\'{a}.''

\item[P.\ \refstepcounter{dummy}\pageref{huygens}:]\label{ahuygens}
Huygens (\citep{1690huygens}, p.\ 20):
``Une autre, \& des plus merveilleuses preprietez de la lumiere
est que, quand il en vient de divers costez, ou mesme d'opposez,
elles font leur effet l'une \`{a} travers l'autre 
sans aucun emp\'{e}chement.''

\setcounter{extrafoot}{\value{footnote}}
\renewcommand{\thefootnote}{\arabic{footnote}}
\setcounter{footnote}{2}

\item[P.\ \refstepcounter{dummy}\pageref{vavilov1}:]\label{avavilov1}
Vavilov (\citep{1928vavilov}, p.\ 555, reprint \citep{1954vavilov1}, p.\ 234):
``\begin{otherlanguage}{russian}
Упрек, который в
\end{otherlanguage}
%{\cyrrm Uprek, kotory\u i v} 
XVIII.\ 
\begin{otherlanguage}{russian}
в.\ часто делали
ньютоновской теории, заключался именно в том, что должны
обнаруживаться соударения световых корпускул, т.\ е.\
на\-рушение суперпозиции. Выходом из затруднения было допущение
крайней малости корпускул: \guillemotleft Знаю, -- писал
$\,$Л$\,$о$\,$м$\,$о$\,$н$\,$о$\,$с$\,$о$\,$в
\end{otherlanguage}
%{\cyrrm v.\ chasto delali 
%n\cprime yutonovsko\u i teorii, zaklyuchalsya imenno v tom, chto dolzhny
%obnaruzhivat\cprime sya soudareniya svetovykh korpuskul, t.\ e.\ 
%na\-rushenie superpozitsii. Vykhodom iz zatrudneniya bylo dopushchenie
%kra\u ine\u i malosti korpuskul: <Znayu, -- pisal 
%$\,$L$\,$o$\,$m$\,$o$\,$n$\,$o$\,$s$\,$o$\,$v}
\footnotemark ${^{\it [K.S.:\ orig.\ }}$ $^{\it footn.]}\, $ 
\footnotetext{$^{\it [K.S.:\ orig.\ footn.]}\,$
\begin{otherlanguage}{russian}
М.\ В.\ Ломоносов, Слово о происхождении света. Собрание разных 
сочинений, ч.\
\end{otherlanguage}
%{\cyrrm M.\ V.\ Lomonosov, Slovo o proiskhozhdenii sveta. Sobranie raznykh 
%sochineni\u i, ch}.\ 
III, 1803, 
\begin{otherlanguage}{russian}
стр.\
\end{otherlanguage}
%{\cyrrm str}.\ 
155. (K.S.: M.\ V.\ Lomonosov, Slovo o 
proiskhozhdenii sveta. Sobranie raznykh sochineni\u{\i}, 
ch.\ (pt.) III, 1803, str.\ (p.) 155 \citep{1756lomonosov})}
\begin{otherlanguage}{russian}
по 
адресу защитников корпускулярной гипотезы, -- что вы разделяете
материю света на толь малые частицы и толь редко
одную по всемирному пространству поставляете, что все оное количество
может сжаться и уместиться в порожних скважинах
одной песчинки\guillemotright.''
\end{otherlanguage}
%{\cyrrm po 
%adresu zashchitnikov korpuskulyarno\u i gipotezy, -- chto vy raz\-delyaete
%materiyu sveta na tol\cprime\ malye chastitsy i tol\cprime\ redko 
%odnuyu po vsemirnomu prostranstvu postavlyaete, chto vse onoe kolichestvo
%mozhet szhat\cprime sya i umestit\cprime sya v porozhnikh skvazhinakh 
%odno\u i peschin\-ki>.}''

\item[P.\ \refstepcounter{dummy}\pageref{vavilov2}:]\label{avavilov2}
Vavilov (\citep{1928vavilov}, \S\ 2, pp.\ 556/557, 
reprint \citep{1954vavilov1}, p.\ 236):
``\begin{otherlanguage}{russian}
Едва ли не самые большие
$\,$м$\,$н$\,$г$\,$о$\,$в$\,$е$\,$н$\,$н$\,$ы$\,$е$\,$
плотности радиации в лабораторных условиях получаются при
помощи света конденсированной искры. Концентрируя
этот свет линзой, легко добиться
мнгновенных значений плотности лучистой энергии,
превышающих соответствующие значения у поверхности
Солнца. Средняя плотность в этом случае мала
вследствие краткой длительности и редкости искр,
но гипотетический эффект \guillemotleft соударений\guillemotright
световых квантов должен быть пропорциональным
$\,$к$\,$в$\,$а$\,$д$\,$р$\,$а$\,$т$\,$у$\,$ мнгновенной
плотности, поэтому искра оказывается значительно
выгоднее, чем, например, дуга. Наблюдения при этом
производились с обычными предосторожностями, на фоне
далеко отстоящих стенок сосуда, отклееных черным бархатом;
для контроля опыты повторялись со светом лампы накаливания,
которая давала такую же
$\,$с$\,$р$\,$е$\,$д$\,$н$\,$ю$\,$ю$\,$ плотность
радиации; в обоих случаях результат был одинаково
отрицательным.''
\end{otherlanguage}
%{\cyrrm Edva li ne samye bol\cprime shie 
%$\,$m$\,$n$\,$g$\,$o$\,$v$\,$e$\,$n$\,$n$\,$y$\,$e$\,$
%plotnosti radiatsii v laboratornykh uslo\-viyakh poluchayutsya pri
%pomoshchi sveta kondensirovanno\u i iskry. Kontsentriruya
%\protect{\`{e}}tot svet linzo\u i, legko dobit\cprime sya
%mngnovennykh znacheni\u i plotnosti luchisto\u i \protect{\`{e}}nergii,
%prevyshayushchikh sootvet\-stvuyushchie znache\-niya u poverkhnosti
%Solntsa. Srednyaya plotnost\cprime\ v \protect{\`{e}}tom sluchae mala
%vsledstvie kratko\u i dlitel\cprime nosti i redkosti iskr, 
%no gipoteti\-cheski\u i \protect{\`{e}}ffekt <soudareni\u i>
%svetovykh kvantov dolzhen byt\cprime\ proportsional\cprime nym
%$\,$k$\,$v$\,$a$\,$d$\,$r$\,$a$\,$t$\,$u$\,$ mngnovenno\u i 
%plotnosti, po\protect{\`{e}}tomu iskra okazyvaet\-sya znachitel\cprime no
%vygodnee, chem, naprimer, duga. Nablyudeniya pri \protect{\`{e}}tom
%proizvodilis\cprime\ s obychnymi predostorozhnostyami, na fone
%daleko ot\-stoyashchikh stenok sosuda, otkleenykh chernym barkhatom;
%dlya kontrolya opyty povtoryalis\cprime\ so svetom lampy nakalivaniya,
%kotoraya davala takuyu zhe
%$\,$s$\,$r$\,$e$\,$d$\,$$\,$n$\,$yu$\,$yu$\,$ plotnost\cprime\
%radiatsii; v oboikh sluchayakh rezul\cprime tat byl odinakovo
%otritsatel\cprime nym.}''

\renewcommand{\thefootnote}{[\alph{footnote}\alph{footnote}]}
\setcounter{footnote}{\value{extrafoot}}

\item[P.\ \refstepcounter{dummy}\pageref{kapitsa}:]\label{akapitsa}
Kapitsa (\citep{1980kapitsa1}, p.\ 31, \citep{1980kapitsa1}, p.\ 40,
reprints: 1. \citep{1980kapitsa2}, p.\ 40.
2. \citep{1981kapitsa}, 3.\ ed.\ p.\ 375, 4.\ ed.\ p.\ 374):
``\begin{otherlanguage}{russian}
В 30-х годах в Кавендишской лаборатории я осуществил
метод получения магнитных полей по силе на порядок
выше, чем до сих пор это было достигнуто. В одной беседе
Эйнштейн пытался меня убедить
экспериментально изучать влияние магнитного
поля на скорость распространения света.
Эти опыты уже делались, никакого эффекта
не было обнаружено. В моих магнитных полях можно было бы
поднять предел точности измерения порядка на два,
поскольку эффект должен был зависеть
от квадрата интенсивности магнитного поля. Я возражал
Эйнштейну, что, согласно существующей
картине электромагнитных явлений, не видно, откуда
можно было бы ждать таково измеримого явления. Не находя
возможности обосновать необходимость таких опытов,
Эйнштейн, наконец, сказал:
\guillemotleft Я думаю, что дорогой господь бог
\end{otherlanguage}
%{\cyrrm V 30-kh godakh v Kavendishsko\u i laboratorii ya osushchestvil
%metod polucheniya magnitnykh pole\u i po sile na poryadok
%vyshe, chem do sikh por \protect{\`{e}}to bylo dostignuto. V odno\u i besede
%\protect{\`{E}}\u inshte\u in pytalsya menya ubedit\cprime\ 
%\protect{\`{e}}ksperimental\cprime no izuchat\cprime\ vliyanie magnitnogo
%polya na skorost\cprime\ rasprostraneniya sveta. 
%\protect{\`{E}}ti opyty uzhe delalis\cprime , nikakogo \protect{\`{e}}ffekta
%ne bylo obnauruzheno. V moikh magnitnykh polyakh mozhno bylo by
%podnyat\cprime\ predel tochnosti izmereniya poryadka na dva, 
%poskol\cprime ku \protect{\`{e}}ffekt dolzhen byl zaviset\cprime\ 
%ot kvadrata intensivnosti magnitnogo polya. Ya vozrazhal 
%\protect{\`{E}}\u inshte\u inu, chto, soglasno sushchestvuyu\-shche\u i
%kartine \protect{\`{e}}lektromagnitnykh yavleni\u i, ne vidno, otkuda
%mozhno bylo by zhdat\cprime\ takovo izmerimogo yavleniya. Ne nakhodya
%vozmozhnosti obosnovat\cprime\ neobkhodimost\cprime\ takikh opytov,
%\protect{\`{E}}\u inshte\u in, nakonets, skazal:
% <Ya dumayu, chto dorogo\u i gospod\cprime\ bog}
(der liebe Gott)
\begin{otherlanguage}{russian}
не мог так создать мир, чтобы магнитное поле не
влияло на скорость света\guillemotright . Конечно, это
аргумент, с которым трудно спорить.''
\end{otherlanguage}
%{\cyrrm ne mog tak sozdat\cprime\ mir, chtoby magnitnoe pole ne 
%vliyalo na skorost\cprime\ sveta>. Konechno, \protect{\`{e}}to
%argument, s kotorym trudno sporit\cprime .}''

\item[P.\ \refstepcounter{dummy}\pageref{delbrueck1}:]\label{adelbrueck1}
Delbr\"uck \citep{1933delbrueck}:
``$\ldots$ da\ss\ es sich um einen {\it Photoeffekt} an einem der unendlich
vielen Elektronen in Zust\"anden negativer Energie handelt, die nach$\,$\ 
D$\,$i$\,$r$\,$a$\,$c$\,$s$\,$\ Theorie den
ganzen Raum mit unendlicher Dichte erf\"ullen und
die zu einem solchen Absorptionsproze\ss\ verm\"oge ihrer Wechselwirkung
mit dem Kern wohl bef\"ahigt w\"aren.''

\item[P.\ \refstepcounter{dummy}\pageref{delbrueck2}:]\label{adelbrueck2}
Delbr\"uck \citep{1933delbrueck}:
``Eine solche Auffassung zwingt dann auch zu der Folgerung,
da\ss\ diese Elektronen negativer Energie $\gamma$-Strahlen zu
{\it streuen} verm\"ogen, und zwar {\it koh\"arent}, analog dem Ph\"anomen
der {\it unverschobenen} Comptonlinie.'' 

\item[P.\ \refstepcounter{dummy}\pageref{heisenberg1}:]\label{aheisenberg1}
Heisenberg (\citep{1985vonmeyenn}, 
letter [374] of June 16, 1934, pp.\ 331-333, specifically p.\ 332):
``Delbr\"uck war neulich hier 
und erz\"ahlte von seinen vergeblichen Versuchen, die koh\"arente 
Streuung zu rechnen. Er w\"are
sicher froh, wenn er die Arbeit gemeinsam mit einem anderen z.B.\ Weisskopf
machen k\"onnte (Sauter ist jetzt in G\"ottingen).''

\item[P.\ \refstepcounter{dummy}\pageref{ebert}:]\label{aebert}
Ebert (\citep{1887ebert}, pp.\ 381-383):

\setcounter{extrafoot}{\value{footnote}}
\renewcommand{\thefootnote}{\arabic{footnote}}
\setcounter{footnote}{0}
\noindent
``Zum Schluss erlaube ich mir noch, das obige Resultat
auf einige astrophysikali\-sche Fragen anzuwenden.

\noindent
Ist die Fortpflanzungsgeschwindigkeit des Lichtes von
seiner Intensit\"at in irgend einem merklichen Grade abh\"angig,
so w\"urde zu dem Umstande, dass die uns am Himmel in
demselben Augenblicke entgegentretenden Erscheinungen in
Wirklichkeit sehr verschiedenen Zeiten angeh\"oren, noch
die weitere Complication hinzutreten, dass diese Zeiten
selbst bei gleicher r\"aumlicher Entfernung f\"ur die verschiedenen
Objecte sehr verschieden w\"aren. Bei den grossen
Intensit\"atsunterschieden, die sich hier vorfinden, und den
grossen Entfernungen, welche die Lichtstrahlen zu durchlaufen
haben, ehe sie zu uns gelangen, konn\-ten sich selbst
kleine Unterschiede in der Fortpflanzungsgeschwindigkeit in
hohem Grade geltend machen. Dies m\"usste z.\ B.\ bei den
physischen Doppelsternen eintreten; hier haben wir zwei
Lichtquellen, die zwar sehr angen\"ahert gleich weit von uns
abstehen, deren Lichtintensit\"at sich aber im allgemeinen sehr
erheblich voneinander unterscheidet. Trotzdem finden wir,
dass, wenn es uns gelungen ist die Bahnen von Doppelsternen
zu berechnen, die beiden Componenten wirklich zu gleicher
Zeit an entsprechenden Punkten ihrer Bahnen erscheinen
und mit ihrem gemeinsamen Schwerpunkte immer auf einer
gera\-den Linien liegen. Diese Thatsache konnen wir also
geradezu als neue St\"utze f\"ur das von uns experimentell festgestellte
Resultat heranziehen. Ein Beispiel m\"oge \"ubrigens
noch zeigen, dass die von mir nachgewiesene Grenze f\"ur die
Unabh\"angigkeit der beiden in Rede stehenden Gr\"ossen f\"ur
die astronomische Praxis jedenfalls v\"ollig ausreicht.

Der von $\,$B$\,$e$\,$s$\,$s$\,$e$\,$l$\,$ 
berechnete, von $\,$C$\,$l$\,$a$\,$r$\,$k$\,$ entdeckte Begleiter
des Sirius besitzt nach den Berechnungen yon 
$\,$A$\,$u$\,$w$\,$e$\,$r$\,$s$\,$ und $\,$P$\,$e$\,$t$\,$e$\,$r$\,$$\,$s$\,$ 
eine ausserordentlich grosse Masse ($\sfrac{1}{2}$ bis $\sfrac{2}{3}$
der Siriusmasse selbst). Gleichwohl erscheint er uns nur
als Stern etwa 9.\ Gr\"osse, seine Helligkeit ist also gegen
die seines Hauptsternes eine sehr geringe, nach 
$\,$S$\,$c$\,$h$\,$\"o$\,$n$\,$f$\,$e$\,$l$\,$d \footnotemark)$^{\it 
[K.S.:\ orig.\ footn.]}\,$
\renewcommand{\thefootnote}{}
\footnotetext{1)$^{\it [K.S.:\ orig.\ footn.]}\,$ 
$\,$S$\,$c$\,$h$\,$\"o$\,$n$\,$f$\,$e$\,$l$\,$d,
Die dunklen Fixsternbegleiter. Mannheimer Verein f\"ur Naturkunde.
30.\ Jahresbericht. (K.S.: cf.\ \citep{1868schoenfeld})}
\renewcommand{\thefootnote}{[\alph{footnote}\alph{footnote}]}nur 
ca.\ $\sfrac{l}{1000}$ von der des Sirius.
Oben wurde f\"ur Strahlen
mittlerer Wellenl\"ange (f\"ur das Natriumlicht) gefunden, dass
bei Abschw\"achung des Lichtes von der Intensit\"at 1 bis zur
Intensit\"at $\sfrac{1}{33}$ die Fortpflanzungsgeschwindigkeit sich noch
nicht um $\sfrac{1}{500\ 000}$ \"andert. Gehen wir von diesem Grenzwerthe
aus, d.\ h.\ nehmen wir an, dass sich bei Helligkeitsdifferenzen
von 33:1 die Fortpflanzungsgeschwindigkeit gerade noch um
diesen Betrag unterscheiden k\"onne, und dass derselbe proportional
mit dem Intensit\"atsunterschiede wachse, so w\"urde
das Licht des Begleiters in dem vorliegenden Beispiele in
jeder Secunde um $30 \times 0,6$ oder 18 km hinter dem Lichte
des Hauptsternes zur\"uckbleiben. Die Entfernung des Siriussystems
betr\"agt in Lichtzeit etwa 30 Jahre. Nehmen wir
an, es seien genau 30 Jahre, so w\"urde das Licht des Begleiters:
\begin{eqnarray*}
\frac{300\ 000}{299\ 982}\ \times\  30 &=& 30,0018
\end{eqnarray*}
Jahre gebrauchen, um zu uns zu gelangen, d.\ h.\ von zwei
gleichzeitig ausgehenden Lichtstrahlen w\"urde der des Begleiters
immer nur etwa 0,7 Tag sp\"ater eintreffen, als der
vom Sirius selbst ausgesandte. Da aber die Umlaufszeit
beider K\"orper um ihren gemeinsamen Schwerpunkt etwa
50 Jahre betr\"agt, so ist jene Zeitdifferenz und die ihr entsprechende
Ortdifferenz zu vernachl\"assigen. Die gemachten
Annahmen sind jedenfalls sehr ung\"unstige; wir k\"onnen also
sicher bei allen Bahnbestimmungen von der etwa noch vorhandenen
Abh\"angigkeit der Fortpflanzungsgeschwindigkeit
von der Intensit\"at, deren Betrag unter den durch das Experiment
festgestellten Grenzen liegt, absehen.

\noindent
Das erhaltene Resultat ist ferner bei den Anwendungen
des $\,$D$\,$o$\,$p$\,$p$\,$l$\,$e$\,$r’schen 
Principes von Bedeutung, wo die relativen
Geschwindigkeiten der Gestirne in der Richtung der Sehlinie
aus Aenderungen in der mittleren Brechbarkeit isolirter
Spectrallinien bestimmt werden. Hier liegen in den einzelnen
F\"allen oft sehr grosse Helligkeitsunterschiede vor, und es
ist folglich f\"ur die Anwendbarkeit dieses Principes wichtig,
direct nachgewiesen zu haben, dass diese Helligkeitsunterschiede
nicht mit in Frage kommen.''

\renewcommand{\thefootnote}{[\alph{footnote}\alph{footnote}]}
\setcounter{footnote}{\value{extrafoot}}

\item[P.\ \refstepcounter{dummy}\pageref{einstein}:]\label{aeinstein}
Einstein (\citep{1969einstein}, letter 69, pp.\ 169/1970):
``Dein Versuch, die Quantenfrage des Feldes von einer neuen Seite
anzugreifen, hat mich sehr interessiert, aber nicht gerade 
\"uberzeugt. Ich glaube immer noch, da\ss\ die 
Wahrscheinlichkeits-Interpretation trotz ihres gro\ss en Erfolges
keine gangbare M\"oglichkeit f\"ur die relativistische 
Verallgemeinerung bildet. Auch die Begr\"undung der Wahl einer 
Hamiltonfunktion f\"ur das elektromagnetische Feld mit der 
Analogie zur spe\-ziel\-len Relativit\"atstheorie hat mich nicht 
\"uberzeugt. Ich f\"urchte, da\ss\ wir alle die wirkliche L\"osung
dieses harten Problems nicht erleben werden.''

\item[P.\ \refstepcounter{dummy}\pageref{born}:]\label{aborn}
Born (\citep{1969einstein}, pp.\ 170/1971):
``Zweierlei Einw\"ande hatte Einstein gegen meine Ideen:
Der erste beruhte auf seiner Ablehnung der 
Wahrscheinlichkeits-In\-terpretation der Quantenmechanik. Dabei geht es
um eine Prinzipienfrage des Denkens, von der noch mehr die Rede sein
wird. Er traf die von Infeld und mir entworfene Theorie eigentlich
nicht, da wir selber mit der Einf\"ugung derselben in die 
Quantenmechanik nicht fertig wurden; er verurteilte unsere 
Bem\"uhungen in dieser Richtung als grunds\"atzlich falsch. Der
zweite Einwand Einsteins bezog sich auf unsere urspr\"ungliche,
\frq klassische\flq\
Feldtheorie, die in sich geschlossen und widerspruchsfrei
war. Sie beruhte auf folgender Analogie: In der speziellen
Relativit\"atstheorie wird die kinetische Energie eines
Teilchens, die in der klassischen Mechanik dem Quadrat der 
Geschwindigkeit proportional ist, durch einen etwas komplizierteren
Ausdruck dargestellt; f\"ur Geschwindigkeiten, die klein gegen die 
des Lichtes sind, geht er in den klassischen Ausdruck \"uber,
weicht aber davon ab, wenn die Geschwindigkeit der des Lichtes 
nahe kommt. In der Maxwellschen Elektrodynamik ist die 
Ener\-giedichte quadratisch in den Feldst\"arken; ich
ersetzte sie durch einen allgemeinen Ausdruck, der in den
klassischen \"ubergeht, wenn die Feldst\"arken klein gegen eine 
\frq absolute Feldst\"arke\flq\
sind, aber davon abweichen, wenn das nicht 
der Fall ist. Daher ergab sich ganz von selbst, da\ss\ die Gesamtenergie
des Feldes einer Punktladung endlich ist, w\"ahrend sie im 
Maxwellschen Feld unendlich gro\ss\ wird. Das absolute Feld mu\ss\
als eine neue Naturkonstante angesehen werden.

Diese Analogie-Konstruktion fand Einstein nicht \"uberzeugend. Infeld
und mir schien sie lange anziehend. Wir gaben die Theorie aus ganz
anderen Gr\"unden auf, n\"amlich weil es uns nicht gelang, sie mit den 
Prinzipien der Quanten-Feldtheorie in Einklang zu bringen.

Jedenfalls war dieser Ansatz der erste Versuch, die Schwierigkeiten
in der Mikro-Physik durch eine nicht-lineare Theorie zu beheben.
Heisenbergs Theorie der Elementarteilchen, die heute viel von sich 
reden macht, ist auch nicht-linear. Aber das geh\"ort nicht hierher.''

\item[P.\ \refstepcounter{dummy}\pageref{heisenberg2}:]\label{aheisenberg2}
Heisenberg (\citep{2001rechenberg}, p.\ 126):
``Das Thema der vorliegenden Arbeit geht auf eine Frage zur\"uck, die
mir Kollege Debye stellte.''

\item[P.\ \refstepcounter{dummy}\pageref{heisenberg3}:]\label{aheisenberg3}
Heisenberg (\citep{1985vonmeyenn}, p.\ 331,
letter of June 17, 1934 to N.\ Bohr):
``Debye hatte die Idee, da\ss\ die Sonnenkorona
durch diese Streuung von Licht entsteht; zu dieser These w\"urde der 
obige Wert [$(e^2/\hbar c)^4 (\hbar/m c)^2$ f\"ur den 
Wirkungsquerschnitt] ausgezeichnet passen. Aber, wie gesagt, man
mu\ss\ noch abwarten, ob alle Rechnungen in Ordnung sind.''

\item[P.\ \refstepcounter{dummy}\pageref{heisenberg4}:]\label{aheisenberg4}
Heisenberg (\citep{1985vonmeyenn}, letter [374]
of June 16, 1934, pp.\ 331-333, specifically p.\ 332):
``Dieser Wert scheint gut zu der Idee von Debye zu passen, 
da\ss\ die Sonnenkorona durch diese Streuung von Licht an Licht entsteht.''

\item[P.\ \refstepcounter{dummy}\pageref{pauli1}:]\label{apauli1}
Pauli (\citep{1993vonmeyenn}, letter [423b]
of December 5, 1935, pp.\ 777/778, specifically p.\ 778):
``Das ,,Euler-Kockel''-Problem f\"ur kurze Wellen 
sieht mir im Moment etwas tr\"ube aus.''

\item[P.\ \refstepcounter{dummy}\pageref{pauli2}:]\label{apauli2}
Pauli (\citep{1993vonmeyenn}, letter [425c]
of January 20, 1936, pp.\ 780-782, specifically p.\ 782):
``Das Problem der Streuung von Licht an Licht bei kurzen Wellen 
habe ich nun, nach Besprechung mit Oppenheimer, definitiv als 
zu kompliziert aufgegeben.''

\item[P.\ \refstepcounter{dummy}\pageref{pauli3}:]\label{apauli3}
Pauli (\citep{1985vonmeyenn}, letter [412] of June 15, 1935, pp.\ 402-405,
specifically p.\ 402):
``Weisskopf ist am Delbr\"uckschen Problem
und die dabei auftretenden Fragen der Subtraktions\-physik werden sehr
unsch\"on.''

\item[P.\ \refstepcounter{dummy}\pageref{akhiezer}:]\label{aakhiezer}
Akhiezer (\citep{1937akhiezer}, p.\ 567):
``\begin{otherlanguage}{russian}
Эффективный поперечник рассеяния
имеет максимум при
\end{otherlanguage}
%{\cyrrm \protect{\`{E}}ffektivny\u i poperechnik rasseyaniya
%imeet maksimum pri} 
$\hbar\omega\sim m c^2$.'' 
%[\`{E}ffektivny\u\i\ poperechnik rasseyaniya
%imeet maksimum pri $\hbar\omega\sim m c^2$].''

\end{itemize}

\pagebreak
\refstepcounter{asubsection}
\refstepcounter{subsection}
\label{appbrillouin}
\section*{Appendix \Alph{asubsection}}
\addcontentsline{toc}{subsection}{Appendix \Alph{asubsection}}

Extract from the book {\it Ondes et Mouvements} (1926) by L.\ de Broglie
(\citep{1926debroglie}, Chap.\ XI, Sec.\ 2, pp.\ 96-98. English
transl.: K.S.. The English translation is appended after the original 
French text.):\\[0.1cm]

\noindent
\begin{otherlanguage}{french}
``{\bf 2.\ Chocs entre atomes de radiation.} --
Examinons une autre question assez curieuse : deux qanta de lumi\`{e}re
peuvent-ils \'{e}changer de l'\'{e}nergie par choc, autrement dit peuvent-ils 
changer de fr\'{e}quence \`{a} la suite d'une collision? L'exp\'{e}rience 
n'a rien r\'{e}v\'{e}l\'{e} de semblable et un tel ph\'{e}nom\`{e}ne est 
tout \`{a} fait \'{e}tranger aux th\'{e}ories classiques. Cependant, il 
existe peut-\^{e}tre une raison de le croire possible au moins en principe. 
Dans la derni\`{e}re partie de ce livre, je
montrerai qu'\`{a} la suite 
des travaux de MM.\ Bose et Einstein et des miens, il est l\'{e}gitime de 
consid\'{e}rer le rayonnement noir comme un gaz d'atomes de lumi\`{e}re;
des raisonnements statistiques appliqu\'{e}s \`{a} nos conceptions
ondulatoires nous conduiront \`{a} la loi Planck, mais elles ne nous
montreront pas par quel m\'{e}canisme cette r\'{e}partition des quanta
entre le diverses valeurs de l'\'{e}nergie se trouve r\'{e}alis\'{e}e et
maintenue. Il parait assez tentant de supposer que l'\'{e}quilibre
r\'{e}sulte des \'{e}changes d'\'{e}nergie et de quantit\'{e} de 
mouvement entre quanta dues \`{a} leur interaction mutuelle, 
\`{a} leurs chocs au sens le plus g\'{e}neral du mot.

Peut-\^{e}tre est-il done int\'{e}ressant d'\'{e}tudier un peu
ce genre de chocs.

Il serait facile de r\'{e}soudre le probl\`{e}me g\'{e}n\'{e}ral
de la rencontre de deux quanta de fr\'{e}quences diff\'{e}rentes comme
nous l'avons fait pour l'effet Compton.

Pour ne pas multiplier les formules, je me contenterai d'envisager
un cas tr\`{e}s simple. Supposons qu'\`{a} l'aide de collimateurs,
on fasse se croiser \`{a} angle droit deux faisceaux monochromatiques
de m\`{e}me fr\'{e}quence, aucune condition de coh\'{e}rence n'\'{e}tant
d'ailleurs exig\'{e}e. Repr\'{e}sentons sch\'{e}matiquement ces deux
faisceaux par des lignes droites AB et ${\rm A}^\prime {\rm B}^\prime$
({\it fig.\ 9}) se croissant en O.

\nopagebreak{
\begin{center}
\hspace{-0.2cm}Fig.\ 9.\\[0.2cm]

\noindent
%\begin{tikzpicture}(140,140)[>=stealth]
\begin{tikzpicture}[>=stealth]
%\draw [help lines] (0,0) grid (5,5);
\draw [->](0,2.5)--(5,2.5);
\draw [->](2.5,5)--(2.5,0);
\draw[dashed] (0.5,4.5)--(4.5,0.5);
\draw[dashed] (2.2,2.5) arc (180:90:0.3cm);
\draw[dashed] (2.5,2.2) arc (270:360:0.3cm);
\draw (-0.05,2.25) node {\sf A};
\draw (2.8,4.95) node {${\sf A}^\prime$};
\draw (4.98,2.25) node {\sf B};
\draw (2.8,-0.05) node {${\sf B}^\prime$};
\draw (4.8,0.45) node {\sf M};
\draw (0.95,4.5) node {${\sf M}^\prime$};
\draw (2.3,2.3) node {\sf 0};
\end{tikzpicture}

\end{center}}

Observons, s'il en existe, la lumi\`{e}re diffus\'{e}e au point O
dans la direction OM bissectrice de l'angle $\widehat{{\rm BOB}^\prime}$.
Si, par suite d'un choc, un quantum est diffus\'{e} suivant OM,
l'autre quantum sera diffus\'{e} suivant ${\rm OM}^\prime$. Soient $\nu$ la
fr\'{e}quence initiale, $\nu_1$ et $\nu_2$ le fr\'{e}quences des quanta
diffus\'{e}s vers M et vers ${\rm M}^\prime$.
Les \'{e}quations de conservation s'\'{e}crivent
\begin{eqnarray*}
h\nu_1\ +\ h\nu_2 &=&2\ h\nu ,\\[0.3cm]
\frac{h\nu_1}{c}\ -\ \frac{h\nu_2}{c}&=&2\ \frac{h\nu}{c}\ \cos 45^\circ\ ;
\end{eqnarray*}
d'o\`{u}
\begin{eqnarray*}
\nu_1&=&\nu (1+\cos 45^\circ)\ =\ 1,7\ \nu ,\\[0.3cm]
\nu_2&=&0,3\ \nu.
\end{eqnarray*}

Si la longueur d'onde des faisceaux employ\'{e}s \'{e}tait 
$\lambda = o^\mu,68$, la radiation observ\'{e}e en M correspondrait 
\`{a} $\lambda_1 = \frac{0,68}{1,7} = o^\mu,4$. En visant le point
de croisement de deux faisceaux rouges, on recueillerait de la
lumi\`{e}re violette. Ce serait un beau ph\'{e}nom\`{e}ne!
Je ne sais s'il existe et, en ce cas, s'il serait observable, mais,
si un jour il \'{e}tait d\'{e}cel\'{e}, sa place serait toute
marqu\'{e}e dans l'ensemble de nos nouvelles vues th\'{e}oretiques
sur les radiations.''\\
\end{otherlanguage}

\noindent
English translation:\\[0.1cm]

\noindent
``{\bf 2.\ Collisions between atoms of radiation.} --
We are considering another interesting problem: Can two light 
quanta exchange energy in a collision, or speaking differently, 
can they change the frequency in course of a collision?
Experience does not reveal anything similar and such a phenomenon 
is rather alien to classical theories. However, there may be a 
reason for considering this to be possible, at least in principle. 
In the last part of this book, I will show, as a result of work done by
Messrs.\ Bose and Einstein and myself, that it is legitimate to 
consider the black radiation in a gas of atoms of light; 
statistical reasoning applied to our wave conceptions leads
us to the Planck law, but it does not explain by what mechanism 
this redistribution of quanta among the different energy levels 
is achieved and maintained. It seems quite tempting to assume that 
the equilibrium results from exchanges of energy and 
momentum among quanta as a result of their mutual interactions, 
their collisions in the most general sense of the word.

Consequently, it is perhaps interesting to somewhat investigate 
this kind of collisions.

It is easy to solve the problem of the encounter of two quanta with 
different frequencies the same way as we did for the case of the 
Compton effect. In order to not multiply formulae, I will only 
consider a very simple case.

Consider, using collimators, intersecting at a right angle 
two monochromatic beams of the same frequency, incidentally, no coherence 
condition is being required. We represent the two beams schematically by
means of the two perpendicular lines AB and ${\rm A}^\prime {\rm B}^\prime$ 
({\it fig.\ 9}) intersecting each other in O.

\begin{center}
Fig.\ 9. (K.S.: see above)
\end{center}

We observe whether any light scatters from the point O into the direction 
OM bisecting the angle $\widehat{{\rm BOB}^\prime}$. 
If, through a collision, a quantum is 
scattered along OM, the other quantum is scattered along ${\rm OM}^\prime$. Let 
$\nu$ be the initial frequency, and $\nu_1$ and $\nu_2$ the frequencies 
of the quanta scattered towards M and towards ${\rm M}^\prime$. The
equations of conservation are
\begin{eqnarray*}
h\nu_1\ +\ h\nu_2 &=&2\ h\nu ,\\[0.3cm]
\frac{h\nu_1}{c}\ -\ \frac{h\nu_2}{c}&=&2\ \frac{h\nu}{c}\ \cos 45^\circ\ ;
\end{eqnarray*}
wherefrom
\begin{eqnarray*}
\nu_1&=&\nu (1+\cos 45^\circ)\ =\ 1,7\ \nu ,\\[0.3cm]
\nu_2&=&0,3\ \nu.
\end{eqnarray*}

If the wavelength of the beams used is $\lambda = o^\mu,68$,
the radiation observed in M would correspond to 
$\lambda_1 = \frac{0,68}{1,7} = o^\mu,4$.
Aiming at the crossing point with two red beams, one would obtain violet light.
This would be a beautiful phenomenon! I do not know if it exists, 
and in that case, whether it is observable, but if one day it was detected 
it would assume a distinguished place within the framework of our new 
theoretical ideas on radiation.''\\

\pagebreak
\refstepcounter{asubsection}
\newcounter{aequation}
\renewcommand{\theequation}{\mbox{\Alph{asubsection}.\arabic{aequation}}}
\refstepcounter{subsection}
\label{appmirror}
\section*{Appendix \Alph{asubsection}}
\addcontentsline{toc}{subsection}{Appendix \Alph{asubsection}}

On the experimental result of Banwell and Farr, Proc.\ Roy.\ Soc.\ London 
\hfill\ \linebreak
{\bf A 175}(1940)1 \citep{1940banwell}:\\

\begin{figure}[ht]
\hspace{1.5cm}
\includegraphics{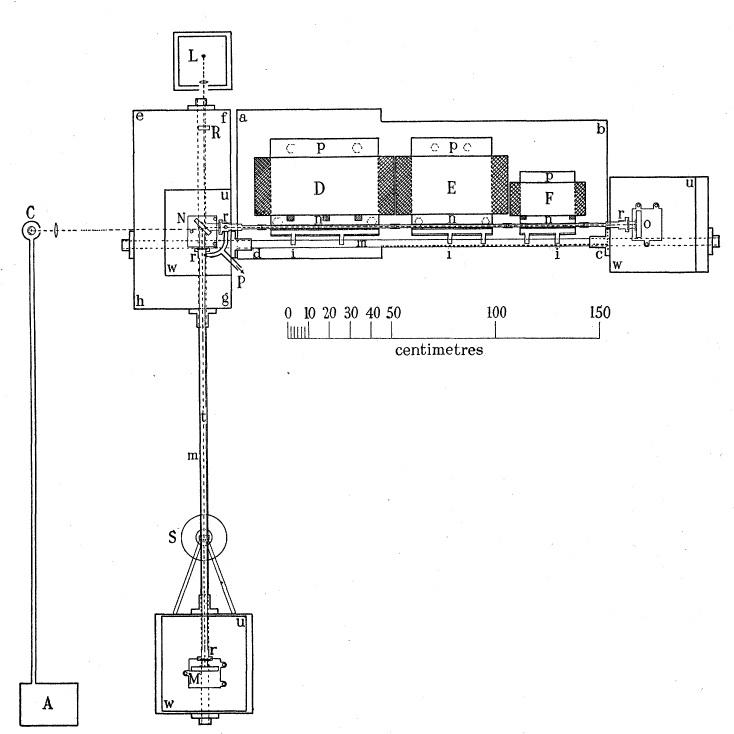}
\caption{\label{banwellfarrfig}Top view of the Michelson interferometer
used by Banwell and Farr (redisplay of 
Fig.\ 1 of \citep{1940banwell}, p.\ 3). $L$ is the light source
while observation is made at $C$ (photoelectric cell). $D$, $E$, $F$ are the 
electromagnets of the arm subjected to the magnetic field, $M$, $O$
are mirrors (silvered flats) while $N$ is a half-silvered flat.}
\end{figure}

\noindent
Given the present-day interest in the type of experiment Banwell and
Farr reported in 1940 \citep{1940banwell} it seems to be interesting
to contemplate the question what influence might have led to the 
non-null result (which Banwell and Farr themselves -- see quote 
in subsec.\ \ref{expconst} in the main text --
considered as likely to be erroneous). In the following we will attempt
to hypothesize about a possible reason for their non-null result for the
refractive index of the vacuum in a magnetic field which is
much larger than the present-day expectation based on the 
Euler-Kockel-Heisenberg effective Lagrangian of QED (and of opposite sign). 
Reading the carefully written article by Banwell and Farr one notices 
that they say very little about one important optical element used 
in their experimental set-up, namely, the mirrors employed in the 
Michelson interferometer. They are only characterized as optical 
flats, silvered or half-silvered (p.\ 1, begin of the second paragraph;
pp.\ 3/4).
The point is that two of the mirrors used by Banwell and Farr in the 
experiment are subject -- at least -- to magnetic stray fields: one mirror
(mirror $O$, cf.\ fig.\ 1 on p.\ 3 of \citep{1940banwell}
redisplayed in our Fig.\ \ref{banwellfarrfig}) at the end 
of the arm of the Michelson interferometer placed in the strong magnetic 
field, and a semi-transparent mirror
(mirror $N$, cf.\ Fig.\ \ref{banwellfarrfig})
used to split the initial light beam into the two arms of the
Michelson interferometer. 
Both mirrors are just located a few 
decimeters away from the coils generating the strong magnetic field.
This aspect of the experimental set-up is
neither mentioned nor discussed by Banwell and Farr.
Let us shortly consider the physical situation
at the mirror (mirror $O$) at the end of the arm of the Michelson
interferometer placed in the strong magnetic field (The
semi-permeable mirror
$N$ is also subjected to stray fields but the light in both arms
of the Michelson interferometer interacts with $N$ in a fairly 
symmetric manner and the mirror $N$ can, therefore, be left outside
the current consideration.). The light from this
arm falls perpendicularly onto the mirror and is perpendicularly
being reflected at the silver layer (by the way, whose thickness is not 
given in the article by Banwell and Farr).
In this metallic reflection (by a metallic medium with 
the complex refractive index
$n = n^\prime + i\ n^{\prime\prime} = n^\prime (1 + i \kappa)$)
the reflected light beam experiences a phase shift $\varphi$ 
relative to the incident light beam given by the 
equation\footnote{\citep{1894drude}, p.\ 86, eq.\ (40); for a modern discussion
see, e.g., \citep{2002dressel}, sec.\ 2.4, p.\ 35, eq.\ (2.4.14);
N.B.\ For simplicity we neglect in the present discussion the 
influence of any intermediate glas layer.}:
\refstepcounter{aequation}
\begin{eqnarray}
\label{drude}
\tan\varphi&=&
\frac{2\; n^\prime \kappa}{{n^\prime}^2 + {n^\prime}^2 \kappa^2 - 1}\ .
\end{eqnarray}
For massive silver at room temperature 
and light of $\lambda = 546.1$ nm (equivalent to a photon energy 
of 2.27 eV) one has
$n^\prime = 0.17$, $n^\prime \kappa = 3.31$ 
(\citep{1985lynch}, Table IX, p.\ 356, interpolated by means of 
\citep{2013RID}) resulting in a phase
shift $\varphi$ of about $0.59$ rad (corresponding to $34^\circ$). 
Now, once the refractive
index of the mirror material (in this case, silver) changes the 
phase shift $\varphi$ also experiences a corresponding change 
$\Delta\varphi$ to some $\bar{\varphi} = \varphi + \Delta\varphi$.
Once this change in the mirror properties (and, consequently,
the change in the phase shift at the mirror) occurs only for one of 
the mirrors at the end of the two arms of the Michelson interferometer
the interferometric pattern observed in the experiment will also
change. If this phenomenon is not properly taken care of the
corresponding change in the interferometric pattern can lead
to a misinterpretation of results. As we will detail further
below it seems to be conceivable that the change in the mirror properties
(mirror $O$) effected by the stray magnetic field may have
led to the non-null result reported by Banwell and Farr.\\

In their Michelson interferometer experiment, Banwell and Farr have
interpreted the observed change in the interference pattern once
the magnetic field acting on one interferometer arm had been switched
on as resulting from a changed vacuum velocity of light 
$\bar{c} = f \bar{\lambda}$ ($f$ is the frequency of the monochromatic
light wave). Hypothesize now that the observed change in the
interference pattern in fact completely originated from a change in 
the refractive index of the silver layer of the mirror $O$
placed in the (stray) magnetic field. Then, according to 
the interpretation applied by Banwell and Farr the optical path
of length $l$ ($= 2\times 1.17\ {\rm m} = 2.34\ {\rm m}$) 
placed in the magnetic field
would be stretched out by $n$ wave cycles of wavelength  
$\bar{\lambda}$ (We also have taken into account the phase shift/delay 
under reflection.)
\refstepcounter{aequation}
\begin{eqnarray}
\label{pathlength1}
l&=&n\ \bar{\lambda}\ -\ \frac{\varphi}{2\pi}\ \bar{\lambda}\ .
\end{eqnarray}
while in reality it should be written as 
\refstepcounter{aequation}
\begin{eqnarray}
\label{pathlength2}
l&=&n\ \lambda\ -\ \frac{\bar{\varphi}}{2\pi}\ \lambda\ .
\end{eqnarray}
From the equations (\ref{pathlength1}), 
(\ref{pathlength2}) follows that the assumed change in the vacuum velocity of 
light $\Delta c = \bar{c} - c$ can be expressed as a change in the 
phase shift under reflection according to the equation
\refstepcounter{aequation}
\begin{eqnarray}
\label{phaseshift1}
\Delta\varphi&=&- 2\pi\ \frac{l}{\lambda}\ \frac{\Delta c}{\bar{c}}\\[0.3cm]
\refstepcounter{aequation}\nonumber\\[-0.5cm]
\label{phaseshift2}
&\simeq&- 2\pi\ \frac{l}{\lambda}\ \frac{\Delta c}{c} 
\end{eqnarray}
where equation (\ref{phaseshift2}) represents the relation
at leading order of the small quantity $\Delta c/c$.
Inserting now the values given by Banwell and Farr into eq.\
(\ref{phaseshift2}) one finds the following. 
The change in the interference
pattern observed in the Michelson interferometer which has been
interpreted by Banwell and Farr as a change in the 
(vacuum) velocity of light in a magnetic field could haven arisen
equally well from a different source in the experimental apparatus.
Let us assume that in the presence of the applied strong magnetic field
stray fields might have modified the reflective properties of 
the silver mirror (mirror $O$) at the end of the arm of the 
Michelson interferometer
placed in the magnetic field. Then, a corresponding change
in the phase shift under reflection experienced by the light beam 
propagating in this arm of the Michelson interferometer 
of the size
\refstepcounter{aequation}
\begin{eqnarray}
\label{phaseshift3}
\Delta\varphi&\simeq&- 2\pi\ (4.7 \pm 2.6) \cdot 10^{-3}. 
\end{eqnarray}
might have modified the interference pattern equivalently. From
eq.\ (\ref{drude}) it is clear that any change in the (complex)
refractive index of the silver layer of the order of 
$\Delta\kappa/\kappa,\ \Delta n^\prime/n^\prime \sim 10^{-2}$
will result in the right order of magnitude of 
change in the phase shift under reflection
given by eq.\ (\ref{phaseshift3}).\\

Let us ponder now the question if it is conceivable that a
magnetorefractive effect might result in a change 
in the refractive index of the silver mirror that has the right magnitude
(We primarily imagine that the magnetic stray field is parallel
to the mirror surface.).
Unfortunately, there is apparently little information available 
in the literature in this respect. 
Gostishchev and Sobol' \citep{1983gostishchev} estimate for low 
temperatures, however, and a magnetic field of the order of $10^5$ oersted 
(corresponding to a magnetic induction of 10 T in vacuo)
that for visible light the change in the reflection coefficient 
(compared to the situation with no magnetic field applied) 
for a metal with a closed Fermi surface is not larger than $10^{-3}$
(the same applies to the refractive index). On one hand side,
the estimate of Gostishchev and Sobol' is somewhat smaller than required
by eq.\ (\ref{phaseshift3}) and on the other hand, they consider 
the case of low temperatures while the experiment of Banwell and
Farr has been performed at room temperature.
To broaden the perspective, let
us mention that the Drude model of metals (free gas of 
conduction electrons) relates the conductivity in a metal to its
refractive index. Consequently, any study of the phenomenon of 
magnetoresistance in a metal has in principle its bearing on 
its magnetorefractive properties. Of course, this connection can 
only be of heuristic value as the reliability of the Drude model  
is restricted to the domain of infrared light, while the experiment of
Banwell and Farr has been performed using visible light.
Magnetoresistance effects are primarily related to two phenomena:
the deviation of the Fermi surface from a spherical shape 
(cf., e.g., \citep{1960ziman}, Sec.\ 12.3, p.\ 490-495, specifically 
p.\ 494) and the influence of size effects as they occur in thin films.
For experimental results see, e.g., \citep{1964fink,1967chopra}.\\

To conclude, in principle magnetorefractive effects in silver mirrors
exist which might lead to a change in the phase shift under reflection
in accordance with eq.\ (\ref{phaseshift3}). The present state of 
research reflected in the literature, however, is insufficient to 
decide the question whether these magnetorefractive effects might have 
the right order of magnitude to explain the non-null result
by Banwell and Farr.\\

\pagebreak
\refstepcounter{asubsection}
\newcounter{bequation}
\renewcommand{\theequation}{\mbox{\Alph{asubsection}.\arabic{bequation}}}
\refstepcounter{subsection}
\label{appstress}
\section*{Appendix \Alph{asubsection}}
\addcontentsline{toc}{subsection}{Appendix \Alph{asubsection}}

Some formulas for the electromagnetic field strength tensor:

\refstepcounter{bequation}
\begin{eqnarray}
\label{fieldstrengthtensor}
F^{\mu\nu}&=&\left(
\begin{array}{*{4}{c}}
0&-\frac{1}{c} E_1&-\frac{1}{c} E_2&-\frac{1}{c} E_3\\[0.3cm] 
\frac{1}{c} E_1&0&- B_3&\ \ B_2\\[0.3cm]
\frac{1}{c} E_2&\ \ B_3&0&- B_1\\[0.3cm]
\frac{1}{c} E_3&- B_2&\ \ B_1&0
\end{array}
\right)
\end{eqnarray}
The (symmetric) stress-energy-momentum tensor of the free
electromagnetic field reads: 
\refstepcounter{bequation}
\begin{eqnarray}
\label{stressenergytensor}
T^{\mu\nu}&=&\ \frac{\displaystyle 1}{\displaystyle \mu_0}
\left[F^{\mu\alpha} F^\nu_{\ \alpha}\ -\ 
\frac{\displaystyle 1}{\displaystyle 4}\ g^{\mu\nu}
F^{\alpha\beta} F_{\alpha\beta}\right]
\end{eqnarray}
After some calculation one finds:
\refstepcounter{bequation}
\begin{eqnarray}
\label{stressenergytensorsquared}
T^{\mu\nu} T_{\mu\nu}&=&2\ \epsilon_0^2
\left[\left({\bf E}^2 - c^2 {\bf B}^2\right)^2\ +\ 
2\ c^2 \left({\bf E}{\bf B}\right)^2\right]\\[0.3cm]
\refstepcounter{bequation}\nonumber\\[-0.5cm]
\label{stressenergytensorsquaredb}
&=&8\ {\cal L}_0^2\ +\ \frac{\displaystyle 4}{\mu_0^2}
\left({\bf E}{\bf B}\right)^2\\[0.3cm]
\refstepcounter{bequation}\nonumber\\[-0.5cm]
\label{stressenergytensorsquaredc}
&=&\ 4\left[\left(T_{00}\right)^2
\ -\ \frac{\displaystyle 1}{\displaystyle c^2}\ {\bf S}^2\right]
\ =\ - 4\  T^{0\nu} T_{0\nu}
\end{eqnarray}
with the free field Lagrange density
\refstepcounter{bequation}
\begin{eqnarray}
\label{freelagrangianappc}
{\cal L}_0&=&-\frac{\displaystyle 1}{\displaystyle 4\mu_0}\ 
F^{\mu\nu} F_{\mu\nu}\ =\
\frac{\displaystyle 1}{\displaystyle 2}\ \epsilon_0 
\left({\bf E}^2 - c^2 {\bf B}^2\right)
\ =\ \frac{\displaystyle 1}{\displaystyle 2}\
\left(\epsilon_0 {\bf E}^2 
- \frac{\displaystyle 1}{\displaystyle\mu_0} {\bf B}^2\right)
\end{eqnarray}
and 
\refstepcounter{bequation}
\begin{eqnarray}
\label{energy}
T_{00}&=&\frac{\displaystyle 1}{\displaystyle 2}\ \epsilon_0 
\left({\bf E}^2 + c^2 {\bf B}^2\right)
\ =\ \frac{\displaystyle 1}{\displaystyle 2}\
\left(\epsilon_0 {\bf E}^2 
+ \frac{\displaystyle 1}{\displaystyle\mu_0} {\bf B}^2\right)\ .
\end{eqnarray}
One recognizes from the eqs.\ (\ref{stressenergytensorsquaredb}),
(\ref{stressenergytensorsquaredc}), that the square of the Poynting vector 
${\bf S} = \frac{1}{\mu_0}\ {\bf E}\times {\bf B}$ is related 
to the field invariant $\left({\bf E}{\bf B}\right)^2$ (By virtue of
the Lagrange identity for three-dimensional vectors ${\bf a}$,
${\bf b}$, ${\bf c}$, ${\bf d}$: 
$({\bf a}\times {\bf b})\cdot ({\bf c}\times {\bf d}) =
({\bf a}\cdot {\bf c})({\bf b}\cdot {\bf d}) -
({\bf b}\cdot {\bf c})({\bf a}\cdot {\bf d})$.).

\pagebreak
\refstepcounter{asubsection}
\renewcommand{\theequation}{\mbox{\Alph{asubsection}.\arabic{equation}}}
\refstepcounter{subsection}
\label{appheisenberg}
\section*{Appendix \Alph{asubsection}}
\setcounter{equation}{0}
\addcontentsline{toc}{subsection}{Appendix \Alph{asubsection}}

Reminiscences of W.\ Heisenberg onto his conversations with
H.\ Euler concerning the scattering of light by light in advance 
of the thesis research of the latter
\citep{1969heisenberg}, chap.\ 13, pp.\ 220-222, 225 (pp.\ 220-222, 225 of 
the reprint, pp.\ 160-162, 164 of the English translation. 
The English translation is appended after the original 
German text.):\\

\noindent
``So kam ich oft mit Euler zusammen, und wir berieten daher \"uber 
die m\"oglichen Konsequenzen der Diracschen Entdeckung und der 
Umwandlung von Energie in Materie.

\guillemotright Wir 
haben doch von Dirac gelernt\guillemotleft, so k\"onnte Euler etwa gefragt 
haben, \guillemotright da\ss\ ein Lichtquant, das an einem
Atomkern vorbeifliegt, sich dabei in ein Paar von Teilchen, 
ein Elektron und ein Positron, verwandeln kann.
Bedeutet das eigentlich, da\ss\ ein Lichtquant aus einem Elektron und 
einem Positron besteht? Dann wäre das Lichtquant so eine Art 
Doppelsternsystem, in dem Elektron und Positron umeinander kreisen. 
Oder ist das eine falsche anschauliche Vorstellung?\guillemotleft

\guillemotright Ich 
glaube nicht, da\ss\ ein solches Bild viel Wahrheit enth\"alt. 
Denn aus diesem Bild w\"urde man doch
schlie\ss en, da\ss\ die Masse eines solchen Doppelsterns nicht viel 
kleiner sein sollte als die Summe der Massen der beiden Teile, aus denen 
es besteht. Und man k\"onnte auch nicht einsehen, warum dieses System sich
immer mit Lichtgeschwindigkeit durch den Raum bewegen mu\ss . Es k\"onnte 
doch auch irgendwo zur Ruhe kommen.\guillemotleft

\guillemotright Was soll man aber dann über das Lichtquant in diesem 
Zusammenhang sagen?\guillemotleft

\guillemotright Man darf vielleicht sagen, da\ss\ das Lichtquant virtuell 
aus Elektron und Positron besteht. Das Wort 
\guilsinglright virtuell\guilsinglleft\
deutet an, da\ss\ es sich um eine M\"oglichkeit handelt. 
Der eben ausgesprochene Satz behauptet dann nur, da\ss\
das Lichtquant sich eben in gewissen Experimenten m\"oglicherweise 
in Elektron und Positron zerlegen l\"a\ss t. Mehr nicht.\guillemotleft

\guillemotright Nun
k\"onnte in einem sehr energiereichen Sto\ss\ ein Lichtquant 
doch vielleicht auch in zwei Elektronen und zwei Positronen verwandelt 
werden. W\"urden Sie dann sagen, da\ss\ das Lichtquant virtuell 
auch aus diesen vier Teilchen besteht?\guillemotleft

\guillemotright Ja, ich glaube, das w\"are konsequent. Das Wort 
\guilsinglright virtuell\guilsinglleft, 
das die M\"oglichkeit bezeichnet, erlaubt ja die Behauptung, da\ss\ das 
Lichtquant virtuell aus zwei oder vier Teilchen besteht. Zwei verschiedene
M\"oglichkeiten schließen sich ja nicht aus.\guillemotleft

\guillemotright Aber was gewinnt man dann noch mit einem solchen
Satz?\guillemotleft\
wandte Euler ein. \guillemotright Dann kann man doch gleich
sagen, da\ss\ jedes Elementarteilchen virtuell aus irgendeiner 
beliebigen Zahl von anderen Elementarteilchen
besteht. Denn bei sehr ener\-giereichen Sto\ss prozessen wird schon 
irgendeine beliebige Zahl von Teilchen entstehen k\"onnen. Das ist 
doch fast keine Aussage mehr.\guillemotleft

\guillemotright Nein, 
so beliebig sind Zahl und Art der Teilchen denn doch nicht. 
Nur solche Konfigurationen von Teilchen werden als m\"ogliche Beschreibung 
des einen darzustellenden Teilchens in Betracht kommen, die die gleiche
Symmetrie haben wie das urspr\"ungliche Teilchen. Statt Symmetrie 
k\"onnte man noch genauer sagen: Transformationseigenschaft
gegen\"uber solchen Operationen, unter denen die Naturgesetze 
unver\"andert bleiben. Wir haben doch schon aus der Quantenmechanik 
gelernt, da\ss\ die station\"aren Zustände eines Atoms durch
ihre Symmetrieeigenschaften charakterisiert sind. So wird es 
eben auch bei den Elementarteilchen sein, die ja auch station\"are 
Zust\"ande aus Materie sind.\guillemotleft
 
Euler war noch nicht so recht zufrieden. \guillemotright Das wird doch 
reichlich abstrakt, was Sie jetzt sagen. Es k\"ame
wohl mehr darauf an, sich Experimente auszudenken, die anders ablaufen, 
als man bisher angenommen h\"atte, und zwar deshalb anders, weil 
die Lichtquanten virtuell aus Teilchenpaaren bestehen. Man w\"urde doch
vermuten, da\ss\ man wenigstens qualitativ vern\"unftige Resultate 
bekommt, wenn man das Bild vom Doppelsternsystem einen Moment ernst nimmt 
und fragt, was nach der fr\"uheren Physik daraus folgen sollte.
Zum Beispiel k\"onnte man sich f\"ur das Problem interessieren, 
ob zwei Lichtstrahlen, die sich im leeren Raum kreuzen, wirklich so 
ungehindert durcheinander hindurchgehen, wie man bisher immer angenommen 
hat und wie die alten Maxwellschen Gleichungen es fordern. Wenn in dem 
einen Lichtstrahl virtuell, das heißt als M\"oglichkeit, Paare von 
Elektronen und Positronen vorhanden sind, so könnte der andere 
Lichtstrahl doch an diesen Teilchen gestreut werden; also m\"u\ss te es 
eine Streuung von Licht an Licht geben, eine gegenseitige
St\"orung der beiden Lichtstrahlen, die man aus der Diracschen Theorie 
ausrechnen k\"onnte und die auch experimentell zu beobachten 
w\"are.\guillemotleft

\guillemotright Ob man so etwas beobachten kann, h\"angt nat\"urlich davon ab, 
wie gro\ss\ diese gegenseitige St\"orung ist. Aber
Sie sollten ihre Wirkung unbedingt ausrechnen. Vielleicht finden die 
Experimentalphysiker dann auch Mittel und Wege, sie 
nach\-zu\-wei\-sen.\guillemotleft

\guillemotright Eigentlich finde ich diese Philosophie des 
\guilsinglright als ob\guilsinglleft, 
die hier betrieben wird, doch sehr merkw\"urdig. Das
Lichtquant verh\"alt sich in vielen Experimenten so, 
›als ob‹ es aus einem Elektron und einem Positron best\"unde. 
Es verh\"alt sich auch manchmal so, 
\guilsinglright als ob\guilsinglleft\ es aus zwei oder 
noch mehr solchen Paaren best\"unde. Scheinbar ger\"at man in eine 
ganz unbestimmte verwaschene Physik hinein. Aber man kann aus der
Diracschen Theorie doch die Wahrscheinlichkeit daf\"ur, da\ss\ ein 
bestimmtes Ereignis eintritt, mit gro\ss er Genauigkeit berechnen, 
und die Experimente werden das Ergebnis schon bestätigen.\guillemotleft

\noindent
$\ldots$

\noindent
Euler berechnete zusammen mit einem anderen Mitglied meines Seminars, 
Kockel, die Streuung von Licht an Licht, und obwohl der experimentelle 
Nachweis hier nicht so direkt gef\"uhrt werden konnte,
besteht heute wohl kein Zweifel mehr daran, daß es die von Euler und 
Kockel behauptete Streuung wirklich gibt.''\\

\noindent
English translation:

\noindent
``And so we would talk about atomic physics instead and, in particular,
about the possible consequences of Dirac's discovery, and the 
transformation of energy into matter.

``Dirac has shown,'' Euler said,``that when a light quantum flies past
an atomic nucleus, it may change into a pair of particles --an electron
and a positron. Does this mean the light quantum itself consists of an
electron and a positron? In that case, it would be a kind of double star,
one in which the electron and positron revolve about each other. Or is
this a false picture?''

``I don't think it's very convincing. You see, the mass of a double
star cannot be much smaller than the sum of the masses of its constituent
parts. Nor would it necessarily have to move through space with the 
velocity of light. There is no reason why it should never come to rest.''

``But, what {\it can} we say about the light quantum in this context?''

``Perhaps that is is virtually made up of an electron and a positron.
The word `virtually' means that we are dealing with a possibility. In 
that case, my assertion means no more than that the light quantum may,
in certain experiments, split up into an electron and a positron--noting 
more.''

``Well, in a very high-energy impact, a light quantum might easily be
transformed into two electrons and two positrons. Does that mean that 
it is virtually made up of these four particles as well?''

``Yes, I believe that would be the consistent view. Since the term
`virtually' denotes possibilities, we are entitled to say that the
light quantum is virtually made up of two of four particles.
Two different possibilities do not necessarily exclude each other.''

``But what is the advantage of this sort of assertion?'' Euler
asked. ``We might equally well say that every elementary particle
is virtually made up of any number of other particles. After all,
any number of particles might be created during high-energy collisions.
In that case our statement says very little indeed.''

``I should not put it like that, for, you see, the number and
type of particles are not as arbitrary as all that. Only such
configurations may be considered possible descriptions of a particular
particle as have the same symmetry as the original particle.
Instead of `symmetry,' we might say more precisely: transformation
characteristics under operations that leave the physical laws
unchanged. After all, quantum mechanics has taught us that the 
stationary states of an atom are characterized by their symmetries.
Things are probably much the same with elementary
particles, which, when all is said and done, are simply stationary
states of matter.''

Euler was still not fully satisfied. ``The whole argument is a bit
too abstract for my liking. What we probably ought to be doing
is to think up experiments that would lead to unexpected results,
and this precisely because light quanta are virtually made up of
pairs of particles. It seems reasonable to assume that we should
obtain at least qualitatively satisfactory results if we stuck to the
model of the double star, and asked what conclusions orthodox
physics would draw. For instance, we could investigate whether
or not two light rays crossing in empty space really pass through
each other with no interaction, as we have assumed until now,
and as the old Maxwellian equations demand. If pairs of electrons
and positrons are virtually present, i.e., contained as a
possibility, in a light ray, then another light ray ought to be
scattered by these particles; hence there would be deflection of
light by light, that is, an interaction of the two light rays. We
ought to be able to demonstrate its existence and to calculate its
extent from Dirac's theory.''

``Whether of not we would be able to observe it would, of 
course, depend on the intensity of the mutual perturbations. But
by all means calculate the effect. Perhaps experimental physicists
will then discover ways and means of corroborating your results.''

``I really think this whole `as if' philosophy is terribly odd.
The light quantum is said to behave in some experiments as if it 
consisted of an electron and a positron. But at other times it
apparently behaves as if it consisted of two or more such pairs. 
The result is a wishy-washy kind of physics. And yet we can use
Dirac's theory to calculate the probability of a certain event with
great precision, and find that experiments will confirm the results.''

\noindent
$\ldots$

\noindent
Meanwhile Euler, together with another of my pupils, 
B.\ Kockel, determined the scattering of light by light, and 
although no experimental verification was possible here,
there is little doubt today that the scattering effect they deduced
is a fact.''

\pagebreak
\refstepcounter{asubsection}
\renewcommand{\theequation}{\mbox{\Alph{asubsection}.\arabic{equation}}}
\refstepcounter{subsection}
\label{apppauli}
\section*{Appendix \Alph{asubsection}}
\setcounter{equation}{0}
\addcontentsline{toc}{subsection}{Appendix \Alph{asubsection}}

\noindent
Extract from a letter of September 27, 1935 from W.\ Pauli to V.\ Weisskopf
(Reprint\-ed as letter [421a] in \citep{1993vonmeyenn}, pp.\ 769-771,
specifically p.\ 770. English transl.: K.S.. The English translation 
is appended after the original German text.):\\

\setcounter{extrafoot}{\value{footnote}}
\renewcommand{\thefootnote}{\arabic{footnote}}
\setcounter{footnote}{2}

\noindent
``$\ldots$Pryce$^{[\ldots]}$ und ich haben beschlossen, 
wenigstens zu versuchen, die H\"aufigkeit des
Streuprozesses zweier Lichtquanten aneinander {\it ohne} 
die Euler-Kockelsche Einschr\"ankung
hinsichtlich Kleinheit der Lichtfrequenzen gegen $m c^2/h$ auszurechnen.

Ich bin mir dar\"uber klar, da\ss\ das Problem nicht leicht ist. 
Nicht nur sind die auftretenden Integrale von einem komplizierteren 
Typus als das erw\"ahnte Heisenberg-Serbersche Integral, sondern vor 
allem bekommt man zun\"achst die praktisch unendlich gro\ss e Zahl von 
$6 \times 24 = 144$ Termen (nach Euler-Kockel).
Dennoch scheint mir das Problem nicht so ganz hoffnungslos zu sein. 
Ich habe den Eindruck, da\ss\ die riesigen Koeffiziententabellen 
von Euler-Kockel nicht so sehr dem Subtraktionsformalismus als solchem 
zur Last zu legen sind als vielmehr dem Umstand, da\ss\ viel zu fr\"uh 
nach den $g$'s entwickelt wurde. Letzteres scheint
mir nicht nur in formaler Hinsicht ungeschickt, sondern es wird auch in
physikalischer Hinsicht durch die Beschr\"ankung auf kleine $v$ 
auch die \"Ahnlichkeit mit der Bornschen 
Theorie\footnote{$^{\it [K.S.:\ orig.\ footn.]}\,$
\label{footborn}Vgl.\ Born
und Infeld (1934). (K.S.: cf.\ \citep{1934born3})}$^{\it 
[K.S.:\ orig.\ }$ $^{\it footn.]}\,$ vorget\"auscht; 
aber diese liefert keine Paarerzeugung!
Das ist so wie wenn man eine Theorie h\"atte, die (f\"ur lange Wellen) 
Dispersion liefert, aber keine Absorption! Nach meiner Meinung haben 
Euler und Kockel in ihrer publizierten 
Note\footnote{$^{\it [K.S.:\ orig.\ footn.]}\,$ \label{footeuler}Euler und
Kockel (1934). (K.S.: cf.\ 
\citep{1935euler})}$^{\it [K.S.:\ orig.\ footn.]}\,$
ein viel zu starkes Gewicht gelegt auf den Vergleich mit
diesem Unget\"um von einer Pseudotheorie!
\setcounter{footnote}{300}

Nun aber zur\"uck zur mathematischen Seite des Problems. 
Was mir eine gewisse Hoffnung gibt, durchzukommen ist der Umstand, 
da\ss\ das Problem sich sehr ver\-ein\-fachen l\"a\ss t durch 
Einf\"uhrung eines speziellen Bezugsystems. Abgesehen von
dem singul\"aren Sonderfall, da\ss\ die beiden Lichtquanten 
im Anfangszustand exakt in derselben Richtung laufen 
(in welchem Sonderfall aus Ihnen sehr bekannten
Gr\"unden die Streuwahrscheinlichkeit ohnehin exakt verschwinden d\"urfte), 
kann man immer ein Bezugsystem einf\"uhren, wo 
$g_l + g_2 = 0$ (d.\ h.\ die Lichtquanten
laufen in entgegengesetzter Richtung aufeinander zu und ihre Frequenzen sind
gleich). Und in diesem Normalkoordinatensystem werden die Formeln viel
einfacher. Vielleicht gelingt es da, die Zahl $6 \times 24$ 
der Terme so weit zu reduzieren, da\ss\ man die Streuwahrscheinlichkeit 
wirklich ausrechnen kann (f\"ur beliebige
Frequenz der Lichtquanten). -- Zun\"achst rechnet Pryce allgemein 
die Spur aus (womit er fast fertig ist) und dann werden wir weiter
sehen, ob wir durchkommen. Princeton ist ein Ort, der sehr geeignet 
ist, um komplizierte Integrale auszurechnen. $\ldots$''\\

\noindent
English translation:

\noindent
``$\ldots$Pryce$^{[\ldots]}$ and I have decided to at least try
to calculate the probability of the scattering process 
of two quanta of light off each other {\it without} the
Euler-Kockel restricion with respect to the smallness of
the light frequency in comparison with $m c^2/h$.

\setcounter{footnote}{2}
I am aware of the fact that the problem is not an easy one.
Not only the integrals occurring are of a more complicated type
than the Heisenberg-Serber integral mentioned but primarily
one gets the practically infinitely large number of 
$6 \times 24 = 144$ terms (according to Euler-Kockel).
Despite this it seems to me that the problem is not
that hopeless. It is my impression that the huge coefficient
tables of Euler-Kockel are not that much due to the subtraction
formalism as such as rather due to the problem that much too
early an expansion in the $g$'s has been applied. To me, the latter 
does not only seem to be awkward but also in physical respect
by virtue of the restriction to small $v$
a similarity to the Born theory\footnote{$^{\it [K.S.:\ orig.\ footn.]}\,$ K.S.: 
see footnote \ref{footborn} on page 
\refstepcounter{dummy}\pageref{footborn}.}$^{\it [K.S.:\ orig.\ footn.]}\,$ is being 
pretended to exist; but it does not 
yield any pair creation! This is the same as if one would have a theory
that (for long waves) yields dispersion but not any absorption!
In my view, in their published note\footnote{$^{\it 
[K.S.:\ orig.\ footn.]}\,$ K.S.: 
see footnote \ref{footeuler} on page 
\refstepcounter{dummy}\pageref{footeuler}.}$^{\it [K.S.:\ orig.\ footn.]}\,$ 
Euler and Kockel have put much to strong emphasis 
on the comparison with this monstrosity of a pseudo-theory.
\setcounter{footnote}{300}

But now back to the mathematical side of the problem.
What provides me with some hope to get through is the fact that
the problem can considerably be simplified by introducing
a special reference system. Disregarding the singular special case
that two quanta of light in the initial state propagate in exactly 
the same direction (in which special case for reasons very well
known to you the scattering probability should vanish exactly anyway),
one can always introduce a reference system where 
$g_l + g_2 = 0$ (i.e., the quanta of light propagate in directions
opposite to each other and their frequencies are the same).
And in this system of normal coordinates the formulas are becoming
much easier. Perhaps one can succeed in reducing the number
$6 \times 24$ of terms that much that one can really calculate the 
scattering probability (for arbitrary frequency of the quanta of light).
-- First Pryce calculates the trace in general (what 
he almost is done with) and then we will see further if we get through.
Princeton is a place which is very much suited for calculating
complicated integrals. $\ldots$''

\renewcommand{\thefootnote}{[\alph{footnote}\alph{footnote}]}
\setcounter{footnote}{\value{extrafoot}}

\pagebreak
\refstepcounter{asubsection}
\renewcommand{\theequation}{\mbox{\Alph{asubsection}.\arabic{equation}}}
\refstepcounter{subsection}
\label{appweisskopf}
\section*{Appendix \Alph{asubsection}}
\setcounter{equation}{0}
\addcontentsline{toc}{subsection}{Appendix \Alph{asubsection}}

Historic recollections by V.\ F.\ Weisskopf related to the 
articles \citep{1936kemmer,1936weisskopf}
(The passages are quoted from \citep{1965weisskopf} with kind
permission of the American Institute of Physics.):\\

\noindent
``{\bf Weisskopf}: $\ldots$
Now I come to the work about the vacuum polarization, which I published 
in the Danish Academy\footnote{K.S.: see ref.\
\citep{1936weisskopf}.}$^{[footn.\ K.S.]}$. 
I very rarely complain about not getting enough 
recognition because I think I've received in my life more recognition 
than I really deserved, but for this paper I don't get enough recognition. 
In my opinion, this paper is really the beginning of re-normalization; 
and when you read it you'll find, it. The first purpose of the work, 
and the one for which it is perhaps best known, was the recalculation 
in a very much simpler way of the Euler-Heisenberg vacuum polarization 
for slowly varying fields, which Wick mentioned today. It was really 
only a recalculation, although with very nice methods suggested by Pauli. 
This is why I didn't publish it in Zeitschrift f\"ur Physik; I thought 
I should have something in the Danish Academy, and so I published this 
there because it was really only a simplification. 

However, this same paper contains a study in which I was not very sure 
of myself, which is also why I published it in the Danish Academy, but 
which excited me very, very much. The study was to show that all the 
infinities that come about in calculation are in fact infinities that 
you cannot measure, namely, infinities of charge, infinities of mass, 
and infinities of what I called there the ``dielectric constant of the 
vacuum.'' As it says there explicitly, one could assume that the total 
result is given by nature and one can forget about these infinities. 
What is given there is the recipe for re-normalization. Again I say 
that if I had had my Sommerfeld training I could have done much more 
with this. In fact I used rather primitive methods there to prove my 
point and perhaps that is another reason that the work was not too well 
known. But you could directly quote from there a recipe for re-normalization. 
There is a paper by Dirac in which he says the same thing for the mass, 
I think, 
but I knew that already. I directly say there are three magnitudes which 
are essentially nowadays the three ``Z's'', the three infinities; and I say 
in there that these are the three infinities; but is characteristic that 
were they finite, you wouldn't be able to notice them.

This is why you can forget them, and I say this explicitly in this paper. 
I did this work rather independently; Pauli was of course interested in 
it and he advised me in many things. By the way, I was always a sloppy 
man, and this paper was one that has the greatest number of calculating 
mistakes of any paper ever written; it's terrific what's wrong in there, 
but in principle it was right. $\ldots$\\

\noindent
$\ldots$\\

\noindent
{\bf Heilbron}:
There's just one paper, isn't there, in the Danish Academy?

\noindent
{\bf Kuhn}:
Let's get that chronology right on the tape now.

\noindent
{\bf Weisskopf}:
The papers are, in order, as follows: '34, the self energy paper ordered by
Pauli\footnote{K.S.: see ref.\ \citep{1934weisskopf}.}$^{[footn.\ K.S.]}$; 
\ \ \ then the boson paper, the Klein-Gordon equation with 
Pauli\footnote{K.S.: see ref.\ \citep{1934pauli}.}$^{[footn.\ K.S.]}$;
\ \ \ and in '36, the polarization of the vacuum and the re-normalization 
proposal\footnote{K.S.: see ref.\ \citep{1936weisskopf}.}$^{[footn.\ K.S.]}$, 
which was
done partially with Pauli but was really written in Copenhagen. That's why it
was published here.\\

\noindent
$\ldots$\\

\noindent
{\bf Weisskopf}: $\ldots$ I did another paper 
with Kemmer on the scattering of light by light. I think this 
was in a letter to Nature\footnote{K.S.: see ref.\
\citep{1936kemmer}.}$^{[footn.\ K.S.]}$. We connected the scattering of light 
by light with the Delbr\"uck scattering. Today it's a triviality; 
one light quantum is replaced by the Coulomb field, but at that 
time it was not so trivial. That thing was in fact the beginning 
of this later paper on the vacuum polarization where very similar 
problems are treated. That just shows that these were the things 
one was worrying about. No, I know what it was about; it had deeper 
significance, that letter to the editor. Euler and Kockel at that 
time, under Heisenberg, calculated the scattering of light by 
light, but had to do a lot of subtracting because there were a 
great many terms that were infinite. They did this in the usual 
clever way and got the result. And Kemmer and I showed that you 
can do the calculation without raking any subtractions, because 
you can show that it is equivalent to the Delbr\"uck scattering, 
replacing one light quantum by the Coulomb field, and the 
Delbr\"uck scattering doesn't diverge. This is a special case 
of what one does now every day if one calculates these things. $\ldots$
''
\pagebreak
\refstepcounter{asubsection}
\renewcommand{\theequation}{\mbox{\Alph{asubsection}.\arabic{equation}}}
\refstepcounter{subsection}
\label{appkemmer}
\section*{Appendix \Alph{asubsection}}
\setcounter{equation}{0}
\addcontentsline{toc}{subsection}{Appendix \Alph{asubsection}}

Historic recollections by N.\ Kemmer \citep{1983kemmer}, pp.\ 171-172
(English transl.: K.S.. The English translation is appended after 
the original German text.):\\

\noindent
``Wei\ss kopf war damals auch in
Z\"urich und arbeitete am Selbstenergieproblem
in diesem neuen Rahmen\footnote{\label{footkemmer}K.S.: i.e., 
hole theory, electron-positron theory 
\citep{1983kemmer}, p.\ 171.}$^{[footn.\ K.S.]}$. Er
fand, da\ss\ die neue Theorie immer noch
an Divergenzen litt, die aber nur schwach
(logarithmisch) waren [11]\footnote{\label{footkemmer11}K.S.: Reference 
in the original article \citep{1983kemmer}: [11]
{\it Weisskopf, V.}, Z.\ Phys.\ {\bf 98}, 27 (1934),
Berichtigung [correction] {\bf 90}, 817 (1934).
(K.S.: cf.\ our ref.\ \citep{1934weisskopf})}$^{[footn.\ K.S.]}$. 
Ich beteiligte mich ein wenig an diesen Rechnungen
und erlebte es mit, als Nachricht von Heisenberg
ankam, da\ss\ zwei seiner Sch\"uler,
Euler und Kockel, aufgrund dieser Theorie
die ,,Streuung von Licht an Licht'' berechnet
hatten [12]\footnote{\label{footkemmer12}K.S.: Reference 
in the original article \citep{1983kemmer}: 
[12] {\it Euler, H.}, u. {\it B.\ Kockel},
Naturwissenschaften {\bf 23}, 246 (1935). (K.S.: cf.\ our ref.\ 
\citep{1935euler}. For the letters of W.\ Heisenberg to 
W.\ Pauli reporting about the  progress of this
calculation see letter [374], pp.\ 331-333, and 
letter [393], pp.\ 358-360, in \citep{1985vonmeyenn}.)}$^{[footn.\ K.S.]}$. 
Dies war ein wohldefinierter,
kleiner, aber im Prinzip beobachtbarer
Effekt, der als Abweichung von
den Maxwellschen Gleichungen beschrieben
werden konnte. Die Autoren mu\ss ten
bis zur vierten Ordnung in der St\"orungstheorie
rechnen und ohne vollst\"andige
Rechtfertigung gewisse divergente Integrale
vernachl\"assigen. Die Berechnung
war enorm lang, obwohl sie nur f\"ur einige
Spezialf\"alle durchgef\"uhrt wurde. Heisenberg
war aber \"uberzeugt, da\ss\ trotz der
Unvollkommenheit der Theorie das Endresultat
relativistisch invariant sein m\"usse.
Unter dieser Annahme konnten Euler
und Kockel ihr Ergebnis in der Form eines
Zusatzes zur Lagrange-Funktion des
Maxwell-Feldes angeben 
\begin{eqnarray*}
\label{appf2}
L&=&\frac{1}{8\pi}\ \left(E^2 - B^2\right)\ +\
\frac{C}{360\pi^2}\ \left[\left(E^2 - B^2\right)^2
+ 7 \left(\vec{E}\vec{B}\right)^2\right]\hspace{2cm}\ (2)
\end{eqnarray*}
(in den damals \"ublichen Einheiten), wo
\begin{eqnarray*}
\label{appf2b}
C&=&\left(\frac{e^2}{\hbar c}\right)^2\cdot
\left(\frac{\hbar}{m c}\right)^3\cdot\frac{1}{m c^2}\ .
\end{eqnarray*}
Unter Wei\ss kopfs Leitung und mit Dr.\
Guido Ludwig als Mitarbeiter berechnete
ich einen verwandten Effekt [13]\footnote{\label{footkemmer13}K.S.: Reference 
in the original article \citep{1983kemmer}: [13] 
{\it Kemmer, N.}, Helv.\ Phys.\ Acta {\bf 10}, 112 (1937)
u. {\it G.\ Ludwig},
Helv.\ Phys.\ Acta {\bf 10}, 182 (1937). (K.S.: cf.\ 
our refs.\ \citep{1937kemmer1,1937kemmer2}.)}$^{[footn.\ K.S.]}$, n\"amlich
die (Delbr\"uck-) Streuung von Licht an einem
elektrostatischen Potential. In der
Abwesenheit invarianter Rechenmethoden
war es nicht offensichtlich, da\ss\ unser
Resultat mit dem Euler-Kockelschen
\"ubereinstimmen w\"urde. Ich fand aber genau
dieselben Koeffizienten wie in (2). So
lernte ich, da\ss\ man Invarianzargumenten
trauen konnte -- auch f\"ur eine ganz unvoll\-kommene
Theorie!''\\

\noindent
English translation:

\noindent
``At that time, Wei\ss kopf also stayed in Zurich
and worked on the self-energy problem within this new
framework\footnote{K.S.: see footnote \ref{footkemmer} on page
\refstepcounter{dummy}\pageref{footkemmer}.}$^{[footn.\ K.S.]}$. 
He discovered that the new theory still suffered from divergencies, 
however weak (logarithmic) ones [11]\footnote{K.S.:
see footnote \ref{footkemmer11} on page 
\refstepcounter{dummy}\pageref{footkemmer11}.}$^{[footn.\ K.S.]}$. 
I took some part in these calculations and witnessed when 
news from Heisenberg arrived that two of his students,
Euler and Kockel, had calculated the ``scattering of light 
by light''  on the basis of this theory
[12]\footnote{K.S.: see footnote \ref{footkemmer12} on page 
\refstepcounter{dummy}\pageref{footkemmer12}.}$^{[footn.\ K.S.]}$. 
This was a well-defined, small, but 
observable in principle effect which could be described as a
deviation from the Maxwell equations. The authors had
to perform calculations up to the fourth order of 
perturbation theory and to neglect certain integrals without complete
justification for doing so. The calculation was enormously long
although it was done for some special cases only. However,
Heisenberg was convinced that the final result had to be 
relativistically invariant despite the imperfection of the theory. 
Relying on this assumption, Euler and Kockel could present their
result in terms of an addition to the Lagrange function of 
the Maxwell field
\begin{eqnarray*}
\label{appf2en}
L&=&\frac{1}{8\pi}\ \left(E^2 - B^2\right)\ +\
\frac{C}{360\pi^2}\ \left[\left(E^2 - B^2\right)^2
+ 7 \left(\vec{E}\vec{B}\right)^2\right]\hspace{2cm}\ (2)
\end{eqnarray*}
(in the units customary then) where
\begin{eqnarray*}
\label{appf2ben}
C&=&\left(\frac{e^2}{\hbar c}\right)^2\cdot
\left(\frac{\hbar}{m c}\right)^3\cdot\frac{1}{m c^2}\ .
\end{eqnarray*}
Under the guidance of Wei\ss kopf and with Dr.\
Guido Ludwig as collaborator I calculated a related effect
[13]\footnote{K.S.: see footnote \ref{footkemmer13} 
on page \refstepcounter{dummy}\pageref{footkemmer13}.}$^{[footn.\ K.S.]}$, namely
the (Delbr\"uck-) scattering of light by an electrostatic
potential. In the absence of invariant calculational methods
it was not obvious that our result would agree with the one
by Euler-Kockel. However, I obtained the same coefficients as in (2).
This way I learned that one can trust invariance arguments --
also for a fairly imperfect theory!''

\pagebreak
\refstepcounter{asubsection}
\renewcommand{\theequation}{\mbox{\Alph{asubsection}.\arabic{equation}}}
\refstepcounter{subsection}
\label{appakhiezer}
\section*{Appendix \Alph{asubsection}}
\setcounter{equation}{0}
\addcontentsline{toc}{subsection}{Appendix \Alph{asubsection}}

Historic recollections by A.\ I.\ Akhiezer 
about his kand.\ diss.\ (Ph.D. Thesis) research leading to the publications
\citep{1936akhiezer}, \citep{1937achieser}
(The following passage is quoted from \citep{1994akhiezer}, 
pp.\ 36-38; for a somewhat 
shorter version also see \citep{1993akhiezer}, pp.\ 107-109, 
pp.\ 77-78 of the English translation.):

\subsubsection*{{\rm ``}Photon-photon scattering}

Alter I passed the {\it theorminimum} Landau gave me a
research subject. I was to study the scattering of light by
light. Landau had earlier given this subject to
Rosenkevitch, who was another of his students. He and
Landau were going to investigate the scattering of light
by light in the low-frequency domain, where the photon
energy is much less than the mass of the electron. 
(Electron-positron pair production makes the electron mass an
important parameter here.) But Rosenkevitch failed to
solve this problem, and besides, there soon appeared a
remarkable paper by Werner Heisenberg and his student
Hans Euler that gave a complete solution. Landau was
upset that the problem ``got away'' from him.

That happened just when I was passing the
{\it theorminimum}. So Landau decided to test my abilities
with this difficult problem of photon-photon scattering,
but this time in the high-frequency domain where the
photon energy exceeds the electron mass. The phenomenon
in question was a fourth-order effect in perturbation
theory. To calculate the scattering probability in that
approximation, one should really use Dirac's relativistic
theory of the electron. But at that time, perturbation
theory had been worked out only in nonrelativistic quantum
mechanics. It was a hard task to calculate the
scattering amplitude, because one had to take into account
numerous intermediate states without overlooking any.

Eventually I had the amplitude written down and
showed it to Landau. And that's when the first, and last,
blowup in my relations with Landau erupted. He didn't
like the nonrelativistic form of the probability amplitude.
Besides, it was written in terms of photon vector potentials
rather than the electromagnetic fields themselves. Therefore
the expression was not gauge invariant. So Landau
started getting angry, but he couldn't make my expression
relativistically and gauge invariant either. Nonetheless I
strongly objected to his assertion that it couldn't be done
within the existing perturbation theory.

This conversation had become rather unpleasant by
the time Rosenkevitch came in. Sizing up the situation,
Rosenkevitch took two candlesticks standing on the desk,
gave one to me and the other to Landau, and said, ``Now
fight it out.'' Landau burst out laughing and said: ``The
hell with you. Do the calculations the way you want.'' I
understood that he couldn't deny my general formula. So
I decided that since the formula was correct, it should
lead eventually to a properly invariant expression.\\

\subsubsection*{`The best gammists'}

It was then that Pomeranchuk began to work with me.
The calculations were horrifying exercises in the manipulation
of gamma matrices. Evgenii Lifshitz, coauthor of
the famous Landau textbooks, joked that Pomeranchuk
and I were ``the best gammists in the Soviet Union.'' (The
joke, which loses something in translation, depends on the
similarity of ``gamma'' to an indecorous Russian word for
excrement.)

Landau was insistent that we check the gauge invariance
of our result. To that end we replaced the vector
potential with the field itself. We got 144 terms, which
had to sum to zero. Chuk (that's what we used to call
Pomeranchuk) and I held our breaths as we did the sum.
I can't remember any other moment when I've been as
happy as I was when I finally saw that the sum did indeed
vanish. We immediately ran to Landau. He was happy
too. Soon we had completed the calculation, getting a
relativistically invariant expression for the scattering cross
section in the high-frequency domain. We also succeeded
in removing possible divergences simply by exploiting the
gauge invariance of the amplitude.

We described our work to Victor Weisskopf, who had
recently arrived in Kharkov. He was very pleased with
it. At Landau's seminar Weisskopf reported on his own
work on the nonlinear electrodynamics of the vacuum, and
he gave me the galley proofs of his paper. Though he
obtained the same result Heisenberg and Euler had gotten
earlier, Weisskopf's method was beautiful. Landau was
full of praise.

Landau suggested that we publish our results in the
British journal {\it Nature}. Pomeranchuk and I wrote a brief
paper, and Fritz Houtermans, a German \'{e}migr\'{e} at our
institute, promptly translated it into English. (For more
on the peripatetic Houtermans see the letter by Victor
Frenkel on page 104 and the article by Iosif Khriplovich
in {\sc PHYSICS TODAY}, July 1992, page 29.) Without bothering
to get official permission, we took the article to the post
office, and soon it was published under our three names.

Then Pomeranchuk and I concerned ourselves with
the problem of the coherent scattering of gamma rays in
the Coulomb field of the nucleus. This was another
problem to be solved in the framework of Dirac theory.
In 1937 there was a nuclear physics conference in Moscow.
Pauli came, and Landau introduced Pomeranchuk and me
to him. We familiarized Pauli with our work on light-light
scattering and gamma scattering off nuclei. He approved.
It was all a brilliant success for Landau's ``school,'' because
it indicated that we were dealing with the most important
theoretical problems of the day.''

\pagebreak
\refstepcounter{asubsection}
\renewcommand{\theequation}{\mbox{\Alph{asubsection}.\arabic{equation}}}
\refstepcounter{subsection}
\label{apptisza}
\section*{Appendix \Alph{asubsection}}
\setcounter{equation}{0}
\addcontentsline{toc}{subsection}{Appendix \Alph{asubsection}}

Historic recollections by L\'{a}sl\'{o} Tisza (MIT, Cambridge, MA, USA)
about the kand.\ diss.\ (Ph.D. Thesis) 
research by A.\ I.\ Akhiezer leading to the publications
\citep{1936akhiezer}, \citep{1937achieser}
(The following passage is quoted from \citep{2003tisza},
pp.\ 307-308.):

\subsubsection*{{\rm ``}III.\ SCATTERING OF LIGHT BY LIGHT}

P.\ A.\ M.\ Dirac published in 1928 a remarkable paper in which
he established a relativistically invariant form for quantum
mechanics. This theory was at first plagued by a curious difficulty,
it called for negative energy states for the electron. Dirac later
showed that instead of reversing the sign of the energy, one might 
reverse the sign of its charge. At first this did not improve the
situation much until the discovery of the positron in 1932 changed 
the difficulty into a most remarkable prediction of any theory. The
use of the Dirac equation still was not obvious. The $4\times 4$
gamma matrices involved in this equation made manipulations difficult.
There was an International Theoretical Conference in May 1934 at the 
UFTI\footnote{K.S.: UFTI = Ukrainian Physico-Technical 
Institute.}$^{[footn.\ K.S.]}$  
where these questions were discussed. By 1935 pair production problems
were rather standard although they remained labor-intensive.

Shura\footnote{K.S.: Shura = A.\ I.\ Akhiezer.}$^{[footn.\ K.S.]}$ 
and myself completed
about the same time our theoretical minimum. We turned to Landau to start
us off in research. The Dirac perturbation theory was ready for use, 
mainly as a result of the so-called Casimir method for handling gamma
matrices; we all started along this line. Shura got the assignment of the
\guillemotleft scattering of light by light\guillemotright . This was a very
difficult problem of fourth-order perturbation calculus. After a while
he and Chuk\footnote{K.S.: Chuk = I.\ Ya.\ Pomeranchuk.}$^{[footn.\ K.S.]}$ 
decided to join 
forces. This was a fortunate decision. I vividly remember the two
sitting side by side at two desks, working through long
sequences of calculations. They were doing the same step
independently and proceeded to the next step only after their 
results checked. They reminded me of the famous cartoon 
characters: Max and Moritz by Wilhelm Busch, two mischievous boys,
one of them with a funny hairdo. This one was clearly Chuk. He was 
always full of ideas, be it something funny, or some important
physics. Landau said that Chuk reminded him of his younger self.
They had both striking ironical faces, but 
Dau\footnote{K.S.: Dau = L.\ D.\ Landau.}$^{[footn.\ K.S.]}$ 
was tall and Chuk was short and 
boyish. Shura has his own benign sense of humor. The two were
sitting at their desks, constantly joking and cursing while doing their
ghastly calculations.

Eventually they finished and convinced Dau that all was right.
(See Akhiezer's paper in Physics Today\footnote{K.S.: see ref.\
\citep{1994akhiezer} (and our App.\ \ref{appakhiezer}).}$^{[footn.\ K.S.]}$
 of 1994.) At the time the foremost expert Victor
Weisskopf was visiting and he endorsed the work as well. When the
secretary received the manuscript for typing, the title 
\guillemotleft Scattering of light by light\guillemotright\ provoked
her perceptive remark: 
\guillemotleft do chevo zhe dumalis!\guillemotright\
In free translation: 
\guillemotleft What will they dream up next?\guillemotright

His hard work paid off, Akhiezer passed his grueling test to
be ready to become Landau's successor in due time.''

\pagebreak
\renewcommand{\refname}{\Large Main list of references\phantomsection
\addcontentsline{toc}{section}{Main list of references}

\noindent
{\rm\small
\begin{minipage}[t]{15cm}
[For references in Cyrillic letters, 
we apply the (new) {\it Mathematical Reviews} transliteration
(transcription) scheme
(to be found at the end of index issues of {\it Mathematical Reviews}).
Georgian references are transliterated on the basis of ISO 9984.
Chinese references are transcribed using Hanyu Pinyin with tone marks.
For Japanese references the modified Hepburn transcription system is used.
Arabic and Persian references are romanized using the ALA-LC standard.]
\end{minipage}}\\[-0.3cm]}

\renewcommand{\refname}{\Large List of full author names\phantomsection
\addcontentsline{toc}{section}{List of full author names}}

\renewcommand{\refname}{\Large Further literature\phantomsection
\addcontentsline{toc}{section}{Further literature}}

\end{document}